\documentclass[nofootinbib,aps,prd,superscriptaddress,preprintnumbers,%
 showpacs,showkeys,floatfix,amssymb,amsfonts]{revtex4-1}  

\usepackage{graphicx}
\usepackage{bm}
\usepackage{amsmath}
\usepackage{amssymb}
\usepackage{amsfonts}
\usepackage{float}
\usepackage{dsfont}  
\usepackage{slashed}  
\usepackage{booktabs}
\usepackage{multirow}
\usepackage{subfigure}
\usepackage[sort&compress]{natbib}
\usepackage{ulem}
\usepackage{empheq}
\usepackage{color}

\newcommand{\be}{\begin{equation}}  
\newcommand{\ee}{\end{equation}}  
\newcommand{\beq}{\begin{eqnarray}} 
\newcommand{\eeq}{\end{eqnarray}}

\newcommand{\bea}{\begin{eqnarray}}
\newcommand{\eea}{\end{eqnarray}}

\DeclareRobustCommand{\eq}[1]{Eq.~\eqref{eq:#1}}

\DeclareRobustCommand{\eqs}[2]{Eqs.~\eqref{eq:#1} and \eqref{eq:#2}}

\DeclareRobustCommand{\eq}[1]{Eq.~(\ref{eq:#1})}
\DeclareRobustCommand{\eqs}[2]{Eqs.~(\ref{eq:#1}) and (\ref{eq:#2})}

\newcommand{\nn}{\nonumber}

\usepackage{xr-hyper}

\usepackage[svgnames,x11names,table]{xcolor}
\usepackage[colorlinks=true,linkcolor=MediumBlue,citecolor=MediumBlue,urlcolor=MediumBlue]{hyperref}
\usepackage{hyperref}

\begin{document}
\title{Generalized Parton Distributions from Lattice QCD \\[1ex] with Asymmetric Momentum Transfer: Unpolarized Quarks}
\author{Shohini Bhattacharya}
\email{sbhattach@bnl.gov}
\affiliation{Physics Department, Brookhaven National Laboratory, Upton, New York 11973, USA}
\author{Krzysztof Cichy}
\affiliation{Faculty of Physics, Adam Mickiewicz University, ul.\ Uniwersytetu Pozna\'nskiego 2, 61-614 Pozna\'{n}, Poland}
\author{Martha Constantinou}
\email{marthac@temple.edu}
\affiliation{Department of Physics,  Temple University,  Philadelphia,  PA 19122 - 1801,  USA}
\author{Jack Dodson}
\affiliation{Department of Physics,  Temple University,  Philadelphia,  PA 19122 - 1801,  USA}
\author{Xiang Gao}
\affiliation{Physics Division, Argonne National Laboratory, Lemont, IL 60439, USA}
\author{Andreas Metz}
\affiliation{Department of Physics,  Temple University,  Philadelphia,  PA 19122 - 1801,  USA}
\author{Swagato Mukherjee}
\affiliation{Physics Department, Brookhaven National Laboratory, Upton, New York 11973, USA}
\author{Aurora Scapellato}
\affiliation{Department of Physics,  Temple University,  Philadelphia,  PA 19122 - 1801,  USA}
\author{Fernanda Steffens}
\affiliation{Institut f\"ur Strahlen- und Kernphysik, Rheinische Friedrich-Wilhelms-Universit\"at Bonn,\\ Nussallee 14-16, 53115 Bonn}
\author{Yong Zhao}
\affiliation{Physics Division, Argonne National Laboratory, Lemont, IL 60439, USA}
%
\begin{abstract}
\,\\[3ex]
Traditionally, lattice QCD computations of generalized parton distributions (GPDs) have been carried out in a {\it symmetric} frame, where the transferred momentum is symmetrically distributed between the incoming and outgoing hadrons. However, such frames are inconvenient since they require a separate calculation for each value of the momentum transfer, increasing significantly the computational cost. In this work, by focusing on the quasi-distribution approach, we lay the foundation for faster and more effective lattice QCD calculations of GPDs exploiting {\it asymmetric} frames, with freedom in the transferred momentum distribution. An important ingredient of our approach is the Lorentz covariant parameterization of the matrix elements in terms of Lorentz-invariant amplitudes, which allows one to relate matrix elements in different frames. We also use this amplitude approach to propose a new definition of quasi-GPDs that is frame-independent and, more importantly, may lead to smaller power corrections in the matching relations to the light-cone GPDs. We demonstrate the efficacy of the formalism through numerical calculations using one ensemble of $N_f$=2+1+1 twisted mass fermions with a clover improvement. The value of the light-quark masses lead to a pion mass of about 260 MeV. Concentrating on the proton, and limiting ourselves to a vanishing longitudinal momentum transfer to the target, we extract the invariant amplitudes from matrix element calculations in both the symmetric and asymmetric frame, and obtain results for the twist-2 light-cone GPDs for unpolarized quarks, that is, $H$ and $E$. 
\end{abstract}
\maketitle


\section{Introduction}
Parton distribution functions (PDFs), which are measurable in processes like inclusive deep-inelastic lepton-nucleon scattering, are key objects containing information about the quark and gluon structure of strongly interacting systems~\cite{Collins:1981uw}.
They provide 1D images of hadrons by describing how the partons are distributed as a function of the momentum fraction $x$ they carry of the hadron's momentum.
PDFs are defined through matrix elements of bi-local quark or gluon operators, with the parton fields having a light-like separation and the operators evaluated for the {\it same} initial and final hadron state.
Generalized parton distributions (GPDs) are generalizations of the concept of PDFs in that the light-like parton operators are computed for {\it different} initial and final states~\cite{Mueller:1998fv, Ji:1996ek, Radyushkin:1996nd}.
GPDs therefore depend on the longitudinal momentum transfer $\xi$ and the invariant momentum transfer $t$ to the target, in addition to their dependence on the parton momentum fraction $x$.
While this makes them complicated multi-variable functions, the information encoded in GPDs is much richer than for PDFs.
In particular, they provide 3D images of hadrons~\cite{Burkardt:2000za, Ralston:2001xs, Diehl:2002he, Burkardt:2002hr}, give access to the angular momenta of partons~\cite{Ji:1996ek}, and have a relation to pressure and shear forces inside hadrons~\cite{Polyakov:2002wz, Polyakov:2002yz, Polyakov:2018zvc}.
The physics of GPDs has been discussed in various review articles~\cite{Goeke:2001tz, Diehl:2003ny, Ji:2004gf, Belitsky:2005qn, Boffi:2007yc, Guidal:2013rya, Mueller:2014hsa, Kumericki:2016ehc}.

Experimental information on GPDs can be obtained from hard exclusive scattering processes such as deep virtual Compton scattering~\cite{Mueller:1998fv, Ji:1996ek, Radyushkin:1996nd, Ji:1996nm, Collins:1998be} and hard exclusive meson production~\cite{Radyushkin:1996ru, Collins:1996fb, Mankiewicz:1997uy}.
But extracting GPDs from such reactions in a model-independent manner is very complicated, mainly because in the observable quantities, like the Compton form factors, the momentum fraction $x$ is integrated over --- see~\cite{Bertone:2021yyz} for a recent discussion and detailed analysis of this issue. 
It is therefore very desirable to obtain information on GPDs from first principles in lattice QCD.
However, lattice QCD calculations of light-cone correlation functions like PDFs and GPDs are challenging due to the time dependence of those objects. 
As a result, for a long time lattice QCD calculations were limited to the lowest Mellin moments of the GPDs~\cite{Hagler:2003jd, QCDSF-UKQCD:2007gdl, Alexandrou:2011nr, Alexandrou:2013joa,Constantinou:2014tga}, with simulations at the physical point available only in recent years~\cite{Green:2014xba,Alexandrou:2017ypw,Alexandrou:2017hac,Hasan:2017wwt,Gupta:2017dwj,Capitani:2017qpc,Alexandrou:2018sjm,Shintani:2018ozy,Jang:2018djx,Bali:2018qus,Bali:2018zgl,Alexandrou:2019ali,Constantinou:2020hdm,Alexandrou:2022dtc}. Despite the progress, their dependence on $x$ remained elusive.  

The quasi-PDF approach, which was proposed in 2013~\cite{Ji:2013dva} and later developed into the large-momentum effective theory~\cite{Ji:2014gla,Ji:2020ect}, opens up the opportunity to directly compute the $x$-dependence of PDFs and related quantities. 
Quasi-PDFs are obtained from spatial equal-time operators that can be studied on Euclidean lattices.
They reduce to their corresponding light-cone PDFs when taking the limit $P^3 = |\vec{P}| \to \infty$ of the hadron momentum, prior to renormalization.
But for lattice QCD studies renormalization is carried out first and $P^3$ is finite, resulting in two sources of discrepancies between quasi-PDFs and light-cone PDFs: 
a different ultraviolet (UV) behavior, as well as higher-twist corrections that are suppressed by powers of ${1}/{P^3}$.
The UV disparities can be dealt with order by order through a matching procedure in perturbative QCD~\cite{Xiong:2013bka,Ma:2014jla, Izubuchi:2018srq}.
We note that other approaches for lattice QCD calculations of the $x$-dependence of light-cone correlation functions exist~\cite{Liu:1993cv,Braun:1994jq,Aglietti:1998ur,Detmold:2005gg,Braun:2007wv,Chambers:2017dov,Radyushkin:2017cyf,Radyushkin:2019mye,Ma:2014jla,Ma:2017pxb,Detmold:2021uru}, some of which are related to the quasi-PDF method.
Encouraging lattice QCD results using such methods were reported for PDFs, see e.g.~\cite{Lin:2014zya,Alexandrou:2015rja,Chen:2016utp,Alexandrou:2016jqi,Alexandrou:2017huk,Chen:2017mzz,Orginos:2017kos,Lin:2017ani,Alexandrou:2018pbm,Lin:2018pvv,Alexandrou:2018eet,Liu:2018uuj,Zhang:2018nsy,Sufian:2019bol,Alexandrou:2019lfo,Izubuchi:2019lyk,Joo:2019jct,Joo:2019bzr,Chai:2020nxw,Joo:2020spy,Bhat:2020ktg,Alexandrou:2020uyt,Alexandrou:2020qtt,Lin:2020ssv,Fan:2020nzz,Gao:2020ito,Lin:2020fsj,Karpie:2021pap,Alexandrou:2021oih,Egerer:2021ymv,HadStruc:2021qdf,Gao:2021dbh,Gao:2022iex,LatticeParton:2022xsd}, including higher twist \cite{Bhattacharya:2020xlt,Bhattacharya:2020jfj,Bhattacharya:2021boh,Bhattacharya:2021moj}, parton distribution amplitudes \cite{Zhang:2017zfe,Zhang:2017bzy,Bali:2018spj,Zhang:2020gaj,Hua:2020gnw,Detmold:2021qln,Hua:2022kcm,Gao:2022vyh} and even transverse-momentum dependent parton distributions \cite{Shanahan:2020zxr,Zhang:2020dbb,Li:2021wvl,Shanahan:2021tst,Schlemmer:2021aij,LPC:2022ibr}. Possible impact on phenomenological studies was also studied, see e.g.~\cite{Cichy:2019ebf,Bringewatt:2020ixn,DelDebbio:2020rgv}. 
The very dynamic progress in this field has been documented in a number of reviews~\cite{Cichy:2018mum,Ji:2020ect,Constantinou:2020pek,Cichy:2021lih,Cichy:2021ewm}.

Applications of these new developments in the case of GPDs are still somewhat sparse in comparison.
Nevertheless, we have seen results for matching~\cite{Ji:2015qla, Xiong:2015nua, Liu:2019urm, Radyushkin:2019owq, Ma:2022ggj}, model studies~\cite{Bhattacharya:2018zxi, Bhattacharya:2019cme, Ma:2019agv, Luo:2020yqj, Shastry:2022obb} and, in particular, the first pioneering lattice QCD calculations for the pion~\cite{Chen:2019lcm} and the nucleon~\cite{Alexandrou:2020zbe,Lin:2020rxa,Alexandrou:2021bbo,CSSMQCDSFUKQCD:2021lkf,Bhattacharya:2021oyr,Lin:2021brq}.
These results are very encouraging, demonstrating explicitly that GPDs can be obtained on the lattice.
But it is rather clear that the full mapping of GPDs with respect to their variables, in particular the momentum transfer $t$ and the skewness $\xi$, is very challenging and computationally much more demanding than for PDFs.
Among the reasons for this is that, so far, they have been computed in symmetric frames of reference, where the momentum transfer is equally split between the source and the sink.
Consequently, every value of the momentum transfer is obtained from a separate and costly calculation.
Here, for the first time, we consider asymmetric frames for the computation of GPDs.
As we will demonstrate below, this allows for more efficient calculations, since different momentum transfers can be obtained in a single calculation.

The paper is organized as follows.
In Section II, we discuss the kinematics of the symmetric and asymmetric frames for our study and how the two frames can be related through a Lorentz transformation.
For both a spin-0 and spin-$\frac{1}{2}$ target, we introduce and discuss the main theoretical tool of Lorentz-invariant amplitudes in terms of which matrix elements that define GPDs can be parameterized.
Based on those amplitudes we also propose a new, frame-independent definition of quasi-GPDs which, in comparison to previously used quasi-GPDs, may converge faster to their respective light-cone GPDs. 
Section III specializes on the Euclidean case and provides decompositions of lattice-calculable matrix elements in terms of these amplitudes, and our lattice setup.
Numerical results are shown in Section IV, where we make a detailed comparison of the symmetric and the asymmetric frames at various stages, in coordinate space and in momentum space.
For the proton and $\xi{ =} 0$ we show, in particular, numerical results for the invariant amplitudes and the twist-2 light-cone GPDs $H$ and $E$.
Section V concludes and discusses future prospects.

\section{Strategy of frame transformation}
\label{sec:strategy}
\subsection{Symmetric and asymmetric frames}
\label{s:frames}

The initial and final momentum of the hadron are frame-dependent, and the most widely used frame of reference to calculate GPDs is the symmetric frame. 
In such a case, the momentum transfer is symmetrically distributed between the incoming ($p_i$) and the outgoing hadrons ($p_f$) (see, e.g., Eqs.~\eqref{eq:pf_symm} - \eqref{eq:pi_symm}). 
In this section, it is convenient to define the momenta in terms of the average momentum $P = \frac{1}{2}(p_i + p_f)$ and the momentum transfer $\Delta = p_f - p_i$,
\begin{align}
\label{eq:pipf_def}
p_i =  P - \frac{\Delta}{2} \, , \quad  p_f = P + \frac{\Delta}{2} \,.
\end{align}
The above expressions are general for any frame, but will differ numerically in each frame. 
An alternative setup to the symmetric one, is an asymmetric frame, where the momentum transfer is not shared between the incoming and outgoing hadrons but is rather entirely applied to the incoming hadron (see, e.g., Eqs.~\eqref{eq:pf_nonsymm} - \eqref{eq:pi_nonsymm}). This frame is of interest for this work and will be used throughout this paper. For completeness, we remind the reader that the energies of the initial and final states, $E_{i/f}=\sqrt{m^2 + (\vec{p}_{i/f})^2}$ (where $m$ is the mass of the hadron), are also different in the two frames. 

While computations in the symmetric frame have been extensively used in model calculations~\cite{Bhattacharya:2018zxi,Bhattacharya:2019cme}, due to the nice symmetry properties of the correlators, they are notoriously difficult to calculate in lattice QCD mainly due to the computational cost to extract a range of values for the momentum transfer.
More specifically, the information on the momentum transfer is present in both the initial and final states. 
As a consequence, every value of $\Delta$ requires a separate calculation. 
Such a constraint places severe limitations on GPD calculations in terms of the range of values of the momentum transfer that can be accessed, and, consequently, skewness. So the question arises whether it is meaningful to calculate GPDs in asymmetric frames, which can be computationally less expensive.
As we will see below, one of the approaches to handle calculations in asymmetric frames is to relate the setup of the symmetric frame to the asymmetric one via an appropriate Lorentz transformation. For instance, a Lorentz transformation along the $z$-direction is not optimal for this purpose because it requires a spatial operator distance (say $z = (0,0_\perp,z^3 \neq 0)$) to develop a non-zero temporal component (that is $z \rightarrow (z^0 \neq 0, 0_\perp, z^3)$) which is problematic for lattice QCD calculations. 
However, any Lorentz transformation transverse to the $z$-direction leaves the operator distances unchanged. 
This means that if one begins with a spatial operator distance in one frame, its counterpart in the other frame of reference remains spatial. Such a transformation is called ``transverse boost". We illustrate this point below by focusing on the simplest case of a transverse boost in the $x$-direction and zero skewness. Note that this method can be generalized for any general transverse boost and for an arbitrary value of skewness. 
Let us begin with relating the \textit{incoming} state in the two frames, $p_i^s = (E_i^s, - \Delta^{1,s}/2,0,P^3)$ and $p_i^a=(E_i^a, -\Delta^{1,a}, 0, P^3)$. 
Lorentz transformation provides $p^s = \Lambda_{\rm LT} \, p^a$,
\begin{align}
\begin{pmatrix}
E_i^s \\[0.1cm]
p_i^{1,s} \\[0.1cm]
p_i^{2,s} \\[0.1cm]
p_i^{3,s}
\end{pmatrix}
& =
\begin{pmatrix}
\gamma & -\gamma \beta & 0 & 0 \\[0.1cm]
-\gamma \beta & \gamma & 0 & 0 \\[0.1cm]
0 & 0 & 1 & 0 \\[0.1cm]
0 & 0 & 0 & 1
\end{pmatrix}
\times 
\begin{pmatrix}
E_i^a \\[0.1cm]
-\Delta^{1,a} \\[0.1cm]
0 \\[0.1cm]
P^3
\end{pmatrix} \, .
\end{align}
This gives,
\begin{align}
    E_i^s & = \gamma (E_i^a+\beta \Delta^{1,a}) \, ,
\label{e:LT1}
\end{align}
and,
\begin{align}
 p_i^{1,s} & = - \gamma (\beta E_i^a +\Delta^{1,a}) \quad \rightarrow \quad  \Delta^{1,s} = 2\gamma (\beta E_i^a +\Delta^{1,a}) \, .
 \label{e:LT2}
\end{align}
Now, we relate the \textit{outgoing} state in the two frames, $p_f^s = (E_f^s, \Delta^{1,s}/2,0,P^3)$ and $p_f^a=(E_f^a, 0, 0, P^3)$. 
(Note that the energies of the incoming and outgoing states are different in the asymmetric frame.) Following the steps outlined above, we find,
\begin{align}
E_i^s & = \gamma E_f^a \, ,
\label{e:LT3}
\end{align}
and,
\begin{align}
 p_f^{1,s} & = -\gamma \beta E_f^a \quad \rightarrow \quad 
 \Delta^{1,s} = -2\gamma \beta E_f^a \, .
 \label{e:LT4}
\end{align}
Using Eqs.~(\ref{e:LT1}) and~(\ref{e:LT3}), we obtain,
\begin{align}
\beta & = - \bigg ( \dfrac{E_i^a-E_f^a}{\Delta^{1,a}} \bigg ) \, .
\label{e:b1}
\end{align}
In turn, using Eqs.~(\ref{e:LT2}) and~(\ref{e:LT4}) yields,
\begin{align}
\beta & = - \dfrac{\Delta^{1,a}}{E_i^a+E_f^a} \, .
\label{e:b2}
\end{align}
The equality of Eqs.~(\ref{e:b1}) and~(\ref{e:b2}) then implies,
\begin{align}
\Delta^{1,a} = \sqrt{(E_i^a)^2-(E_f^a)^2} \, .
\end{align}
Hence, $\beta$ can be written more compactly as,
\begin{align}
\beta = - \sqrt{\dfrac{E_i^a-E_f^a}{E_i^a+E_f^a}} <0 \, .
\label{e:beta}
\end{align}
This implies $\Delta^{0,a} < 0$, and 
\begin{align}
\gamma = \dfrac{1}{\sqrt{1-\beta^2}} = \sqrt{\dfrac{E_i^a+E_f^a}{2E_f^a}} \, .
\label{e:gamma}
\end{align}
Therefore, by using the expressions for $(\beta, \gamma)$, we can write down uniquely what the symmetric frame variables $(E_i^s, \Delta^{1,s})$ are supposed to be in terms of the asymmetric frame variables $(E_i^a, E_f^a, \Delta^{1,a})$. The energy should be,
\begin{align}
E_i^s = \gamma E_f^a = \sqrt{\dfrac{E_f^a(E_i^a+E_f^a)}{2}} \, ,
\end{align}
and the transverse-momentum transfer,
\begin{align}
\Delta^{1,s} & = - 2\gamma \beta E_f^a \, ,\nonumber \\[0.2cm]
\text{or,} \quad \Delta^{1,s} & = 2\sqrt{\dfrac{E_f^a(E_i^a-E_f^a)}{2}} = 2 \sqrt{\dfrac{E_f^a}{2(E_i^a+E_f^a)}} \,  \Delta^{1,a} \, .
\end{align}
The above method can be straightforwardly generalized for $\vec{\Delta}_{\perp} = (\Delta^1, \Delta^2)$ as well as for an arbitrary value of skewness.

\subsection{Spin-$\boldsymbol{0}$ particles}
\label{s:spin0}
In this section, we study spin-0 particles such as the pion, where the method can also be generalized to spin-$1/2$ (next subsection) and higher-spin particles. 
The (unpolarized) pion GPD is defined from the matrix element
\begin{align}
F^{\mu}(z, P, \Delta) & = \langle p_f | \bar{q} (-\tfrac{z}{2}) \gamma^\mu \, {\cal W}(-\tfrac{z}{2}, \tfrac{z}{2})  q (\tfrac{z}{2}) | p_i\rangle \, ,
\label{e:mat}
\end{align}
where ${\cal W}$ is a straight Wilson line that makes the correlator gauge invariant. Light-cone GPD $H$ can be defined in a Lorentz-invariant manner (see below), whereas one can think of different definitions for quasi-GPDs based on the approach that one wants to take to perform their calculations in different frames.
In the first approach, and as discussed in Sec.~\ref{s:frames}, one can perform a calculation of a quasi-GPD in the symmetric frame, but for this purpose make use of the asymmetric frame. In this case, it is not only crucial to understand how the kinematic variables transform between frames (see Sec.~\ref{s:frames}), but it is equally crucial to understand how the matrix elements underlying quasi-GPDs themselves transform between frames. Historically, (unpolarized) quasi-GPDs have been defined through the operator $\gamma^0$~\cite{Liu:2019urm, Radyushkin:2019owq,Chen:2019lcm,Alexandrou:2020zbe,Lin:2020rxa} (see also the next section). By using a transverse boost, we find that the matrix element $\langle .. \gamma^0 .. \rangle $ in the symmetric frame can be written as a linear combination of matrix elements of different operators $\langle .. (\gamma^0 + \gamma^1 ) .. \rangle $ in the asymmetric frame,
\begin{equation}
\hspace{-0.2cm}
\langle p_f | \bar{q} (-\tfrac{z}{2}) \gamma^0 \, {\cal W}(-\tfrac{z^3}{2}, \tfrac{z^3}{2}) \, q (\tfrac{z}{2}) |p_i \rangle^s = \gamma \langle p_f | \bar{q} (-\tfrac{z}{2}) \gamma^0 \, {\cal W}(-\tfrac{z^3}{2}, \tfrac{z^3}{2}) \,  q (\tfrac{z}{2}) |p_i \rangle^a - \gamma \beta \langle p_f | \bar{q} (-\tfrac{z}{2}) \gamma^1 \, {\cal W}(-\tfrac{z^3}{2}, \tfrac{z^3}{2}) \, q (\tfrac{z}{2}) |p_i \rangle^a .
\label{e:LT_matrix_elements}
\end{equation}
This equation essentially shows how the $0^{\rm th}$ component of a 4-vector changes under a Lorentz transformation. This implies that a transverse boost that uniquely fixes $(\beta,\gamma)$ (Eqs.~(\ref{e:beta}) and~(\ref{e:gamma})) allows for an exact calculation of quasi-GPDs in the symmetric frame through matrix elements of the asymmetric frame. However, Eq.~(\ref{e:LT_matrix_elements}) also makes it clear that a quasi-GPD defined through the operator $\gamma^0$ is intrinsically Lorentz non-invariant. In the limit of a large momentum, though, we recover
\begin{align}
\lim_{P^{3} \rightarrow \infty} \langle .. \gamma^0 .. \rangle^s &\, \approx\, \langle .. \gamma^0 .. \rangle^a + \mathcal{O} \bigg ( \dfrac{1}{P^{3}} \bigg ) \langle .. \gamma^1 .. \rangle^a  \,\rightarrow\, \langle .. \gamma^0 .. \rangle^a \, ,
\end{align}
implying that the contribution from the matrix element $\langle .. \gamma^1 .. \rangle $ is essentially a power correction at finite values of momentum $P^3$. 

We now illustrate the Lorentz non-invariance of the (above) historic definition of quasi-GPD through an altogether different approach and then motivate a new definition for quasi-GPDs that is frame independent, and more importantly, may potentially reduce power corrections. We call this the second approach or the amplitude approach. As a first step, we build a Lorentz-covariant decomposition of the matrix element in Eq.~(\ref{e:mat}) in terms of the available vectors $(P^{\mu}, z^{\mu}, \Delta^\mu)$,
\begin{align}
F^{\mu}(z, P, \Delta) & = P^\mu A_1 (z\cdot P, z \cdot \Delta, \Delta^2, z^2) + z^\mu m^2A_2 (z\cdot P, z \cdot \Delta, \Delta^2, z^2) + \Delta^\mu A_3 (z\cdot P, z \cdot \Delta, \Delta^2, z^2) \,,
\label{e:spin0_decomp}
\end{align}
where $m$ is the mass of the target.
Here, $A_i$'s are the Lorentz-invariant (and, thus, frame-independent) amplitudes whose arguments are functions of Lorentz scalars\footnote{In the literature, the amplitudes have also been called generalized Ioffe time distributions (ITDs)~\cite{Radyushkin:2019owq}.}. The light-cone GPD $H$ in both symmetric and asymmetric frames is defined from the correlator
\begin{align}
F^{+}(z,P,\Delta) & \equiv P^+ H (z, P, \Delta) \nonumber \\[1ex]
& = P^+ A_1(z\cdot P, z \cdot \Delta, \Delta^2,0) + \Delta^+ A_3(z\cdot P, z \cdot \Delta, \Delta^2, 0) \, ,\nonumber \\[3ex]
\text{or,} \quad H (z, P, \Delta) & = A_1(z\cdot P, z \cdot \Delta, \Delta^2,0) +   {\Delta^+\over P^+} A_3(z\cdot P, z \cdot \Delta, \Delta^2,0) \, ,\label{eq:LC0}
\end{align}
where $z^\mu=(0,z^-,0_\perp)$. Note that $\Delta^+/P^+=z\cdot \Delta / z\cdot P$, so the above GPD is the same in both frames as long as $z\cdot P$, $z\cdot \Delta$ and $\Delta^2$ are held to be the same.
The light-cone GPD $H$ in the momentum space is defined,
\begin{align}
	H(x,\xi, t) &= P^+\int {dz^-\over 4\pi}\ e^{ixP^+z^-}H(z, P, \Delta) \,,
\end{align}
where the skewness parameter $\xi = - \Delta^+/(2P^+)$.
In the literature, the light-cone GPD has also been defined in the symmetric frame as~\cite{Ji:1996ek,Ji:1996nm}
\begin{align}
	H(x,\xi,t) &={1\over n\cdot P} \int {d\lambda \over 4\pi}\ e^{ix\lambda }\langle p_f | \bar{q} (-\tfrac{\lambda n}{2}) \slashed n \, {\cal W}(-\tfrac{\lambda n}{2}, \tfrac{\lambda n}{2})  q (\tfrac{\lambda n}{2}) | p_i\rangle
	\label{e:GPD_cov_def}
\end{align}
for a lightlike vector $n^\mu\propto (1,0,0,-1)$. In the symmetric frame, the average momentum $P$ has its dominant component along the light-cone direction that is anti-collinear to $n$. The above expression allows us to generalize $H$ as
\begin{align}
	H(z \cdot P, z \cdot \Delta, \Delta^2,0) &= {1\over  z\cdot P}\langle p_f | \bar{q} (-\tfrac{z}{2}) \slashed z \, {\cal W}(-\tfrac{z}{2}, \tfrac{z}{2})  q (\tfrac{z}{2}) | p_i\rangle\nn\\[1ex]
& = A_1(z\cdot P, z \cdot \Delta, \Delta^2,0)  +  {z\cdot \Delta \over z\cdot P} A_3(z\cdot P, z \cdot \Delta, \Delta^2,0)  \,,
\label{e:H_LKpin0}
\end{align}
which is independent of the orientation of $z^\mu$ and equivalent to \eq{LC0} in the coordinate system where $z^\mu=(0,z^-,0_\perp)$.
Therefore, $H$ is Lorentz invariant as long as the scalars $z \cdot P, z \cdot \Delta, \Delta^2$ are fixed, and the $H$ GPD in the momentum space is the Fourier transform by integrating along a fixed direction in the $(z \cdot P, z \cdot \Delta)$ plane with $z\cdot \Delta = -2\xi (z\cdot P)$, i.e.,
\begin{align}
	H(x,\xi, t) &= \int {d(z\cdot P)\over 4\pi}\ e^{ixz\cdot P}H\big(z \cdot P, -2\xi(z \cdot P), t,0\big)\,.
\end{align}
(Note that $x$ is the Fourier conjugate of $z\cdot P$.)

Now, we turn to the quasi-GPD ${\cal{H}}$ which in the coordinate space is connected to the light-cone GPD $H$ through the matching formula~\cite{Radyushkin:2019owq}, 
\begin{align}\label{eq:match}
{\cal{H}} \big(z \cdot P, -2\xi(z\cdot P), \Delta^2, z^{2}, \mu^2 \big) & = \int_{-1}^1 du\, \bar{C}\, (u, z \cdot P, \xi, z^2, \mu^2) \, H \big(u(z \cdot P), -2u\xi(z\cdot P), \Delta^2, \mu^2\big) \, ,
\end{align}
where $\bar{C}$ is the short-distance matching coefficient that can be calculated perturbatively~\cite{Radyushkin:2019owq,Liu:2019urm,Ma:2022ggj}, and $\mu$ is the renormalization scale in the $\overline{\rm MS}$ scheme. (We will revisit the derivation of the matching equation towards the end of this section.)
At leading order in $\alpha_s$, the above formula indicates that ${\cal{H}}$ collapses to $H$ in the light-cone limit $z^2\to0$,
\begin{align}
\lim_{z^2\to 0} {\cal{H}} (z\cdot P, z \cdot \Delta, \Delta^2, z^2) = H (z\cdot P, z \cdot \Delta, \Delta^2,0) + {\cal O}(\alpha_s) \, .
\label{e:spin0_LO_arg}
\end{align}
Therefore, a natural candidate for a frame independent quasi-GPD is the generalization of Lorentz-invariant ${\cal{H}}$ to include $z^2 \neq 0$, i.e.,
\begin{align}
{\cal{H}} (z \cdot P, z \cdot \Delta, \Delta^2, z^2) & \equiv A_1(z\cdot P, z \cdot \Delta, \Delta^2, z^2) +   {z\cdot \Delta \over z\cdot P} A_3(z\cdot P, z \cdot \Delta, \Delta^2, z^2)\, .
\label{e:H_LC_improvedLO}
\end{align}
Note that this result in the forward limit agrees with the quasi-PDF definition using the $\gamma^0$ matrix element~\cite{Radyushkin:2017cyf}.
Since both sides of Eq.~(\ref{e:spin0_LO_arg}) are Lorentz invariant (recall also Eq.~(\ref{e:H_LKpin0})), at finite $z^2$ the difference between ${\cal{H}}$ and $H$ are frame-independent subleading power corrections in $A_i$'s.
Correspondingly, the quasi-GPD ${\cal H}$ is defined as
\begin{align}
	{\cal H}(x,\xi, P^3, t) &=  \int {d(z\cdot P)\over 4\pi}\ e^{ixz\cdot P}{\cal{H}} \big(z \cdot P, -2\xi(z \cdot P), t,z^2\big)\,,
\end{align}
where the measure $d(z\cdot P) = -P^3 dz^3$ with fixed $P^3$.

A direct implication of \eq{match} is that the skewness of the GPD $H$, $\xi= -\Delta^+/(2P^+)$, is equal to the quasi-skewness $\tilde \xi= -\Delta^3/(2P^3)$ of the corresponding GPD ${\cal{H}}$, as they both are given by $-z\cdot \Delta/(2z\cdot P)$. To better understand this, let us recall the derivation of the factorization formula for the quasi-GPD. At short $z^2$, the matrix element in Eq.~(\ref{e:mat}) has an operator product expansion (OPE)~\cite{Liu:2019urm},
\begin{align}\label{eq:cope}
	 \langle p_f | \bar{q} (-\tfrac{z}{2}) \slashed n \, {\cal W}_{\rm Q}(-\tfrac{z}{2}, \tfrac{z}{2})  q(\tfrac{z}{2}) | p_i\rangle &= \sum_{n=0}^\infty C_n(\mu^2z^2) {\cal F}_n(-iz) \sum_{m=0}^{[n/2]} {\cal B}_{n,m}(\mu)\nn\\[1ex]
	 &\quad \times n_{\mu_0} n_{\mu_1}\ldots n_{\mu_n} (i\partial^{\mu_{n-2m+1}})\ldots (i\partial^{\mu_{n}}) \langle p_f | {\mathbf O}^{\mu_0\mu_1\ldots \mu_{n-2m}} |p_i\rangle \,,
\end{align} 
where $C_n$ are Wilson coefficients, ${\cal F}_n$ is a special polynomial series, ${\mathbf O}^{\mu_0\mu_1\ldots \mu_{n}}$ are the conformal operators, and ${\cal B}_{n,m}(\mu)$ are perturbative coefficient functions that diagonalizes the anomalous dimension matrix of the operators that mix with ${\mathbf O}^{\mu_0\mu_1\ldots \mu_{n}}$. The conformal operator is defined as~\cite{Braun:2003rp,Efremov:1978rn}
\begin{align}
	n_{\mu_0} n_{\mu_1}\ldots n_{\mu_n} {\mathbf O}^{\mu_0\mu_1\ldots \mu_{n}} 
	&= \sum_{m=0}^{[n/2]}C_{n,m}^{3/2} \ (in\cdot \partial)^{n-2m}\bar{q} \slashed n (in\cdot \overleftrightarrow{D})^{2m} q - \mbox{traces}\,,
\end{align}
where $C_n^{3/2}(x) = \sum_{m=0}^{[n]/2}C_{n,m}^{3/2} x^{2m} $ is a Gegenbauer polynomial in $x$.
The off-forward matrix element of ${\mathbf O}^{\mu_0\mu_1\ldots \mu_{n}}$ is the Gegenbauer moment, which, according to Lorentz covariance, can be parameterized as
\begin{align}
	\langle p_f | n_{\mu_0} n_{\mu_1}\ldots n_{\mu_n} {\mathbf O}^{\mu_0\mu_1\ldots \mu_{n}} |p_i\rangle &= \sum_{m=0}^{[n/2]}C_{n,m}^{3/2} \left[  (-n\cdot \Delta)^{2m} (2n\cdot P)^{n+1-2m} - \mbox{traces}\right]\phi_{n,m}(t)\,,
\end{align}
where $\phi_{n,m}(t)$ are frame-independent form factors.

For $\mu_0=\mu_1=\ldots= \mu_n=+$,
\begin{align}\label{eq:ggb1}
	\langle p_f | {\mathbf O}^{++\ldots +} |p_i\rangle &= \sum_{m=0}^{[n/2]}C_{n,m}^{3/2} \  (-\Delta^+)^{2m} (2P^+)^{n+1-2m} \phi_{n,m}(t)= (2P^+)^{n+1}  \sum_{m=0}^{[n]/2}C_{n,m}^{3/2} \xi^{2m}\phi_{n,m}(t)\,.
\end{align}
And for $\mu_0=\mu_1=\ldots= \mu_n=3$,
\begin{align}\label{eq:ggb2}
	\langle p_f | {\mathbf O}^{zz\ldots z} |p_i\rangle &= (-1)^{n+1}\sum_{m=0}^{[n/2]}C_{n,m}^{3/2} \left[  (\Delta^3)^{2m} (-2P^3)^{n+1-2m}  - \mbox{traces}\right]\phi_{n,m}(t) \nn\\
	& = (2P^3)^{n+1} \sum_{m=0}^{[n]/2}C_{n,m}^{3/2} \tilde \xi^{2m}\phi_{n,m}(t) - \mbox{traces}\,.
\end{align}
On the other hand, from the operator definition we have
\begin{align}
	\langle p_f | {\mathbf O}^{++\ldots +} |p_i\rangle &=(2P^+)^{n+1} \xi^n \int_{-1}^1 dy\ C_n^{3/2}({y\over \xi}) H(y,\xi, t)\,,
\end{align}
then according to the Lorentz covariance of \eqs{ggb1}{ggb2} we have
\begin{align}\label{eq:cfm2}
	\langle p_f | {\mathbf O}^{zz\ldots z} |p_i\rangle &= (2P^3)^{n+1} \tilde \xi^n \int_{-1}^1 dy\ C_n^{3/2}({y\over \tilde\xi}) H(y,\tilde\xi, t) - \mbox{traces} \,.
\end{align}
Therefore, following the footsteps of Ref.~\cite{Liu:2019urm}, we can plug \eq{cfm2} into the OPE formula \eq{cope} and derive the exact matching formula for the quasi-GPD,
\begin{align}
	{\cal H}(x,\tilde\xi,P^3,t,\mu) &= \int {dy\over |\tilde\xi|} C\left({x\over \tilde \xi}, {y\over \tilde\xi}, {\mu\over \tilde\xi P^3} \right) H(y,\tilde\xi, t,\mu) + \mbox{power corrections}\,,
	\label{e:matching_equation}
\end{align}
where $C$ is the matching kernel.

\subsection{Spin-$\boldsymbol{1/2}$ particles}
\label{sec:spin_1_2}
In this section, we turn our attention to spin-$1/2$ particles, such as the proton. 
As in the case of spin-0 particles, it is crucial to derive a Lorentz-covariant decomposition of the vector matrix element for spin-$1/2$ particles.
It turns out that constraints from parity alone are sufficient to write down the general structure of the vector matrix element. One ends up in finding eight linearly-independent Dirac structures multiplied by eight Lorentz-invariant (frame-independent) amplitudes, 
\begin{align}
\label{eq:parametrization_general}
F^{\mu} (z,P,\Delta) & = \bar{u}(p_f,\lambda') \bigg [ \dfrac{P^{\mu}}{m} A_1 + m z^{\mu} A_2 + \dfrac{\Delta^{\mu}}{m} A_3 + i m \sigma^{\mu z} A_4 + \dfrac{i\sigma^{\mu \Delta}}{m} A_5 \nonumber \\[1ex]
& \hspace{5cm} + \dfrac{P^{\mu} i\sigma^{z \Delta}}{m} A_6 + m z^{\mu} i\sigma^{z \Delta} A_7 + \dfrac{\Delta^{\mu} i\sigma^{z \Delta}}{m} A_8  \bigg ] u(p_i, \lambda) \, ,
\end{align}
where
$\sigma^{\mu \nu} \equiv \tfrac{i}{2} (\gamma^\mu \gamma^\nu - \gamma^\nu \gamma^\mu)$,  
$\sigma^{\mu z} \equiv \sigma^{\mu \rho} z_\rho$, 
$\sigma^{\mu \Delta} \equiv \sigma^{\mu \rho} \Delta_\rho$, $\sigma^{z \Delta} \equiv \sigma^{\rho \tau} z_\rho \Delta_\tau$, 
$z \equiv (z^0 = 0, z_\perp = 0_\perp, z^3 \neq 0)$ with a summation implied for repeated indices. Also, we use the compact notation $A_i \equiv A_i (z\cdot P, z \cdot \Delta, \Delta^2, z^2)$. 
The steps involved in the derivation of Eq.~(\ref{eq:parametrization_general}) are outlined in Appendix~\ref{s:ITDs_derivation}. 
This derivation parallels the steps presented in Ref.~\cite{Meissner:2009ww}. (See also Ref.~\cite{Rajan:2017cpx} where this matrix element was parameterized in momentum space and for a straight Wilson line.) Note that the amplitudes $A_1,\,A_2,$ and $A_3$ are analogous to the spin-0 case. We also note that one can choose to work with a basis of different parametrization other than Eq.~(\ref{eq:parametrization_general}). 
However, the number of amplitudes will remain the same and, hence, one would always require eight independent lattice matrix elements to disentangle the amplitudes.
Therefore, there is no obvious gain in computational cost if a different parametrization is used.%

For spin-$1/2$ particles, there are two (vector) light-cone GPDs $H$ and $E$ defined through~\cite{Diehl:2002he},
\begin{align}
F^{+} (z, P^{s/a}, \Delta^{s/a}) & = \bar{u}^{s/a}(p_f^{s/a}, \lambda ') \bigg [\gamma^{+} H(z, P^{s/a}, \Delta^{s/a})  + \frac{i\sigma^{+\mu}\Delta^{s/a}_{\mu}}{2m} E(z, P^{s/a}, \Delta^{s/a}) \bigg ] u^{s/a}(p^{s/a}_i, \lambda) \, .
\label{e:GPD_para_spin1/2}
\end{align}
After using $\mu =+$ in Eq.~(\ref{eq:parametrization_general}), we can perform a change of basis of the resulting expression to map the $A_i$'s onto the GPDs in Eq.~(\ref{e:GPD_para_spin1/2}),
\begin{align}
\label{eq:H}
H (z,P^{s/a},\Delta^{s/a}) & = A_1 + \dfrac{\Delta^{+,s/a}}{P^{+,s/a}} A_3 \, ,\\[3ex]
\label{eq:E}
E (z,P^{s/a},\Delta^{s/a}) & = - A_1 - \dfrac{\Delta^{+,s/a}}{P^{+,s/a}} A_3 + 2 A_5 + 2P^{+,s/a}z^- A_6 + 2 \Delta^{+,s/a} z^- A_8 \, ,
\end{align}
where the arguments of the $A_i$'s have no dependence on $z^{2}$.
We can make the above expressions formally Lorentz invariant as (see Sec.~\ref{s:spin0}),
\begin{align}
\label{eq:H_improved}
H  (z\cdot P^{s/a}, z \cdot \Delta^{s/a}, (\Delta^{s/a})^2)& = A_1 + \dfrac{\Delta^{s/a} \cdot z}{P^{s/a} \cdot z} A_3 \, , \\[3ex]
\label{eq:E_improved}
E(z\cdot P^{s/a}, z \cdot \Delta^{s/a}, (\Delta^{s/a})^2) & = - A_1 - \dfrac{\Delta^{s/a} \cdot z}{P^{s/a} \cdot z} A_3 + 2 A_5 + 2 P^{s/a} \cdot z A_6 + 2\Delta^{s/a} \cdot z A_8 \, .
\end{align}
We emphasize that one can arrive at Eqs.~(\ref{eq:H_improved}) and~(\ref{eq:E_improved}) by contracting both sides of Eq.~(\ref{eq:parametrization_general}) with $z_\mu$ (where $z_\mu$ is an arbitrary light-like vector) and by ensuring that $z^2=0$ (recall Eq.~(\ref{e:GPD_cov_def}) and Eq.~(\ref{e:H_LKpin0})). This implies that light-cone GPDs are frame-independent as long as the Lorentz scalars such as $(z \cdot P^{s/a}, z \cdot \Delta^{s/a}, (\Delta^{s/a})^2)$ are taken to be the same in the two frames.

We now turn to quasi-GPDs. As emphasized in Sec.~\ref{s:spin0}, the essence of the matching equation is the equivalence of the quasi-GPDs and the light-cone GPDs at the leading order. Therefore, a natural way to define the quasi-GPDs ${\cal{H}}$ and ${\cal{E}}$ is through a Lorentz-invariant generalization of the light-cone definitions in Eqs.~(\ref{eq:H_improved}) and~(\ref{eq:E_improved}) to $z^{2} \neq 0$, i.e.,
\begin{align}
\label{eq:Hq_improved}
{\cal H}  (z\cdot P^{s/a}, z \cdot \Delta^{s/a}, (\Delta^{s/a})^2, z^{2})& = A_1 + \dfrac{\Delta^{s/a} \cdot z}{P^{s/a} \cdot z} A_3 \, , \\[3ex]
\label{eq:Eq_improved}
{\cal E}(z\cdot P^{s/a}, z \cdot \Delta^{s/a}, (\Delta^{s/a})^2 ,z^2) & = - A_1 - \dfrac{\Delta^{s/a} \cdot z}{P^{s/a} \cdot z} A_3 + 2 A_5 + 2 P^{s/a} \cdot z A_6 + 2 \Delta^{s/a} \cdot z  A_8 \, ,
\end{align}
where now the arguments of the $A_i$'s have a non-zero dependence on $z^{2}$. We expect that the definitions in Eqs.~(\ref{eq:Hq_improved})-(\ref{eq:Eq_improved}) may have a faster convergence to the light-cone GPDs at the leading order, although such a statement needs a rigorous justification from the theory side.~\footnote{Our argument parallels Ref.~\cite{Radyushkin:2017cyf}, where similar arguments where made for the quasi-PDFs. See the next paragraph for a discussion on the convergence of the various definitions.} 
Furthermore, these definitions of quasi-GPDs differ from their (respective) light-cone GPDs by frame-independent power corrections beyond the leading order.

Historically, quasi-GPDs have been defined through the $\gamma^0$ operator as,
\begin{align}
F^{0} (z, P^{s/a},\Delta^{s/a}) & = \langle p_f^{s/a}, \lambda' | \bar{q} (-\tfrac{z}{2}) \gamma^0 q (\tfrac{z}{2}) | p^{s/a}_i, \lambda\rangle \nonumber \\[1ex]
& = \bar{u}^{s/a}(p_f^{s/a}, \lambda ') \bigg [\gamma^{0} {\cal{H}}^{s/a}_0 (z,P^{s/a},\Delta^{s/a}) + \frac{i\sigma^{0\mu}\Delta^{s/a}_{\mu}}{2m} {\cal{E}}^{s/a}_0 (z,P^{s/a},\Delta^{s/a}) \bigg ] u^{s/a}(p^{s/a}_i, \lambda) \, .
\label{e:historic}
\end{align}
After using $\mu =0$ in Eq.~(\ref{eq:parametrization_general}), we can perform a change of basis of the resulting expression to map the $A_i$'s onto the quasi-GPDs in Eq.~(\ref{e:historic}). The relations in the symmetric frame read,
\begin{align}
\label{eq:quasiH_symm}
{\cal{H}}^{s}_0 (z,P^{s},\Delta^{s}) & = A_1 + \dfrac{\Delta^{0,s}}{P^{0,s}} A_3 - \dfrac{m^{2} \Delta^{0,s} z^3}{2P^{0,s} P^{3,s}} A_4 + \bigg [ \dfrac{(\Delta^{0,s})^{2} z^{3}}{2P^{3,s}}  - \dfrac{\Delta^{0,s} \Delta^{3,s} z^3 P^{0,s}}{2(P^{3,s})^2} - \dfrac{z^{3} (\Delta^{s}_\perp)^2}{2P^{3,s}} \bigg ] A_6 \nonumber \\[1ex]
& +  \bigg [ \dfrac{(\Delta^{0,s})^{3} z^{3}}{2P^{0,s} P^{3,s}}  - \dfrac{(\Delta^{0,s})^2 \Delta^{3,s} z^3}{2(P^{3,s})^2} - \dfrac{\Delta^{0,s} z^{3} (\Delta^{s}_\perp)^2}{2P^{0,s} P^{3,s}} \bigg ] A_8 \, ,
\\[3ex]
\label{eq:quasiE_symm}
{\cal{E}}^{s}_0 (z,P^{s},\Delta^{s}) & = - A_1 - \dfrac{\Delta^{0,s}}{P^{0,s}} A_3 + \dfrac{m^2 \Delta^{0,s} z^{3}}{2P^{0,s} P^{3,s}} A_4 + 2 A_5 + \bigg [ - \dfrac{(\Delta^{0,s})^{2}z^{3}}{2P^{3,s}}  + \dfrac{P^{0,s} \Delta^{0,s} \Delta^{3,s} z^{3}}{2 (P^{3,s})^{2}} + \dfrac{z^{3} (\Delta^{s}_\perp)^2}{2 P^{3,s}} - \dfrac{2z^{3}(P^{0,s})^{2}}{P^{3,s}} \bigg ] A_6 \nonumber \\[1ex]
& + \bigg [ - \dfrac{(\Delta^{0,s})^{3}z^{3}}{2P^{0,s} P^{3,s}}  + \dfrac{(\Delta^{0,s})^2 \Delta^{3,s} z^{3}}{2(P^{3,s})^{2}} + \dfrac{\Delta^{0,s} z^{3}(\Delta^{s}_\perp)^2}{2P^{0,s} P^{3,s}} - \dfrac{ 2z^{3}P^{0,s} \Delta^{0,s}}{P^{3,s}} \bigg ] A_8 \, .
\end{align}
On the other hand, the relations in the asymmetric frame read, 
\begin{align}
\label{eq:quasiH_nonsymm}
{\cal{H}}^{a}_0 (z,P^{a},\Delta^{a}) & = A_1 + \dfrac{\Delta^{0,a}}{{P}^{0,a}} A_3 - \bigg [ \dfrac{m^2 \Delta^{0,a} z^3}{2{P}^{0,a} {P}^{3,a}} - \dfrac{1}{(1+\tfrac{\Delta^{3,a}}{2{P}^{3,a}})} \dfrac{m^2 \Delta^{0,a} \Delta^{3,a} z^3}{4 {P}^{0,a} ({P}^{3,a})^2} \bigg ] A_4 \nonumber \\[1ex]
& + \bigg [ \dfrac{(\Delta^{0,a})^2 z^3}{2{P}^{3,a}} - \dfrac{1}{(1+\tfrac{\Delta^{3,a}}{2{P}^{3,a}})} \dfrac{(\Delta^{0,a})^2 \Delta^{3,a} z^3}{4 ({P}^{3,a})^2} - \dfrac{1}{(1+\tfrac{\Delta^{3,a}}{2{P}^{3,a}})} \dfrac{{P}^{0,a} \Delta^{0,a} \Delta^{3,a} z^3}{2 ({P}^{3,a})^2} - \dfrac{z^3 (\Delta^{a}_\perp)^2}{2 {P}^{3,a}} \bigg ] A_6
\nonumber\\[1ex]
& + \bigg [ \dfrac{(\Delta^{0,a})^3 z^3}{2{P}^{0,a} {P}^{3,a}} - \dfrac{1}{(1+\tfrac{\Delta^{3,a}}{2{P}^{3,a}})} \dfrac{(\Delta^{0,a})^3 \Delta^{3,a} z^3}{4 {P}^{0,a} ({P}^{3,a})^2} - \dfrac{1}{(1+\tfrac{\Delta^{3,a}}{2{P}^{3,a}})} \dfrac{(\Delta^{0,a})^2 \Delta^{3,a} z^3}{2({P}^{3,a})^2} - \dfrac{z^3 (\Delta^{a}_\perp)^2 \Delta^{0,a}}{2 {P}^{0,a} {P}^{3,a}} \bigg ] A_8 \, ,
\\[3ex]
\label{eq:quasiE_nonsymm}
{\cal{E}}^{a}_0 (z,P^{a},\Delta^{a}) & = - A_1 - \dfrac{\Delta^{0,a}}{{P}^{0,a}} A_3 -  \bigg [ - \dfrac{m^2 \Delta^{0,a} z^3}{2 {P}^{0,a} {P}^{3,a}} - \dfrac{1}{(1 + \tfrac{\Delta^{3,a}}{2 {P}^{3,a}})} \bigg ( \dfrac{m^2 z^3}{{P}^{3,a}} - \dfrac{m^2 \Delta^{0,a} \Delta^{3,a} z^3}{4 {P}^{0,a} ({P}^{3,a})^2} \bigg ) \bigg ] A_4 + 2A_5 \nonumber \\[1ex]
& + \bigg [ - \dfrac{(\Delta^{0,a})^2 z^{3}}{2 {P}^{3,a}} - \dfrac{1}{(1 +\tfrac{\Delta^{3,a}}{2{P}^{3,a}})} \bigg ( \dfrac{{P}^{0,a} \Delta^{0,a} z^3}{ {P}^{3,a}} - \dfrac{(\Delta^{0,a})^2 \Delta^{3,a} z^3}{4 ({P}^{3,a})^2} \bigg ) - \dfrac{1}{(1+\tfrac{\Delta^{3,a}}{2 {P}^{3,a}})} \bigg ( \dfrac{2z^3 ({P}^{0,a})^2}{{P}^{3,a}} 
\nonumber\\[1ex]
& - \dfrac{{P}^{0,a} \Delta^{0,a} \Delta^{3,a} z^3}{2 ({P}^{3,a})^2} \bigg ) + \dfrac{z^3 (\Delta^{a}_\perp)^2}{2{P}^{3,a}} \bigg ] A_6 + \bigg [ - \dfrac{(\Delta^{0,a})^3 z^{3}}{2 {P}^{0,a}{P}^{3,a}} - \dfrac{1}{(1 +\tfrac{\Delta^{3,a}}{2{P}^{3,a}})} \bigg ( \dfrac{ (\Delta^{0,a})^2 z^3}{{P}^{3,a}} - \dfrac{(\Delta^{0,a})^3 \Delta^{3,a} z^3}{4 \overline{P}^{0,a} ({P}^{3,a})^2} \bigg ) \nonumber \\[1ex]
& - \dfrac{1}{(1+\tfrac{\Delta^{3,a}}{2 {P}^{3,a}})} \bigg ( \dfrac{2z^3 {P}^{0,a} \Delta^{0,a}}{{P}^{3,a}} - \dfrac{(\Delta^{0,a})^2 \Delta^{3,a} z^3}{2 ({P}^{3,a})^2} \bigg ) + \dfrac{z^3 (\Delta^{a}_\perp)^2 \Delta^{0,a}}{2 {P}^{0,a} {P}^{3,a}} \bigg ] A_8 \, .
\end{align}
Several comments are in order: For finite values of the momentum, the above expressions contain additional amplitudes that are not present in the light-cone expressions. 
Thus, contrary to their forward limit where arguments are made in favor of $\gamma^0$ because of reduced amplitudes~\cite{Radyushkin:2017cyf}, here we find that additional amplitudes are found in $\gamma^0$ in the off-forward limit. (Note that the different definitions of quasi-GPDs preserve the norm; see also Ref.~\cite{Bhattacharya:2019cme}.) 
Second, the intrinsic Lorentz non-invariance associated with the historical definitions of quasi-GPDs (formally) implies that the basis vectors $(\gamma^0, i\sigma^{0\Delta^{s/a}})$ do not form a complete basis for a spatially-separated bi-local operator for finite values of momentum. 
Therefore, for defining quasi-GPDs in a Lorentz-invariant manner, one needs to have a different basis other than just $(\gamma^0, i\sigma^{0\Delta^{s/a}})$. 
From this perspective, one can infer that our Lorentz-invariant definition of quasi-GPDs is essentially a redefinition of quasi-GPDs in terms of a suitable linear combination of operators (which turns out to be $\gamma^{1/2}$) that reduces the additional amplitudes present in the historic definitions. This may potentially provide a faster convergence to the light-cone GPD at the leading-order, an argument that requires further theoretical investigation. In Sec.~\ref{sec:res_quasi}, we will study numerically three definitions of quasi-GPDs:
\begin{itemize}
    \item[--] Definition in the symmetric frame via the $\gamma^0$ operator (${\cal{H}}^{s}_0 (A_i; z) \, , {\cal{E}}^{s}_0 (A_i; z)$), Eqs.~(\ref{eq:quasiH_symm})-(\ref{eq:quasiE_symm}),
    \item[--] Definition in the asymmetric frame via the $\gamma^0$ operator (${\cal{H}}^{a}_0 (A_i; z) \, , {\cal{E}}^{a}_0 (A_i; z)$), Eqs.~(\ref{eq:quasiH_nonsymm})-(\ref{eq:quasiE_nonsymm}),
    \item[--] Lorentz-invariant definition (${\cal{H}} (A_i; z) \, , {\cal{E}} (A_i; z)$), Eqs.~(\ref{eq:Hq_improved}) - (\ref{eq:Eq_improved}).
\end{itemize}
As previously stated, the three definitions are not equivalent; they differ in terms of the amplitudes that contribute and power corrections. Thus, it is interesting to numerically compare the convergence of these definitions and also get an idea about the relative size of power corrections.

\section{Lattice setup}

\subsection{Matrix elements parametrization}
\label{sec:Lat_setup_A}

One of the goals of this work is to calculate in lattice QCD the Lorentz-invariant amplitudes defined in Eq.~\eqref{eq:parametrization_general}.
To this end, we perform a proof-of-concept analysis based on two calculations of the vector matrix elements, as outlined in Sec.~\ref{sec:strategy}.
Two separate calculations are performed, one in the symmetric and one in an asymmetric frame, which allows us to compare the estimates for $A_i$.
For self-consistency, in this section we present the setup in Euclidean space, where we use lower indices in $P$ and $\Delta$ to avoid confusion in the equations presented. 
The notation for the symmetric frame is 
\bea
\label{eq:pf_symm}
\vec{p}^{\,s}_f&=&\vec{P} + \frac{\vec{\Delta}}{2} = \left(+\frac{\Delta_1}{2},+\frac{\Delta_2}{2},P_3\right)\,,\\[3ex]
\label{eq:pi_symm}
\vec{p}^{\,s}_i&=&\vec{P} - \frac{\vec{\Delta}}{2}= \left(-\frac{\Delta_1}{2},-\frac{\Delta_2}{2},P_3\right)\,,
\eea
and for the asymmetric frame, in which all the momentum transfer is assigned to the initial state, is
\bea
\label{eq:pf_nonsymm}
\vec{p}^{\,a}_f&=&\vec{P} =  \left(0,0,P_3\right) \,,\\[3ex]
\label{eq:pi_nonsymm}
\vec{p}^{\,a}_i&=&\vec{P} - \vec{\Delta} =  \left(-\Delta_1,-\Delta_2,P_3\right)\,.
\eea
In the above equations, a factor of $\frac{2\pi}{L}$ ($L$: spatial extent of the lattice) is included in $\Delta_1$, $\Delta_2$, and $P_3$. 
As can be seen, the setup corresponds to zero skewness, that is $(p_i)_3=(p_f)_3=P_3$.
Numerical calculations of the matrix elements in the two frames at the same value of the arguments of the $A_i$ must be in line with the Lorentz invariance of the $A_i$. Such a numerical confirmation is a highly non-trivial check of the numerical calculations and the underlying equations, which relate the matrix elements and the amplitudes in the two frames. As mentioned previously, the matrix elements are frame-dependent, and, in general, decompose into different sets of $A_i$ in the two frames. 
This is demonstrated in Eqs.~\eqref{eq:Pi0G0_s} - \eqref{eq:Pi3G3_s} and Eqs.~\eqref{eq:Pi0G0_ns} - \eqref{eq:Pi3G3_ns} below\,\footnote{All equations in this section are given in Euclidean space.}.
The analysis takes into consideration four matrix elements of the vector operator, that is, $\gamma_0$, $\gamma_1$, $\gamma_2$, and $\gamma_3$.
The matrix element of $\gamma_3$ is needed only for the extraction of $A_2$ and $A_7$.
We note that the operator $\gamma_3$ has a finite mixing under renormalization for lattice regularizations with chiral symmetry breaking~\cite{Constantinou:2017sej,Alexandrou:2017ypw,Chen:2017mie,Green:2017xeu}.
However, for twisted mass fermions, which we use in this work, the mixing is between $\gamma_3$ and $\gamma_5$; the latter has a vanishing physical matrix elements in the forward limit. 

To disentangle $A_1$ - $A_8$, we need eight independent matrix elements, which can be obtained using the unpolarized ($\Gamma_0$) and three polarized ($\Gamma_k$) parity projectors defined as
\begin{eqnarray}
\Gamma_0 &=& \frac{1}{4} \left(1 + \gamma_0\right)\,, \\
\Gamma_k &=& \frac{1}{4} \left(1 + \gamma_0\right) i \gamma_5 \gamma_k\,, \quad k=1,2,3\,.
\end{eqnarray}
The $\Gamma_0$ projector corresponds to an unpolarized proton, while the three $\Gamma_k$ to a polarized projector in the $k$ direction.
The parity projectors are applied to the right-hand-side of Eq.~\eqref{eq:parametrization_general}, along with the spinor normalization.
Finally, a trace is taken to mimic the procedure of extracting the lattice matrix elements, that is
\begin{equation}
 {\rm Tr}\Big[\Gamma_\kappa \, \Lambda(p) \, F_\mu \Lambda(p) \Big]\,, \quad \mu, \kappa=0,1,2,3\,,
 \label{eq:tr} 
\end{equation}
with the following normalization for the spinor sum
\begin{equation}
\Lambda_{eucl}(p) = \left(\frac{-ip \hspace*{-0.2cm}\slash +m}{2m} \right) \,.
\end{equation}
The trace is performed analytically, and the obtained expressions correspond to the decomposition of the matrix elements, with the ground state denoted by ${\Pi_\mu(\Gamma_\kappa)}$. 
The matrix elements, $\Pi^{s/a}_\mu(\Gamma_\kappa)$, for each operator $\gamma_\mu$ and projector $\Gamma_\kappa$ combination are given in Eqs.~\eqref{eq:Pi0G0_s} - \eqref{eq:Pi3G3_s} for the symmetric frame (superscript $s$), and Eqs.~\eqref{eq:Pi0G0_ns} - \eqref{eq:Pi3G3_ns} for the asymmetric case (superscript $a$). For simplicity of the presentation, we adopt the expressions at zero skewness. The general equations for $\xi\ne0$ can be found in Appendix~\ref{app:C}. 

\newpage
\begin{eqnarray}
\label{eq:Pi0G0_s}
  \Pi^s_0(\Gamma_0) &=& K \,  \left(
  \frac{   E  
    \left( E   ( E  +m)- P_3^2\right)}{2 m^3} \,A_1
    +\frac{    ( E  +m)
    \left(-E ^2+m^2+ P_3^2\right)}{m^3}\,A_5
    +\frac{  E   P_3  \left(-E ^2+m^2+ P_3^2\right)
    z}{m^3}\,A_6
  \right) \hspace*{0.75cm}\\[3ex]
\label{eq:Pi0G1_s}
  \Pi^s_0(\Gamma_1) &=& i\, K \,     \left(
 \frac{\ E     P_3  \Delta_2 }{4
    m^3}\,A_1 -\frac{  ( E  +m)  P_3  \Delta_2 }{2
    m^3}\,A_5 -\frac{  E  \left( P_3^2+m
    ( E  +m)\right) z \Delta_2 }{2 m^3}\,A_6
    \right)\\[3ex]
\label{eq:Pi0G2_s}
  \Pi^s_0(\Gamma_2) &=& i\,K\,  \left(
 -\frac{ E     P_3 
     \Delta_1 }{4 m^3}\,A_1  + \frac{ ( E  +m)
     P_3   \Delta_1 }{2 m^3}\,A_5+\frac{  E  \left( P_3^2+m
    ( E  +m)\right) z  \Delta_1 }{2 m^3}\,A_6
   \right)\\[3ex]
  \Pi^s_0(\Gamma_3) &=& 0\\[3ex]
    \label{eq:Pi1G0_s}
  \Pi^s_1(\Gamma_0) &=& i\,K\,     \left(
  -\frac{ \left( E  
    ( E  +m)- P_3^2\right)  \Delta_1 }{2 m^3}\,A_3+\frac{
P_3  \Delta_1 z  }{4 m}\,A_4-\frac{   P_3 
    \left(-E ^2+m^2+ P_3^2\right) z
     \Delta_1 }{m^3}\,A_8
    \right)\\[3ex]
\label{eq:Pi1G1_s}
  \Pi^s_1(\Gamma_1) &=& K\,  \left(
      \frac{ P_3   \Delta_1  \Delta_2  }{4
    m^3}\,A_3+\frac{ \Delta_1 \Delta_2  z  }{8
    m}\,A_4 -\frac{  \left( P_3^2+m ( E  +m)\right)  \Delta_1  
    \Delta_2 z }{2 m^3}\,A_8
    \right)\\[3ex]
\label{eq:Pi1G2_s}
  \Pi^s_1(\Gamma_2) &=& K\,  \left(
  -\frac{ P_3 
     \Delta_1^ 2}{4 m^3}\,A_3 +\frac{ \left(4
     E   ( E  +m)- \Delta_1^ 2\right) z}{8 m}\,A_4+\frac{  \left( P_3^2+m
    ( E  +m)\right)  \Delta_1^ 2  z}{2 m^3}\,A_8
    \right)\\[3ex]
\label{eq:Pi1G3_s}
    \Pi^s_1(\Gamma_3) &=& K\,  
     \frac{ ( E  +m) \Delta_2  }{2 m^2}\,A_5\\[3ex]
  \label{eq:Pi2G0_s}
  \Pi^s_2(\Gamma_0) &=& i\,K\,  \left(
  -\frac{ \left( E   ( E  +m)- P_3^2\right) \Delta_2  }{2
    m^3}\,A_3   +\frac{ P_3  \Delta_2  z }{4 m}\,A_4-\frac{
   P_3  \left(-E ^2+m^2+ P_3^2\right)
    \Delta_2   z}{m^3}\,A_8 
  \right)\\[3ex]
  \label{eq:Pi2G1_s}
    \Pi^s_2(\Gamma_1) &=& K\, \left(
    \frac{
     P_3   \Delta_2^ 2}{4 m^3}\,A_3-\frac{
    \left( 4  E   ( E  +m)- \Delta_2^ 2\right) z}{8
    m}\,A_4-\frac{  \left( P_3^2+m
    ( E  +m)\right)  \Delta_2^ 2 z}{2 m^3}\,A_8
    \right)\\[3ex]
\label{eq:Pi2G2_s}
  \Pi^s_2(\Gamma_2) &=& K\,  \left(
  -\frac{  P_3 
     \Delta_1  \Delta_2  }{4 m^3}\,A_3-\frac{
     \Delta_1 \Delta_2   z }{8 m}\,A_4+\frac{  \left( P_3^2+m
    ( E  +m)\right)  \Delta_1 \Delta_2   z }{2 m^3}\,A_8
  \right)\\[3ex]
\label{eq:Pi2G3_s}
    \Pi^s_2(\Gamma_3) &=& - K\, \frac{ ( E  +m) \Delta_1 }{2 m^2}\,A_5 \\[3ex]
\label{eq:Pi3G0_s}
  \Pi^s_3(\Gamma_0) &=& i\,K\, \Bigg(
 \frac{
    \left( P_3^2- E   ( E  +m)\right)  P_3 }{2
    m^3}\,A_1 +\frac{ \left( E  
    ( E  +m)- P_3^2\right) z}{2 m}\,A_2-\frac{ 
    \left(-E ^2+m^2+ P_3^2\right)  P_3 }{m^3}\,A_5 \nonumber \\ [1ex]
    &&\qquad-\frac{
   \left(-E ^2+m^2+ P_3^2\right)
     P_3^2 z}{m^3}\,A_6+\frac{ 
    \left(-E ^2+m^2+ P_3^2\right)
     P_3  z^2}{m}\,A_7 
  \Bigg)\\[3ex]
\label{eq:Pi3G1_s}
  \Pi^s_3(\Gamma_1) &=& K\, \Bigg(
  \frac{ \Delta_2  
     P_3^2}{4 m^3}\,A_1 -\frac{  \Delta_2  P_3  z  }{4 m}\,A_2
     -\frac{
    \left( P_3^2+m ( E  +m)\right) \Delta_2  }{2 m^3}\,A_5
     -\frac{ \left( P_3^2+m
    ( E  +m)\right) \Delta_2   P_3  z }{2 m^3}\,A_6 \nonumber\\[1ex]
    && \,\,\qquad +\frac{ \left( P_3^2+m
    ( E  +m)\right) \Delta_2   z^2}{2 m}\,A_7
    \Bigg)
\end{eqnarray}
\begin{eqnarray}
\label{eq:Pi3G2_s}
  \Pi^s_3(\Gamma_2) &=& K\,  \Bigg(
  -\frac{  \Delta_1   P_3^2}{4 m^3}\,A_1 +\frac{  \Delta_1 P_3  z   }{4 m}\,A_2 
  +\frac{ \left( P_3^2+m ( E  +m)\right)  \Delta_1 }{2
    m^3}\,A_5 +\frac{    \left( P_3^2+m ( E  +m)\right)  \Delta_1  P_3  z }{2
    m^3}\,A_6 \qquad \qquad \nonumber\\[1ex]
    && \,\,\qquad-\frac{
    \left( P_3^2+m ( E  +m)\right)  \Delta_1  z^2}{2
    m}\,A_7
  \Bigg)\\[3ex]
\label{eq:Pi3G3_s}
  \Pi^s_3(\Gamma_3) &=& 0\,.
\end{eqnarray}
$K$ is a kinematic factor that depends on the normalization of the proton state, 
\begin{equation}
K = \frac{2 m^2}{\sqrt{E_f E_i (E_f + m) (E_i + m)}}\,.
\end{equation}
In fact, $K$ takes a simpler form in the symmetric frame, that is $2 m^2/(E(E+m))$, due to $E_i=E_f\equiv E$.
The above equations are elegant, which can be attributed to the zero skewness and the simplification of $K$. 
A general feature of the set of equations is that some of them express physically-equivalent matrix elements corresponding to momentum transfer along the two different transverse directions. 
For instance, Eqs.~\eqref{eq:Pi0G1_s} and \eqref{eq:Pi0G2_s} describe the same physical situation, but once with momentum transfer along the $2$-direction and once along the $1$-direction, respectively.
Similar situation occurs for the pairs of Eqs.~\eqref{eq:Pi1G0_s}, \eqref{eq:Pi2G0_s}, Eqs.~\eqref{eq:Pi1G1_s}, \eqref{eq:Pi2G2_s},  Eqs.~\eqref{eq:Pi1G2_s}, \eqref{eq:Pi2G1_s}, Eqs.~\eqref{eq:Pi1G3_s}, \eqref{eq:Pi2G3_s}, and Eqs.~\eqref{eq:Pi3G1_s}, \eqref{eq:Pi3G2_s}. 
Therefore, these equations can be combined, leading to a set of eight independent equations that can disentangle $A_1$ - $A_8$. 
Another characteristic of the symmetric frame is that there are three sets of equations to disentangle certain $A_i$. 
In particular, Eqs.~\eqref{eq:Pi0G0_s} - \eqref{eq:Pi0G2_s} together with Eqs.~\eqref{eq:Pi1G3_s}, \eqref{eq:Pi2G3_s}  are needed to disentangle $A_1$ $A_5$, and $A_6$, while Eqs.~\eqref{eq:Pi1G0_s} - \eqref{eq:Pi2G2_s} disentangle $A_3$, $A_4$, and $A_8$.
Finally, Eqs.~\eqref{eq:Pi3G0_s} - \eqref{eq:Pi3G2_s} can be combined with $A_1$, $A_5$, and $A_6$ to extract $A_2$ and $A_7$.

The decomposition of Eq.~\eqref{eq:parametrization_general} in the asymmetric frame defined in Eqs.~\eqref{eq:pf_nonsymm} - \eqref{eq:pi_nonsymm} leads to more complicated kinematic coefficients mainly because $E_i \ne E_f$. 
Also, the kinematic factor $K$ is canceled by the coefficients of $A_i$ due to its simple structure, leading to more elegant expressions. 
The parametrization for each operator and parity projector is
\begin{eqnarray}
\label{eq:Pi0G0_ns}
+  \Pi^a_0(\Gamma_0) &=& K\,  \Bigg(
 -\frac{  ( E_f + E_i )
    ( E_f - E_i -2 m) ( E_f +m)}{8 m^3}\,A_1 
    -\frac{
    ( E_f - E_i -2 m) ( E_f +m) ( E_f - E_i )}{4
    m^3}\,A_3 \nonumber \\[1ex]
   && \qquad  +\frac{
    ( E_i - E_f )  P_3  z}{4 m}\,A_4  +\frac{ ( E_f + E_i )
    ( E_f +m) ( E_f - E_i )}{4 m^3}\,A_5
     +\frac{  E_f  ( E_f + E_i )  P_3  
    ( E_f - E_i )z}{4 m^3}\,A_6\nonumber \\[1ex]
   && \qquad 
    + \frac{ E_f   P_3
    ( E_f - E_i )^2  z }{2 m^3}\,A_8 
  \Bigg) \\[3ex]
  \Pi^a_0(\Gamma_1) &=& i\,K\,     \Bigg(
\frac{ ( E_f + E_i )
       P_3  \Delta_2  }{8 m^3}\,A_1 
    +\frac{ ( E_f - E_i )  
     P_3  \Delta_2  }{4 m^3}\,A_3
     +\frac{ 
    ( E_f +m)  \Delta_2 z }{4 m}\,A_4
     -\frac{  ( E_f + E_i +2 m)  P_3 
    \Delta_2  }{4 m^3}\,A_5 \nonumber \\[1ex]
   && \qquad  -\frac{ E_f  ( E_f + E_i )  
    ( E_f +m) \Delta_2  z }{4 m^3}\,A_6  -\frac{  E_f 
    ( E_f - E_i )   ( E_f +m) \Delta_2 z  }{2
    m^3}\,A_8 
    \Bigg)\\[3ex]
\label{eq:Pi0G2_ns}
  \Pi^a_0(\Gamma_2) &=& i\,K\,  \Bigg(
 -\frac{ ( E_f + E_i )
       P_3   \Delta_1 }{8 m^3}\,A_1
    -\frac{ ( E_f -E_i )  
     P_3   \Delta_1 }{4 m^3}\,A_3
     -\frac{ 
    ( E_f +m)  \Delta_1 z }{4 m}\,A_4 
    + \frac{   ( E_f + E_i +2 m)  P_3 
     \Delta_1 }{4 m^3}\,A_5  \nonumber \\[1ex]
   && \qquad 
     +\frac{  E_f  ( E_f + E_i )  
    ( E_f +m)   \Delta_1 z}{4 m^3}\,A_6
    +\frac{   E_f 
    ( E_f - E_i )   ( E_f +m)   \Delta_1 z}{2
    m^3}\,A_8
   \Bigg)\\[3ex]
  \Pi^a_0(\Gamma_3) &=& 0 
\end{eqnarray}
\begin{eqnarray}
  \Pi^a_1(\Gamma_0) &=& i\,K\,     \Bigg(
 -\frac{  
    ( E_f - E_i -2 m) ( E_f +m)  \Delta_1 }{8 m^3}\,A_1
    +\frac{
     ( E_f - E_i -2 m) ( E_f +m)  \Delta_1 }{4
    m^3}\,A_3
    +\frac{  P_3 \Delta_1   z }{4 m}\,A_4  \nonumber \\[1ex]
   && \qquad 
+     \frac{ ( E_f - E_i ) 
    ( E_f +m)  \Delta_1 }{4 m^3}\,A_5 +\frac{
     E_f  ( E_f - E_i )   P_3   \Delta_1 z }{4
    m^3}\,A_6+\frac{   E_f  ( E_i - E_f )   P_3
     \Delta_1   z}{2 m^3}\,A_8
    \Bigg)\\[3ex]
  \Pi^a_1(\Gamma_1) &=& K\,  \Bigg(
  -\frac{  P_3   \Delta_1 
    \Delta_2  }{8 m^3}\,A_1 +\frac{ P_3   \Delta_1  \Delta_2  }{4
    m^3}\,A_3 
    +\frac{ P_3   \Delta_1  \Delta_2  }{4 m^3}\,A_5 
    +\frac{
     E_f  ( E_f +m)  \Delta_1\Delta_2    z }{4 m^3}\,A_6 
     -\frac{
     E_f  ( E_f +m)  \Delta_1\Delta_2    z }{2
    m^3}\,A_8 
    \Bigg) \qquad \\[3ex]
  \Pi^a_1(\Gamma_2) &=& K\,  \Bigg(
  \frac{  P_3 \Delta_1^ 2}{8 m^3}\,A_1 -\frac{  P_3   \Delta_1^ 2}{4
    m^3}\,A_3
        +\frac{( E_f + E_i )
    ( E_f +m) z}{4 m}\,A_4 
        +\frac{P_3  \left(2 ( E_f - E_i )
    m- \Delta_1^ 2\right)}{4 m^3}\,A_5  \nonumber \\[0.8ex]
   && \qquad
    -\frac{  E_f  ( E_f +m) \Delta_1^ 2 z  }{4
    m^3}\,A_6 
+  \frac{E_f  ( E_f +m)   \Delta_1^ 2 z}{2
    m^3}\,A_8   
    \Bigg)\\[2.8ex]
    \Pi^a_1(\Gamma_3) &=& K\,  \Bigg(
    \frac{  P_3  z \Delta_2  }{4
    m}\,A_4 + \frac{
    ( E_f +m) \Delta_2  }{2 m^2}\,A_5  \Bigg) \\[2.8ex]
  \Pi^a_2(\Gamma_0) &=& i\,K\,  \Bigg(
  -\frac{ 
    ( E_f - E_i -2 m) ( E_f +m) \Delta_2  }{8 m^3}\,A_1
    +\frac{
     ( E_f - E_i -2 m) ( E_f +m) \Delta_2  }{4
    m^3}\,A_3
    +\frac{   P_3 \Delta_2  z  }{4 m}\,A_4\nonumber \\[0.8ex]
   && \qquad
    +  \frac{( E_f - E_i )
     ( E_f +m) \Delta_2  }{4 m^3}\,A_5  +\frac{
     E_f  ( E_f - E_i )   P_3  \Delta_2 z  }{4
    m^3}\,A_6+\frac{   E_f  ( E_i - E_f )   P_3 
    \Delta_2  z }{2 m^3}\,A_8
  \Bigg)\\[2.8ex]
  \Pi^a_2(\Gamma_1) &=& K\,  \Bigg(
  -\frac{ P_3   \Delta_2^ 2}{8
    m^3}\,A_1  +\frac{ P_3   \Delta_2^ 2}{4 m^3}\,A_3 
    -\frac{
    ( E_f + E_i ) ( E_f +m) z}{4 m}\,A_4
    -\frac{  P_3 
    \left(2 ( E_f - E_i ) m- \Delta_2^ 2\right)}{4 m^3}\,A_5\nonumber \\[0.8ex]
   && \qquad
    +\frac{
     E_f  ( E_f +m) \Delta_2^ 2 z }{4 m^3}\,A_6-\frac{   E_f 
    ( E_f +m)\Delta_2^ 2 z  }{2 m^3}\,A_8
  \Bigg)\\[2.8ex]
  \Pi^a_2(\Gamma_2) &=& K\,  \Bigg(
  \frac{   P_3 
     \Delta_1  \Delta_2  }{8 m^3}\,A_1
     -\frac{ P_3   \Delta_1 
    \Delta_2  }{4 m^3}\,A_3 
    -\frac{ P_3   \Delta_1  \Delta_2  }{4
    m^3}\,A_5 
    -\frac{  E_f  ( E_f +m)  \Delta_1  \Delta_2  z }{4
    m^3}\,A_6  
    +\frac{   E_f  ( E_f +m)  \Delta_1 \Delta_2  z  }{2
    m^3}\,A_8
  \Bigg)\\[2.8ex]
\label{eq:Pi2G3_ns}
    \Pi^a_2(\Gamma_3) &=& K\,  \Bigg(
   -\frac{ P_3 \Delta_1  z  }{4
    m}\,A_4 
 -\frac{
    ( E_f +m)  \Delta_1 }{2 m^2}\,A_5    \Bigg) \\[2.8ex]
  \Pi^a_3(\Gamma_0) &=& i\,K\, \Bigg(
      \frac{ 
    ( E_f - E_i -2 m) ( E_f +m)  P_3 }{4
    m^3}\,A_1
        -\frac{ 
    ( E_f - E_i -2 m) ( E_f +m) z}{4 m}\,A_2
         +\frac{ E_f  ( E_i - E_f )
     P_3 }{2 m^3}\,A_5 \nonumber \\[0.8ex]
   && \qquad
+  \frac{  E_f  ( E_i - E_f )
   P_3^2  z }{2 m^3}\,A_6
    +\frac{ E_f 
    ( E_f - E_i )  P_3 z^2 }{2 m}\,A_7 
  \Bigg)\\[2.8ex]
  \Pi^a_3(\Gamma_1) &=& K\, \Bigg(
  \frac{  \Delta_2    P_3^2}{4
    m^3}\,A_1
        -\frac{  \Delta_2 P_3    z }{4 m}\,A_2
            -\frac{  E_f 
    ( E_f +m) \Delta_2  }{2 m^3}\,A_5
    -\frac{ E_f  ( E_f +m) \Delta_2  P_3    z}{2
    m^3}\,A_6 
    +\frac{  E_f 
    ( E_f +m) \Delta_2   z^2}{2 m}\,A_7
    \Bigg)\\[2.8ex]
  \Pi^a_3(\Gamma_2) &=& K\,  \Bigg(
  -\frac{ \Delta_1   P_3^2}{4 m^3}\,A_1  
     +\frac{
     \Delta_1 P_3   z }{4 m}\,A_2 
          +\frac{  E_f  ( E_f +m)  \Delta_1 }{2
    m^3}\,A_5
      +\frac{
     E_f  ( E_f +m)  \Delta_1P_3   z  }{2 m^3}\,A_6
     -\frac{ E_f  ( E_f +m)
     \Delta_1  z^2}{2 m}\,A_7 
  \Bigg) \qquad \\[2.8ex]
\label{eq:Pi3G3_ns}
  \Pi^a_3(\Gamma_3) &=& 0
\end{eqnarray}
The matrix elements above involve more $A_i$ compared to the symmetric frame, and, for instance, Eqs.~\eqref{eq:Pi0G0_ns} - \eqref{eq:Pi0G2_ns} contain $A_1$, $A_3$, $A_4$, $A_5$, $A_6$, and $A_8$. 
Consequently, $(A_1,\,A_5,\,A_6)$ and $(A_3,\,A_4,\,A_8)$ are not decoupled in the asymmetric frame, contrary to the symmetric one.
In fact, in the symmetric frame, the coefficients of $A_3$, $A_4$, $A_8$ in the $\gamma_0$ matrix elements vanish due to $E_i=E_f$. 
Similarly, $A_4$, $A_5$, $A_6$, drop out the matrix elements of $\gamma_1$ and $\gamma_2$ except for the projector $\Gamma_3$.
The above simplifications lead to the aforementioned decoupling and to more compact expressions in the symmetric frame.
Similarly to the symmetric frame, $A_2$ and $A_7$ appear only in the decomposition of $\Pi_3$.

To summarize, Eqs.~\eqref{eq:Pi0G0_s} - \eqref{eq:Pi3G3_s} and Eqs.~\eqref{eq:Pi0G0_ns} - \eqref{eq:Pi3G3_ns}, consist the basis of required matrix elements for the calculation of the Lorentz invariant amplitudes, and are used to present numerical results for the $A_i$ for the setup of Table~\ref{tab:stat}.

\subsection{Decomposition of Lorentz-invariant amplitudes}

The Lorentz-invariant amplitudes can be disentangled using the parametrizations given in the previous subsection, which requires the inversions of Eqs.~\eqref{eq:Pi0G0_s} - \eqref{eq:Pi3G3_s} and Eqs.~\eqref{eq:Pi0G0_ns} - \eqref{eq:Pi3G3_ns}, in the symmetric and asymmetric frame, respectively.
As mentioned previously, the matrix elements of the operators $\gamma_0,\,\gamma_1,\,\gamma_2$ are sufficient to disentangle $A_1$, $A_3$, $A_4$, $A_5$, $A_6$, and $A_8$, in both frames. 
$A_1$, $A_5$, and $A_6$ can then be incorporated to the parametrization of the matrix elements of $\gamma_3$ to extract $A_2$ and $A_7$.

The analytic inversion of the equations can, potentially, lead to expressions with complicated kinematic coefficients. 
For simplicity in the presentation, we show the expressions for $\vec{\Delta}= (\Delta,0,0)$.
We use a subscript $s$ and $a$ in the matrix elements to differentiate between the two frames; $A_i$ are frame-independent and do not carry such an index.
The expressions for the symmetric frame take the form
\begin{eqnarray}
\label{eq:A1_s} 
A_1 &=&  \frac{ \left(m
    (E +m)+P_3^2\right)}{E  (E +m)} \Pi^s_0(\Gamma_0) 
    -i\, \frac{P_3  \Delta   }{2 E  (E +m)} \Pi^s_0(\Gamma_2)
    -\frac{ \Delta }{2 E } \Pi^s_2(\Gamma_3)\,,   \\[3ex]
A_2 &=& 
\frac{P_3  \left( m (E +m)+P_3^2\right)}{2 m^4 z} \Pi^s_0(\Gamma_0) 
-i\, \frac{P_3^2  \Delta  }{4 m^4 z} \Pi^s_0(\Gamma_2)
-\frac{P_3  \Delta   (E +m)}{4 m^4 z}\Pi^s_2(\Gamma_3) \nonumber \\[1ex]
&& -i\,\frac{ \left(m (E +m)+P_3^2\right)}{m^2 z(E +m)}\Pi^s_3(\Gamma_0) 
    -\frac{P_3  \Delta  }{2 m^2 z (E +m)}\Pi^s_3(\Gamma_2)
    \,,    \\[3ex]
A_3 &=&  i\, \frac{ \left(m (E +m)+P_3^2\right)}{ \Delta  (E +m)}\Pi^s_1(\Gamma_0)  
+\frac{P_3}{2 (E +m)} \Pi^s_1(\Gamma_2) \,,   \\[3ex]
A_4 &=&  -\frac{ 1 }{m z} \Pi^s_2(\Gamma_1)\,,  \\[3ex]
A_5 &=&   -\frac{E  }{ \Delta }  \Pi^s_2(\Gamma_3)\,,   \\[3ex]
A_6 &=& 
\frac{P_3 }{2 E  z (E +m)}\Pi^s_0(\Gamma_0)
+i\,\frac{\left(P_3^2-E  (E +m)\right)}{E   \Delta  z(E +m)}\Pi^s_0(\Gamma_2) 
+\frac{P_3  }{E   \Delta  z}\Pi^s_2(\Gamma_3) \,,    \\[3ex]
A_7 &=& 
+\frac{P_3^2 }{4 m^4 z^2}\Pi^s_0(\Gamma_0)
+i\, \frac{ P_3 \left(P_3^2-E  (E +m)\right)}{2 m^4  \Delta  z^2} \Pi^s_0(\Gamma_2) -\frac{ (E +m) (E -P_3)(E +P_3)}{2 m^4  \Delta  z^2}\Pi^s_2(\Gamma_3)
\nonumber \\[1ex]
&& -i\,\frac{ P_3 }{2 m^2 z^2 (E +m)}\Pi^s_3(\Gamma_0) 
+\frac{ \left(P_3^2-E  (E +m)\right)}{ \Delta  m^2 z^2  (E +m)}\Pi^s_3(\Gamma_2)
\,,    \\[3ex]
\label{eq:A8_s}
A_8 &=&  \frac{\,i\, P_3 }{2  \Delta  z (E +m)}\Pi^s_1(\Gamma_0) 
    +\frac{\left(E  (E +m)-P_3^2\right)}{ \Delta ^2 z (E +m)} \Pi^s_1(\Gamma_2)  
    +\frac{m}{ \Delta ^2 z} \Pi^s_2(\Gamma_1)  \,.
\end{eqnarray}
Below, we give $A_i$ in the asymmetric frame for $\vec{\Delta}=(\Delta,0,0)$, which, as anticipated, has more complicated expressions than the symmetric frame.
\begin{eqnarray}
\label{eq:A1_ns} 
A_1 &=& 
\frac{2 m^2}{E_f(E_i+m)} \frac{\Pi^a_0(\Gamma_0)}{K}
+i\, \frac{2 (E_f-E_i) P_3 m^2}{E_f (E_f+m) (E_i+m)  \Delta }\frac{\Pi^a_0(\Gamma_2)}{K} 
 +\frac{2 (E_i-E_f) P_3  m^2}{E_f (E_f+E_i) (E_f+m) (E_i+m)}\frac{\Pi^a_1(\Gamma_2)}{K}  \nonumber \\[1.2ex]
&& + i\,\frac{2 (E_i-E_f) m^2}{E_f (E_i+m)  \Delta }\frac{\Pi^a_1(\Gamma_0)}{K} 
+\frac{2 (E_i-E_f) P_3  m^2}{E_f (E_f+E_i) (E_f+m) (E_i+m)}\frac{\Pi^a_2(\Gamma_1)}{K}
+\frac{2 (E_f-E_i) m^2}{E_f (E_i+m)  \Delta } \frac{\Pi^a_2(\Gamma_3)}{K}  \,,  \\[4.5ex]
A_2 &=& 
 \frac{2 P_3 }{E_f (E_i+m) z}\frac{\Pi^a_0(\Gamma_0)}{K} 
 + i\,\frac{2 (E_f-E_i) (E_f-m) }{E_f (E_i+m)  \Delta  z} \frac{\Pi^a_0(\Gamma_2)}{K}
 +\frac{2 (E_i-E_f) \,i\, P_3 }{E_f (E_i+m)  \Delta  z} \frac{\Pi^a_1(\Gamma_0)}{K}\nonumber \\[1.2ex]
&&  -\frac{2 (E_f-E_i) (E_f-m)}{E_f (E_f+E_i) (E_i+m) z}\frac{\Pi^a_1(\Gamma_2)}{K} -\frac{2 (E_f-E_i) (E_f-m)}{E_f (E_f+E_i)(E_i+m) z}  \frac{\Pi^a_2(\Gamma_1)}{K} 
 +\frac{2 P_3 (E_f-E_i)}{E_f (E_i+m)  \Delta  z} \frac{\Pi^a_2(\Gamma_3)}{K}\nonumber \\[1.2ex]
&&  - i\,\frac{2}{(E_i+m) z}\frac{\Pi^a_3(\Gamma_0)}{K} 
 +\frac{2 (E_f-E_i) P_3 }{(E_f+m) (E_i+m)  \Delta  z}\frac{\Pi^a_3(\Gamma_2)}{K}    \,,    \\[4.5ex]
A_3 &=&  
\frac{ m^2}{E_f(E_i+m)} \frac{\Pi^a_0(\Gamma_0)}{K} 
+i\,\frac{(E_f-E_i)  P_3 m^2}{E_f (E_f+m) (E_i+m)  \Delta }\frac{\Pi^a_0(\Gamma_2)}{K}  
+i\, \frac{(E_f+E_i) m^2}{E_f (E_i+m)  \Delta } \frac{\Pi^a_1(\Gamma_0)}{K} \nonumber \\[1.2ex]
&& +\frac{P_3 m^2}{E_f (E_f+m) (E_i+m)}   \frac{\Pi^a_1(\Gamma_2)}{K}\,,   \\[4.5ex]
A_4 &=&  
 \frac{2
    (E_f-E_i) m P_3 }{E_f (E_f+m)(E_i+m)  \Delta  z} \frac{\Pi^a_2(\Gamma_3)}{K} 
    -\frac{2 m }{E_f (E_i+m) z}  \frac{\Pi^a_2(\Gamma_1)}{K}  \,,  \\[4.5ex]
A_5 &=&   
 \frac{m^2 P_3 }{E_f (E_f+m) (E_i+m)}\frac{\Pi^a_2(\Gamma_1)}{K}
 -\frac{(E_f+E_i) m^2  }{E_f (E_i+m)  \Delta } \frac{\Pi^a_2(\Gamma_3)}{K} \,,   \\[4.5ex]
A_6 &=& 
\frac{P_3 m^2}{E_f^2 (E_f+m) (E_i+m) z} \frac{\Pi^a_0(\Gamma_0)}{K} 
+i\,\frac{(E_f-E_i-2 m) m^2}{E_f^2 (E_i+m)  \Delta  z}\frac{\Pi^a_0(\Gamma_2)}{K}  
+i\,\frac{(E_i-E_f)  P_3 m^2}{E_f^2 (E_f+m) (E_i+m)  \Delta  z} \frac{\Pi^a_1(\Gamma_0)}{K} \nonumber \\[1.2ex]
&&+\frac{(-E_f+E_i+2 m) m^2}{E_f^2 (E_f+E_i) (E_i+m) z} \frac{\Pi^a_1(\Gamma_2)}{K} 
+\frac{2 (m-E_f) m^2}{E_f^2 (E_f+E_i) (E_i+m) z} \frac{\Pi^a_2(\Gamma_1)}{K} 
+\frac{2 P_3  m^2}{E_f^2 (E_i+m) \Delta  z} \frac{\Pi^a_2(\Gamma_3)}{K} \,,    \\[4.5ex]
A_7 &=& 
\frac{(E_f-m)}{E_f^2 (E_i+m) z^2} \frac{\Pi^a_0(\Gamma_0)}{K} 
+i\, \frac{(E_f-E_i-2 m) P_3}{E_f^2 (E_i+m)  \Delta  z^2} \frac{\Pi^a_0(\Gamma_2)}{K}  
-i\,\frac{(E_f-E_i)  (E_f-m) }{E_f^2 (E_i+m)  \Delta  z^2} \frac{\Pi^a_1(\Gamma_0)}{K} \nonumber \\[1.2ex]
&&
+\frac{(-E_f+E_i+2 m) P_3}{E_f^2 (E_f+E_i) (E_i+m) z^2} \frac{\Pi^a_1(\Gamma_2)}{K} 
+\frac{\left(2 m^2+E_f (E_i-E_f)\right) P_3}{E_f^2 (E_f+E_i) (E_f+m) (E_i+m)
    z^2} \frac{\Pi^a_2(\Gamma_1)}{K} \nonumber \\[1.2ex]
&& 
+\frac{\left(E_f(E_f-E_i)-2 m^2\right) }{E_f^2 (E_i+m)  \Delta  z^2} \frac{\Pi^a_2(\Gamma_3)}{K}
-i\, \frac{P_3 }{E_f (E_f+m)(E_i+m) z^2}\frac{\Pi^a_3(\Gamma_0)}{K} 
+\frac{(E_f-E_i-2 m)}{E_f (E_i+m)  \Delta  z^2} \frac{\Pi^a_3(\Gamma_2)}{K}\,,    \\[4.5ex]
\label{eq:A8_ns}
A_8 &=& 
\frac{P_3 m^2}{2 E_f^2 (E_f+m) (E_i+m) z} \frac{\Pi^a_0(\Gamma_0)}{K}
+i\,\frac{(E_f-E_i-2 m) m^2}{2 E_f^2 (E_i+m)  \Delta  z}\frac{\Pi^a_0(\Gamma_2)}{K}
+ i\,\frac{(E_f+E_i) P_3 m^2}{2 E_f^2 (E_f+m) (E_i+m)  \Delta  z} \frac{\Pi^a_1(\Gamma_0)}{K} \nonumber \\[1.2ex]
&&
+\frac{(E_f-E_i-2 m) m^2}{2 E_f^2 (E_f-E_i) (E_i+m) z} \frac{\Pi^a_1(\Gamma_2)}{K}
-\frac{m^3}{E_f^2 (E_f-E_i) (E_i+m) z}\frac{\Pi^a_2(\Gamma_1)}{K} 
+\frac{P_3 m^3}{E_f^2 (E_f+m)(E_i+m)  \Delta  z}\frac{\Pi^a_2(\Gamma_3)}{K} \,. \qquad \qquad
\end{eqnarray}

\newpage
\subsection{$H$ and $E$ GPDs}

\label{sec:FH_FE_setup}

One of the main motivations of this work is to extract the twist-2 light-cone GPDs for unpolarized quarks, that is $H$ and $E$. 
We begin by constructing the quasi GPDs in coordinate space from the $\gamma^0$, ${\cal H}_0$ and ${\cal E}_0$, using the Lorentz-invariant amplitudes, $A_i$.
With the $A_i$ being frame-independent, one can relate ${\cal H}_0$ and ${\cal E}_0$ to the matrix elements of either frame; this is a powerful relation, as a calculation in the asymmetric frame requires less computational resources (see Sec.~\ref{subsec:calc_params}). With this in mind, and using the definition of Eq.~\eqref{e:historic} for the quasi-GPDs in the symmetric frame, we adopt Eqs.~\eqref{eq:quasiH_symm} - \eqref{eq:quasiE_symm}, which for zero skewness simplify to
\begin{eqnarray}
\label{eq:FHs}
{\cal H}_0^s(A_i^s;z) &=&
A_1 + \frac{z (\Delta_1^2 + \Delta^2_2)}{2 P_3}  {{A_6}}\,,  \\[3ex]
\label{eq:FEs}
{\cal E}_0^s(A_i^s;z)  &=&
- A_1 - \frac{m^2 z }{P_3}  {{A_4}} + 2 A_5 - \frac{z \left( 4 E^2 + \Delta_1^2 + \Delta^2_2\right)}{2P_3}   {{A_6}}   \,.
\end{eqnarray}
For simplicity we suppresse the remaining arguments of the quasi-GPDs and the arguments of the amplitudes. It is useful to rewrite Eqs.~\eqref{eq:FHs} - \eqref{eq:FEs} in terms of matrix elements in the symmetric frame that, for $\vec{\Delta}=(\Delta,0,0)$, leads to
\begin{eqnarray}
\label{eq:FHs_Pi}
{\cal H}_0^s(A_i^s;z)  &=& \Pi^s_0(\Gamma_0)-i\,\frac{ \Delta}{2 P_3}  \Pi^s_0(\Gamma_2)\,,  \\[3ex]
\label{eq:FEs_Pi}
{\cal E}_0^s(A_i^s;z)  &=& -\frac{m}{E +m}  \Pi^s_0(\Gamma_0)  -i\,\frac{2  m \left(m (E +m)+P_3^2\right)}{P_3 \Delta (E +m)}\Pi^s_0(\Gamma_2)\,.
\,
\end{eqnarray}
As expected, Eqs.~\eqref{eq:FHs_Pi} - \eqref{eq:FEs_Pi} are the usual expressions extracted from the matrix elements of operator $\gamma_0$ previously used for the unpolarized GPDs~\cite{Alexandrou:2020zbe}. 
We also find that the $z$-dependence in the kinematic factors cancels out. 
While the above observations validate the methodology of relating ${\cal H}_0$ and ${\cal E}_0$ to $A_i$, using Eqs.~\eqref{eq:FHs_Pi} - \eqref{eq:FEs_Pi} is computationally costly and not optimal for lattice QCD calculations to extract ${\cal H}_0^s(A_i^s;z)$ and ${\cal E}_0^s(A_i^s;z)$. 
Instead, one may still employ the historically-used symmetric frame definition for the unpolarized quasi-GPDs (Eqs.~\eqref{eq:FHs} - \eqref{eq:FEs}), but perform a calculation of the matrix elements in the asymmetric frame for the extraction of the amplitudes $A_i$.
We define such a case by ${\cal H}_0^s(A_i^a;z)$ and ${\cal E}_0^s(A_i^a;z)$.
Such a possibility is due to the frame invariance of the amplitudes $A_i$. 
As outlined in Sec.~\ref{sec:strategy}, the kinematic coefficients between the two frames are related via a Lorentz boost transformation.

\medskip
An alternative approach is to define the quasi-GPDs in the asymmetric frame using only the matrix elements of $\gamma_0$ (Eq.~\eqref{e:historic}). In such a case, one can relate ${\cal H}_0^a$ and ${\cal E}_0^a$ to the $A_i$ as
\begin{eqnarray}
\label{eq:FHa}
{\cal H}_0^a(A_i^a;z)  & = & 
A_1 + \frac{\Delta_0}{P_0} {{A_3}} + \frac{m^2 z \Delta_0}{2P_0P_3}  {{A_4}}+ \frac{z (\Delta_0^2 + \Delta_1^2+ \Delta_2^2)}{2 P_3}  {{A_6}} + \frac{z (\Delta_0^3 + \Delta_0 (\Delta_1^2+ \Delta_2^2))}{2P_0 P_3}   {{A_8}} \,,\\[3ex]
\label{eq:FEa}
{\cal E}_0^a(A_i^a;z)   &= &
- A_1 - \frac{\Delta_0}{P_0}  {{A_3}} - \frac{m^2 z (\Delta_0+2P_0)}{2P_0P_3}  {{A_4}} + 2 A_5 - \frac{z \left(\Delta_0^2 + 2 P_0 \Delta_0 + 4 P_0^2 + \Delta_1^2+ \Delta_2^2 \right)}{2P_3}   {{A_6}} \nonumber\\[1ex] 
&&  - \frac{z \Delta_0 \left(\Delta_0^2 + 2 \Delta_0 P_0 + 4 P_0^2 + \Delta_1^2+ \Delta_2^2\right)}{2P_0 P_3}   {{A_8}}\,.
\end{eqnarray}
Substituting $A_i$ in the asymmetric frame for $\vec{\Delta}=(\Delta,0,0)$ (Eqs.~\eqref{eq:A1_ns} - \eqref{eq:A8_ns}), we find
\begin{eqnarray}
\label{eq:FHa_Pi}
{\cal H}_0^a(A_i^a;z)  &=& i\, \frac{2 m^2(E_f-E_i)}{KP_3 \Delta  (E_i+m)}\Pi^a_0(\Gamma_2)
+\frac{2 m^2 (E_f+E_i+2 m)}{K(E_f+E_i) (E_f+m) (E_i+m)} \Pi^a_0(\Gamma_0) \,, \\[3ex]
\label{eq:FEa_Pi}
{\cal E}_0^a(A_i^a;z)  &=&  -\frac{4 m^3}{K(E_f+E_i) (E_f+m)(E_i+m)}\Pi^a_0(\Gamma_0)
- i\,\frac{4 m^3}{KP_3 \Delta  (E_i+m)} \Pi^a_0(\Gamma_2)\,.
\end{eqnarray}
We note that (${\cal H}_0^s$, ${\cal E}_0^s$) differ from (${\cal H}_0^a$, ${\cal E}_0^a$) due to their Lorentz non-invariant definition. 
However, in the infinite momentum limit both approach the correct light-cone limit.
While working at finite momentum boost, a different functional form in the two frames is found.  
Here, we compare ${\cal H}_0$ and ${\cal E}_0$ between the two frames for pedagogical reasons, as exact agreement between them is not anticipated. 
Theoretically, there is no preference in using ${\cal H}_0^a$ and ${\cal E}_0^a$ versus ${\cal H}_0^s$ and ${\cal E}_0^s$. 
Historically, the latter is employed, and one convenient approach is the extraction of ${\cal H}_0^s(A_i^a;z) $ and ${\cal E}_0^s(A_i^a;z)$ which uses matrix elements in asymmetric frames. 

Another aspect of this work follows a different approach from the one outlined above. In particular, we propose a Lorentz-invariant quasi-GPDs ${\cal H}$ and ${\cal E}$ definition, as given in Eqs.~\eqref{eq:Hq_improved} - \eqref{eq:Eq_improved}. Such definitions may have faster convergence to light-cone GPDs. 
However, further theoretical and numerical investigation is required to reach concrete conclusions.
The expressions of Eqs.~\eqref{eq:Hq_improved} - \eqref{eq:Eq_improved} simplify for zero skewness, that is
\begin{eqnarray}
\label{eq:FH_impr}
{\cal H}(A_i^{s/a};z)   &= & A_1 \,,\\[3ex]
\label{eq:FE_impr}
{\cal E}(A_i^{s/a};z)   &= & - A_1 + 2 A_5 + 2 z P_3 A_6 \,.
\end{eqnarray}
Being Lorentz invariant, the above definitions are equivalent in both frames, that is, ${\cal H}(A_i^s;z) ={\cal H}(A_i^a;z)$ and ${\cal E}(A_i^s;z) ={\cal E}(A_i^a;z)$. 
For completeness, we provide the expressions of ${\cal H}$ and ${\cal E}$ using matrix elements in each frame. 
As above, we use as an example the case $\vec{\Delta}=(\Delta,0,0)$, which may be written in terms of matrix elements in the symmetric frame
\begin{eqnarray}
{\cal H}(A_i^s;z)  &=& \frac{ \left(m
    (E +m)+P_3^2\right)}{E  (E +m)} \Pi^s_0(\Gamma_0) 
    -i\, \frac{P_3 \Delta   }{2 E  (E +m)} \Pi^s_0(\Gamma_2)
    -\frac{\Delta }{2 E } \Pi^s_2(\Gamma_3) \,,  \\[3ex]
{\cal E}(A_i^s;z)   &=& -\frac{m}{ E }\Pi^s_0(\Gamma_0) - i\, \frac{2 m  P_3 }{ E   \Delta  }\Pi^s_0(\Gamma_2)-\frac{2
    m^2 }{ E   \Delta  }\Pi^s_2(\Gamma_3)\,,
\end{eqnarray}
or, alternatively in the asymmetric frame
\begin{eqnarray}
{\cal H}(A_i^a;z)  &=& \frac{2 m^2}{E_f(E_i+m)} \frac{\Pi^a_0(\Gamma_0)}{K}
+i\, \frac{2 (E_f-E_i) P_3 m^2}{E_f (E_f+m) (E_i+m) \Delta }\frac{\Pi^a_0(\Gamma_2)}{K} 
 +\frac{2 (E_i-E_f) P_3  m^2}{E_f (E_f+E_i) (E_f+m) (E_i+m)}\frac{\Pi^a_1(\Gamma_2)}{K}  \nonumber \\[1ex]
&& + i\,\frac{2 (E_i-E_f) m^2}{E_f (E_i+m) \Delta }\frac{\Pi^a_1(\Gamma_0)}{K} 
+\frac{2 (E_i-E_f) P_3  m^2}{E_f (E_f+E_i) (E_f+m) (E_i+m)}\frac{\Pi^a_2(\Gamma_1)}{K}
+\frac{2 (E_f-E_i) m^2}{E_f (E_i+m) \Delta } \frac{\Pi^a_2(\Gamma_3)}{K}  \,,   \\[3ex]
{\cal E}(A_i^a;z)  &=& 
-\frac{2 m^3}{ E_f ^2 ( E_i +m)}\frac{\Pi_0^a(\Gamma_0)}{K} 
-i\, \frac{2   m^3  P_3 ( E_f + E_i +2
    m)}{ E_f ^2  \Delta   ( E_f +m) ( E_i +m)} \frac{\Pi^a_0(\Gamma_2) }{K} 
   +\frac{2 m^3  P_3 
    ( E_f + E_i +2 m)}{ E_f ^2
    ( E_f + E_i ) ( E_f +m) ( E_i +m)}\frac{\Pi^a_1(\Gamma_2)}{K} \qquad \nonumber \\[1ex]
    &&
     + i\,\frac{2  
    m^3  ( E_f - E_i )}{ E_f ^2  \Delta  
    ( E_i +m)}\frac{\Pi^a_1(\Gamma_0)}{K} 
    +\frac{4 m^4  P_3  }{ E_f ^2
    ( E_f + E_i ) ( E_f +m) ( E_i +m)}\frac{\Pi^a_2(\Gamma_1)}{K} -\frac{4 m^4
    }{ E_f ^2  \Delta   ( E_i +m)} \frac{\Pi^a_2(\Gamma_3)}{K} \,.
\end{eqnarray}
We note that the definition of ${\cal H}$ and ${\cal E}$ can be interpreted as the construction of a new operator that is a combination of $\gamma_0$, $\gamma_1$ and $\gamma_2$.
We emphasize that it is important to provide a comparison of the ${\cal H}$ and ${\cal E}$ GPDs in the two frames at the same value of $t$. This requires $\Delta^{a}_\perp \neq \Delta^{s}_\perp$. Such a relation is $P^3$-dependent, but for the values of $P^3$ employed in this work (see Sec.~\ref{subsec:calc_params}), are numerically similar ($t^s=-0.69$ GeV$^2$, $t^a=-0.64$ GeV$^2$).

\subsection{Renormalization and Matching}
\label{s:renorm_match}
A schematic structure of the historical definition of a quasi-GPD (say ${\cal{H}}$) is,
\begin{align}
    {\cal{H}}_0 & \rightarrow c_0 \langle .. \gamma^0 .. \rangle \, ,
    \label{e:non_cov}
\end{align}
while, as we will show in Sec.~\ref{sec:FH_FE_setup}, a schematic structure of the Lorentz-invariant definition is,
\begin{align}
   {\cal{H}} & \rightarrow c_0 \langle .. \gamma^0 .. \rangle + c_1 \langle .. \gamma^1 .. \rangle + c_2 \langle .. \gamma^2 .. \rangle \, .
    \label{e:cov}
\end{align}
Here, $(c_0,c_1,c_2)$ are frame-dependent kinematic factors. Specifically, the linear combination $c_1 \langle .. \gamma^1 .. \rangle + c_2 \langle .. \gamma^2 .. \rangle$ is such that it eliminates the additional amplitudes (with respect to the light-cone case) present in $c_0 \langle .. \gamma^0 .. \rangle$ alone, thereby (potentially) projecting the resulting quasi-GPD faster to the light-cone GPD at the leading order, which is Lorentz-invariant. The question that we want to discuss here concerns the strategy to renormalize the linear combination of $(\gamma^0, \gamma^1, \gamma^2)$ in Eq.~(\ref{e:cov}). Since the UV divergence of the quark bilinear operator is multiplicative and independent of the Dirac $\Gamma$ matrix~\cite{Ji:2017oey,Ishikawa:2017faj,Green:2017xeu}, one can just use the renormalization factor for $\gamma^0$ to renormalize the quasi-GPD. Besides, since $\gamma^0$ and $\gamma^{1,2}$ are free from $O(1)$ operator mixings due to chiral symmetry breaking on the lattice~\cite{Constantinou:2017sej,Chen:2017mie}, this issue is also avoided in the Lorentz-invariant definition. Note that for our numerical results, we will use the matching coefficient from Ref.~\cite{Liu:2019urm}. It is known that the GPD matching coefficient for the operator $\gamma^0$ reduces to that for the corresponding PDF when $\xi =0$, even if $t \neq 0$~\cite{Radyushkin:2019owq,Liu:2019urm}. The PDF matching coefficient for $\gamma^0$ is for the amplitude $A_1$, which is also the only contributing amplitude to the LI definition of the GPD when $\xi =0$. Therefore, the matching coefficients for the $\gamma^0$ and the LI definitions of the GPDs are equal. We will elaborate this point more, including the general case of $\xi \neq 0$, in a forthcoming publication, along with exploring hybrid renormalization~\cite{Ji:2020brr} for quasi-GPDs, as well as an improved RI/MOM-type~\cite{Constantinou:2017sej,Stewart:2017tvs} scheme to eliminate unwanted finite lattice contributions~\cite{Constantinou:2022aij}.

\subsection{Calculation parameters}
\label{subsec:calc_params}

We calculate the proton matrix elements of the non-local vector operator containing spatially-separated quark fields, in the $\hat{z}$ direction, connected by a Wilson line.
The proton states are momentum-boosted with nonzero momentum transfer between the initial and final state,
\begin{equation}
\label{eq:ME}
h^\mu_V(\Gamma_\kappa,z,p_f,p_i)\equiv \langle N(p_f)|\bar\psi\left(z\right) \gamma_j {\cal W}(0,z)\psi\left(0\right)|N(p_i)\rangle\,, \quad \mu, \kappa: 0,1,2,3\,.
\end{equation}
$|N(P_i)\rangle$ and $|N(P_f)\rangle$ are the initial (source) and final (sink) state of the proton.
The remaining variables are defined previously.
We use momentum smearing~\cite{Bali:2016lva} to improve the overlap with the proton ground state and suppress gauge noise; Ref.~\cite{Alexandrou:2016jqi} demonstrated that the method is essential for non-local operators.
More relevant to the present analysis, it was found that the statistical noise is $z$-dependent and reduces by a factor of 4-5 in the real part, and 2-3 in the imaginary part of the unpolarized GPDs~\cite{Alexandrou:2020zbe}.
The matrix element $h^\mu_V$ is extracted from the ratio
\begin{equation}
R^\mu (\Gamma_\kappa, p_f, p_i; t_s, \tau) = \frac{C^{\rm 3pt}_\mu (\Gamma_\kappa, p_f, p_i; t_s, \tau)}{C^{\rm 2pt}(\Gamma_0, p_f;t_s)} \sqrt{\frac{C^{\rm 2pt}(\Gamma_0, p_i, t_s-\tau)C^{\rm 2pt}(\Gamma_0, p_f, \tau)C^{\rm 2pt}(\Gamma_0, p_f, t_s)}{C^{\rm 2pt}(\Gamma_0, p_f, t_s-\tau)C^{\rm 2pt}(\Gamma_0, p_i, \tau)C^{\rm 2pt}(\Gamma_0, p_i, t_s)}}\,,
 \end{equation}
where $C^{\rm 2pt}$ and $C^{\rm 3pt}$, are the two- and three-point correlators. $\tau$ is the current insertion time and $t_s$ is the source-sink time separation; the source is taken at zero timeslice.
We extract the ground-state contribution to $h^\mu_V$ from $R^\mu$ by taking a plateau fit with respect to $\tau$ in a region of convergence, indicated by $\Pi^j(\Gamma_\kappa)$. 
For simplicity, the dependence on $z$, $p_f$, and $p_i$ is not shown explicitly in the matrix elements $\Pi^j(\Gamma_\kappa)$.

The calculation is performed on a gauge ensemble of $N_f=2+1+1$ twisted-mass fermions including clover improvement, and Iwasaki-improved gluons~\cite{Alexandrou:2018egz}.
The quark masses correspond to a pion with mass 260 MeV.
The ensemble has a volume of $32^3 \times 64$ and lattice spacing $a=0.093$ fm.
Several of the matrix elements beyond the commonly used $\gamma_0$, have small and noisy signal. 
To keep the statistical noise under control, we use a source-sink separation of $t_s=10 a = 0.93$ fm.
The study of excited states via calculations of multiple time separations is beyond the scope of this work.
In future precision calculations, we will include excited states studies.
In Table~\ref{tab:stat}, we give the statistics for the calculation in the symmetric and the asymmetric frame. The Lorentz-invariant amplitudes have definite symmetry with respect to $P_3 \to -P_3$, $z\to -z$ and $\vec{\Delta} \to -\vec{\Delta}$ (see Appendix~\ref{sec:AppB}) and, therefore, we combine all data contributing to the same value of momentum transfer squared, $t$. 
We remind the reader that for the kinematic setup in the two frames as given in Table~\ref{tab:stat}, $t^s$ and $t^a$ are not the equal, but are sufficiently close to each other for a comparison between the two frames to be meaningful. We emphasize that the asymmetric frame is computationally more efficient, as one can obtain more than one value of $\vec{\Delta}$ within the same computational cost.
\begin{table}[h!]
\begin{center}
\renewcommand{\arraystretch}{1.9}
\begin{tabular}{lcccc|cccc}
\hline
frame & $P_3$ [GeV] & $\quad \mathbf{\Delta}$ $[\frac{2\pi}{L}]\quad$ & $-t$ [GeV$^2$] & $\quad \xi \quad $ & $N_{\rm ME}$ & $N_{\rm confs}$ & $N_{\rm src}$ & $N_{\rm tot}$\\
\hline
symm      & $\pm$1.25 &($\pm$2,0,0), (0,$\pm$2,0)  &0.69   &0   &8   &249 &8  &15936 \\
asymm  & $\pm$1.25 &($\pm$2,0,0), (0,$\pm$2,0)  &0.64   &0   &8   &269 &8  &17216\\
\hline
\end{tabular}
\caption{Statistics for the symmetric and asymmetric frame at zero skewness and $t_s=10 a$. $N_{\rm ME}$, $N_{\rm confs}$, $N_{\rm src}$ and $N_{\rm total}$ is the number of matrix elements, configurations, source positions per configuration and total statistics, respectively.}
\label{tab:stat}
\end{center}
\end{table}

\vspace*{-0.35cm}
\section{Lattice results}

\subsubsection{Matrix elements and Lorentz-invariant amplitudes}

In this section, we present selected matrix elements in the two frames. 
We point out that the matrix elements in the symmetric frame have definite symmetries with respect to $P_3 \to -P_3$, $z \to -z$, and $\vec{\Delta} \to -\vec{\Delta}$. 
However, such symmetries are, in general, not present in the asymmetric frame, which prevents one from taking averages of the matrix elements for $\pm P_3$, $\pm z$, or $\pm \vec{\Delta}$ before extracting the $A_i$; the amplitudes have definite symmetries that are outlined in Appendix~\ref{sec:AppB}. 
For consistency in the analysis, the averaging over, e.g., $\pm z$ is done at the level of $A_i$ for both frames.

\begin{figure}[h!]
    \centering
    \includegraphics[scale=0.23]{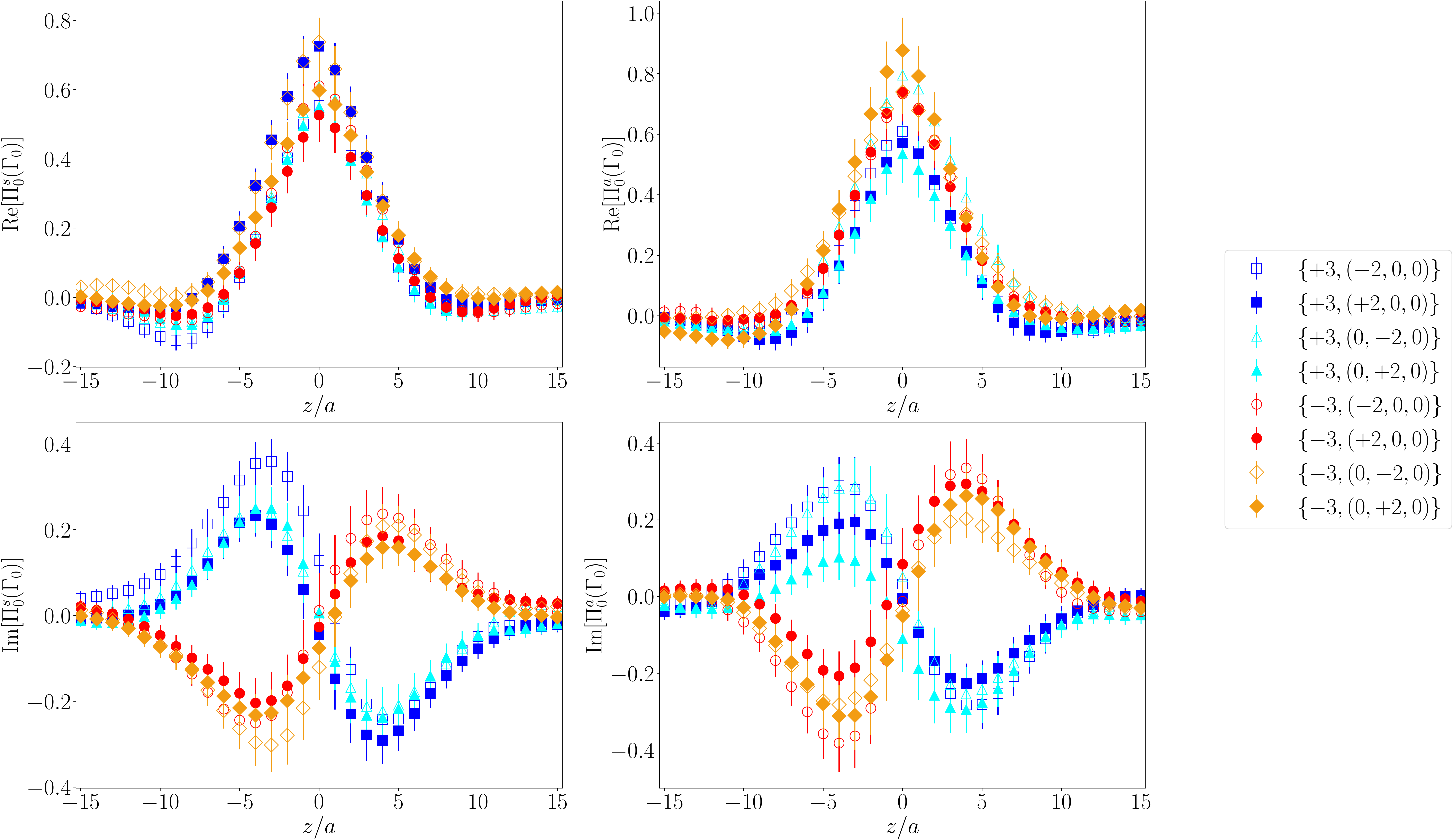}
    \vspace*{-0.35cm}
    \caption{Bare matrix element $\Pi_0(\Gamma_0)$ in the symmetric frame (left) and in the asymmetric frame (right), for $|P_3|=1.25$ GeV and $t=-0.69$ GeV$^2$ ($t=-0.64$ GeV$^2$) for the symmetric (asymmetric) frame. The top (bottom) panel corresponds to the real (imaginary) part. The notation in the legend is $\{P_3,\vec{\Delta}\}$ in units of $2\pi/L$. }
    \label{fig:Pi0G0}
\end{figure}

\begin{figure}[h!]
    \centering
\includegraphics[scale=0.23]{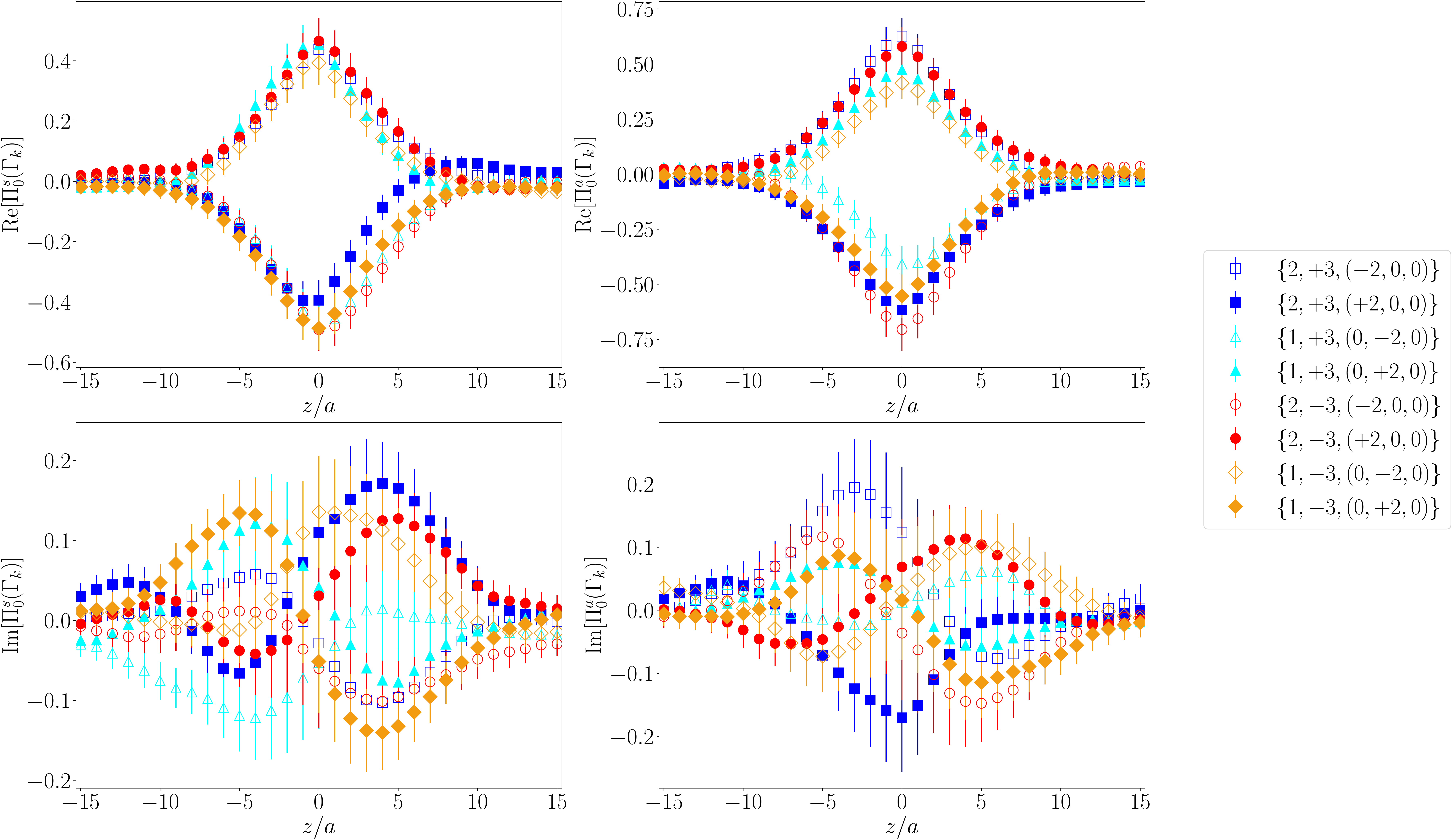}
    \vspace*{-0.35cm}
    \caption{Bare matrix elements $\Pi_0(\Gamma_1)$ and $\Pi_0(\Gamma_2)$ in the symmetric frame (left) and in the asymmetric frame (right). The legend indicates $\{\kappa,P_3,\vec{\Delta}\}$ in units of $2\pi/L$, with $\kappa$ corresponding to $\Pi_0(\Gamma_\kappa)$.  The remaining notation is the same as Fig.~\ref{fig:Pi0G0}.}
    \label{fig:Pi0G1}
\end{figure}

\begin{figure}[h!]
    \centering
    \includegraphics[scale=0.23]{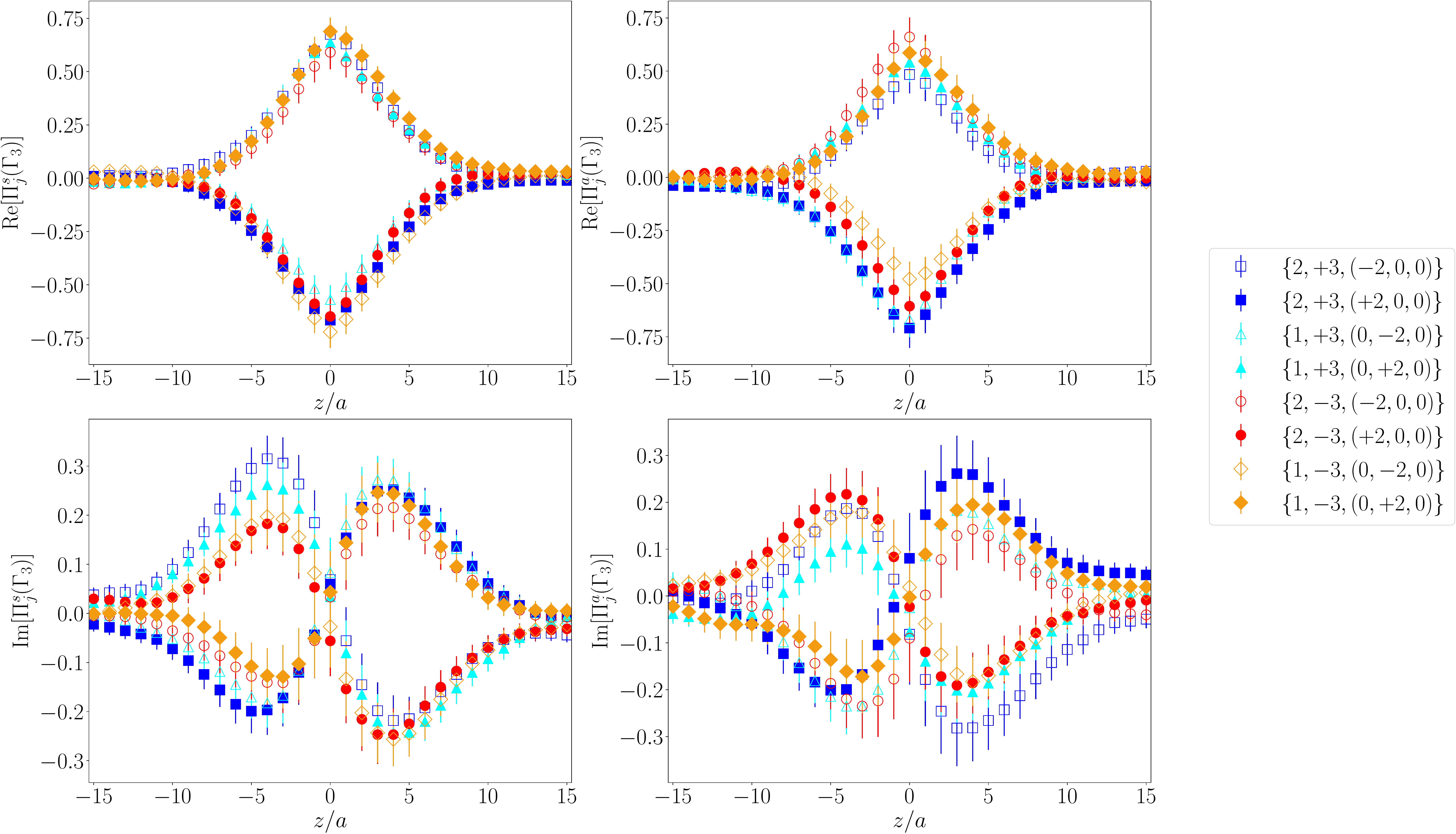}
    \vspace*{-0.35cm}
    \caption{Bare matrix elements $\Pi_2(\Gamma_3)$ and $\Pi_1(\Gamma_3)$ in the symmetric frame (left) and in the asymmetric frame (right). The legend indicates $\{j, P_3,\vec{\Delta}\}$ in units of $2\pi/L$, with $j$ corresponding to $\Pi_j(\Gamma_3)$.  The remaining notation is the same as Fig.~\ref{fig:Pi0G0}.  }
    \label{fig:Pi2G3}
\end{figure}

\begin{figure}[h!]
    \centering
    \includegraphics[scale=0.23]{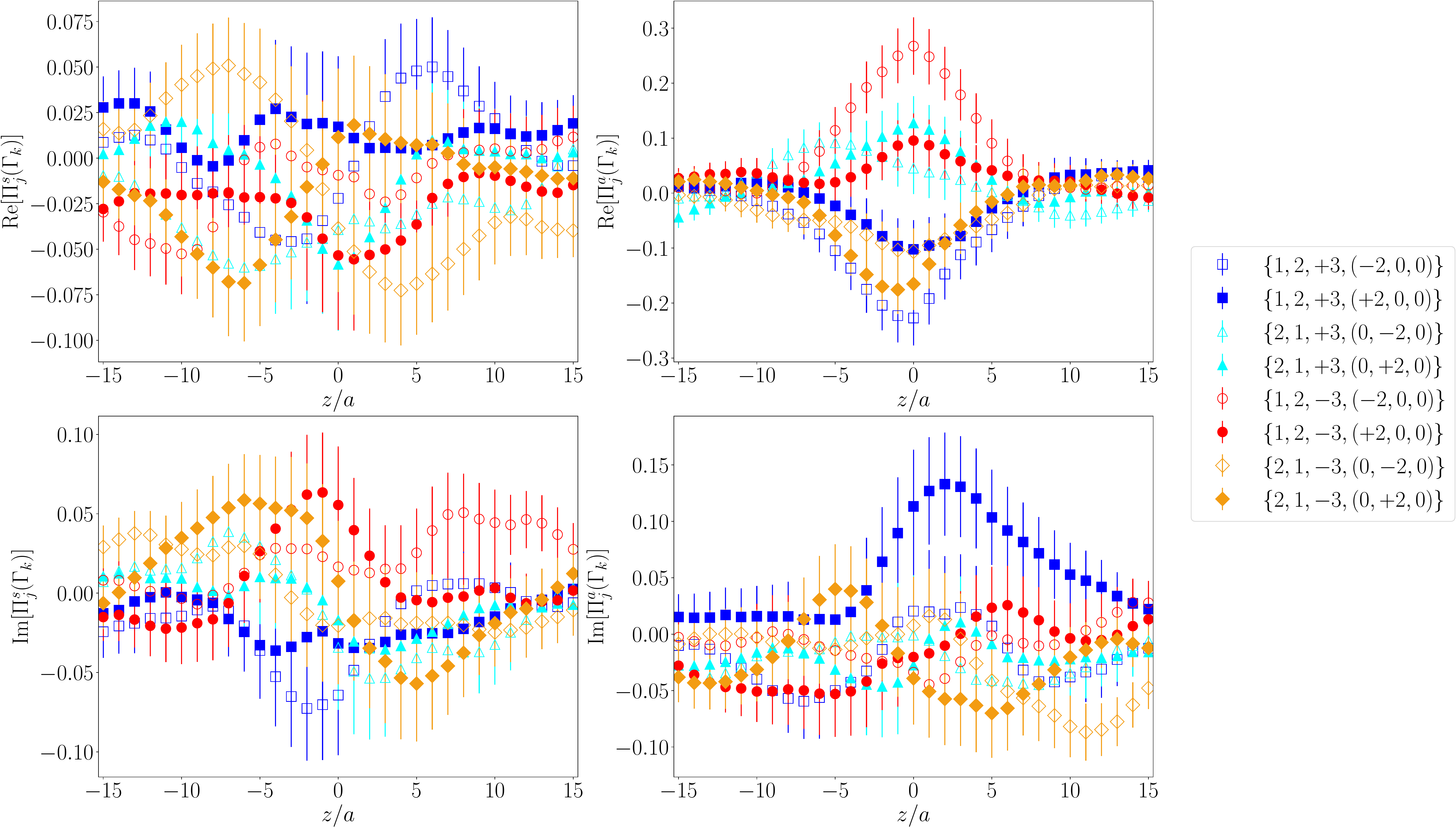}
    \vspace*{-0.35cm}
    \caption{Bare matrix elements $\Pi_1(\Gamma_2)$ and $\Pi_2(\Gamma_1)$ in the symmetric frame (left) and in the asymmetric frame (right). The legend indicates $\{j, \kappa,P_3,\vec{\Delta}\}$ in units of $2\pi/L$, with $j,\kappa$ corresponding to $\Pi_j(\Gamma_\kappa)$.  The remaining notation is the same as Fig.~\ref{fig:Pi0G0}. }
    \label{fig:Pi1G2}
\end{figure}

In Fig.~\ref{fig:Pi0G0}, we show the real and imaginary parts of bare $\Pi_0(\Gamma_0)$ in both frames for the eight combinations of $\pm P_3$ and $\pm \vec{\Delta}$ given in Table~\ref{tab:stat}. 
Similarly,  Fig.~\ref{fig:Pi0G1}, Fig.~\ref{fig:Pi2G3} and Fig.~\ref{fig:Pi1G2} show $\Pi_0(\Gamma_j)$, $\Pi_j(\Gamma_3)$ and $\Pi_j(\Gamma_{\kappa})$ ($j,\,\kappa=1,\,2$, $j\ne \kappa$), respectively, for the $P_3$ and $\vec{\Delta}$ combinations that lead to non-vanishing matrix elements.
There are several comments and observations for the behavior of the matrix elements.
First, $\Pi^s_\mu$ and $\Pi^a_\mu$ are not equivalent and, thus, not directly comparable due to their frame dependence. 
For instance, their frame dependence can be seen in Eqs.~\eqref{eq:Pi0G0_s} -  \eqref{eq:Pi3G3_ns} in Sec.~\ref{sec:Lat_setup_A}, where the matrix elements parametrize in different combinations of $A_i$ with frame-dependent kinematic coefficients. Also, the numerical values of some of the kinematic factors, e.g., $E_i$, depend on the frame.
For example, $\Pi^s_0(\Gamma_0)$ contains information on $A_1$, $A_5$, and $A_6$, while $\Pi^a_0(\Gamma_0)$ contains information on $A_1$, $A_3$, $A_4$, $A_5$, $A_6$, and $A_8$. 
Second, the matrix elements in the symmetric frame have definite symmetries, which are validated in the data shown in Figs.~\ref{fig:Pi0G0} - \ref{fig:Pi1G2}. 
For instance, Eq.~\eqref{eq:Pi0G0_s} has a symmetric real part and an anti-symmetric imaginary part with respect to $P_3\to-P_3$ at fixed $z$, 
which can be traced back to the symmetries of $A_i$ ($A_1^\star(-z\cdot P)=A_1(z\cdot P)$, $A_5^\star(-z\cdot P)=A_5(z\cdot P)$, $A_6^\star(-z\cdot P)=-A_6(z\cdot P)$) combined with the symmetry properties of the kinematic factors.
For simplicity, we do not show all arguments of $A_i$.

Third, the lack of definite symmetries in the asymmetric frame appears to be a small effect in $\Pi^a_0(\Gamma_0)$ and $\Pi^a_2(\Gamma_3)$, and comparable to the current statistical uncertainties.
However, the effect in $\Pi^a_1(\Gamma_2)$ is more significant, particularly when $\vec{\Delta} \to -\vec{\Delta}$.
Finally, some of the matrix elements, e.g., $\Pi^a_1(\Gamma_2)$, are theoretically nonzero but are suppressed in magnitude. 
This has implications in the signal of some of the $A_i$, as we will see below.

The next step in the analysis is to decompose the matrix elements into the corresponding Lorentz-invariant amplitudes. 
The fact that the $A_i$ have definite symmetries makes them interesting to isolate and study. 
This is done for each kinematic setup of Table~\ref{tab:stat} (total of eight). 
The $A_i$ from different kinematic setups can be combined according to their symmetries, by symmetrizing or antisymmetrizing with respect to $\pm P_3 z$, based on the findings given in Appendix~\ref{sec:AppB}.
The frame independence of $A_i$ is a major advantage of the proposed parametrization, which we observe numerically using our lattice QCD calculation. 
Such a test should not be understood as a check of the theoretical formulation, but rather a consistency check of the lattice estimates for $A_i$. The extent of agreement in the two frames provides an estimates of systematic effects arising from non-vanishing lattice spacing.

\begin{figure}[h!]
    \centering
    \includegraphics[scale=0.23]{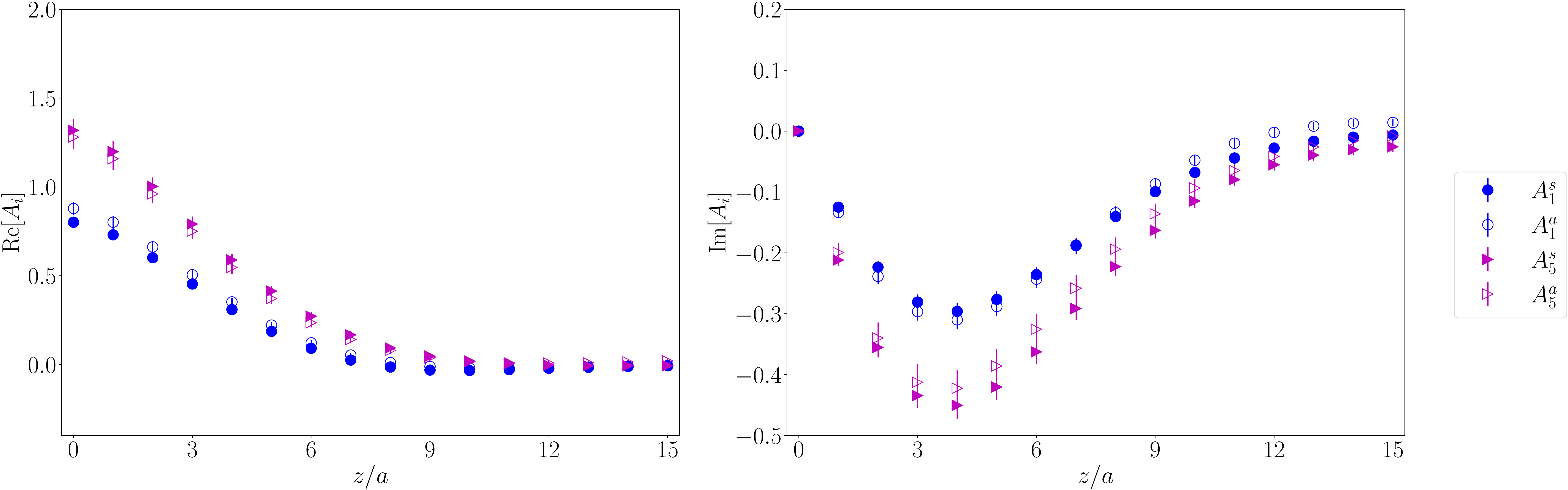}
    \vspace*{-0.3cm}
    \caption{Comparison of bare values of $A_1$ and $A_5$ in the symmetric (filled symbols) and asymmetric (open symbols) frame. The real (imaginary) part of each quantity is shown in the left (right) column. The data correspond to $|P_3|=1.25$ GeV and $t=-0.69$ GeV$^2$ ($t=-0.64$ GeV$^2$) for the symmetric (asymmetric) frame.}
    \label{fig:A156}
\end{figure}
\begin{figure}[h!]
    \centering
 \includegraphics[scale=0.23]{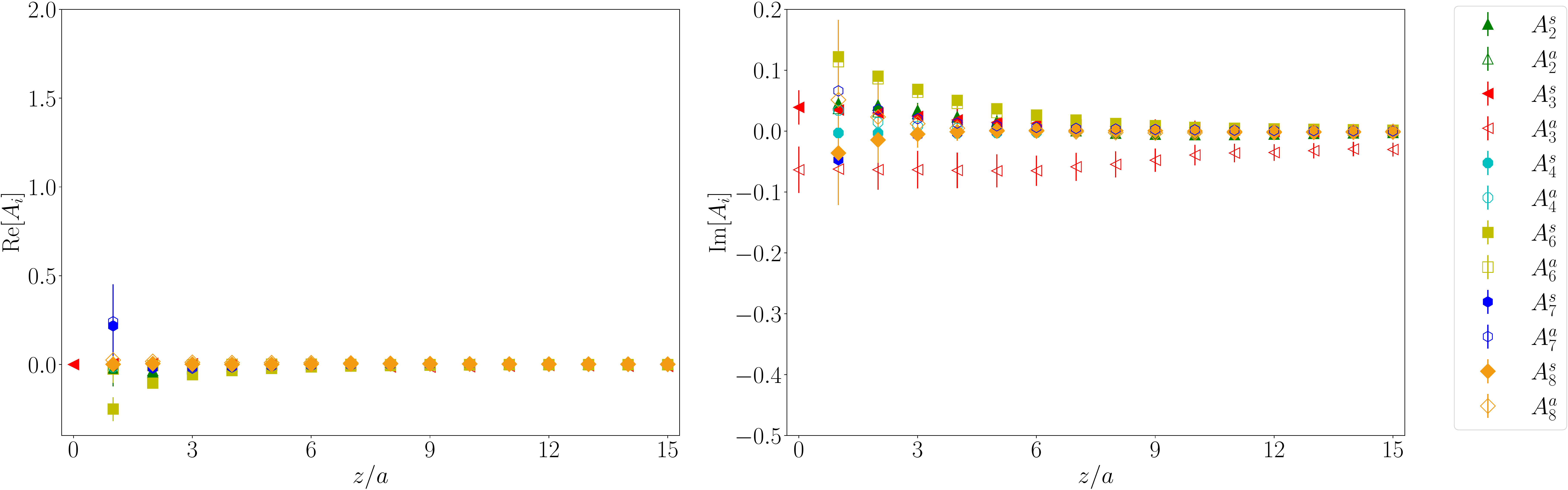}
    \vspace*{-0.3cm} 
    \caption{Comparison of bare values of $A_2$, $A_3$, $A_4$, $A_6$, $A_7$, $A_8$ in the symmetric (filled symbols) and asymmetric (open symbols) frame. The notation is the same as Fig.~\ref{fig:A156}.}
    \label{fig:A348}
\end{figure}

In Figs.~\ref{fig:A156} - \ref{fig:A348}, we present the bare $A_i$ after combining all values of  $\vec{P}_3$ and $\vec{\Delta}$. 
The amplitudes $A_2$, $A_4$, $A_6$, $A_7$ and $A_8$ are accompanied with an explicit factor of $z^n$ ($n=1,2$) in Eqs.~\eqref{eq:Pi0G0_s} - \eqref{eq:Pi3G3_ns}, and therefore, cannot be accessed at $z=0$.
One may extrapolate their $z$ dependence to obtain $A_i(z=0)$.
In the presentation of Figs.~\ref{fig:A156} - \ref{fig:A348}, we keep the same range in the $y$-axis for all the amplitudes for a better quantitative comparison.

We find that $A_5$ has the largest magnitude both in the real and the imaginary parts, followed by $A_1$. 
The remaining amplitudes are found to be very small or negligible, which can be traced back to the small signal for some of the matrix elements, such as $\Pi_1(\Gamma_2)$. 
Overall, we find very good agreement between the two frames for each $A_i$, as expected theoretically. We remind the reader that there is no exact match of the momentum transfer in the two frames ($t^s=-0.69$ GeV$^2$, $t^s=-0.64$ GeV$^2$) and small differences may be attributed to the$~\sim 7\%$ change in $t$, as well as unquantified systematic uncertainties. Such a difference between $t^s$ and $t^a$ is, in general, not an obstacle in the proposal of Sec.~\ref{sec:FH_FE_setup}, where a Lorentz boost transformation can relate the momentum transfer between the two frames, without ambiguity in the extracted light-cone GPDs.

\subsubsection{Lorentz invariant and non-invariant quasi-GPDs}
\label{sec:res_quasi}

\begin{figure}[h!]
    \centering
    \includegraphics[scale=0.25]{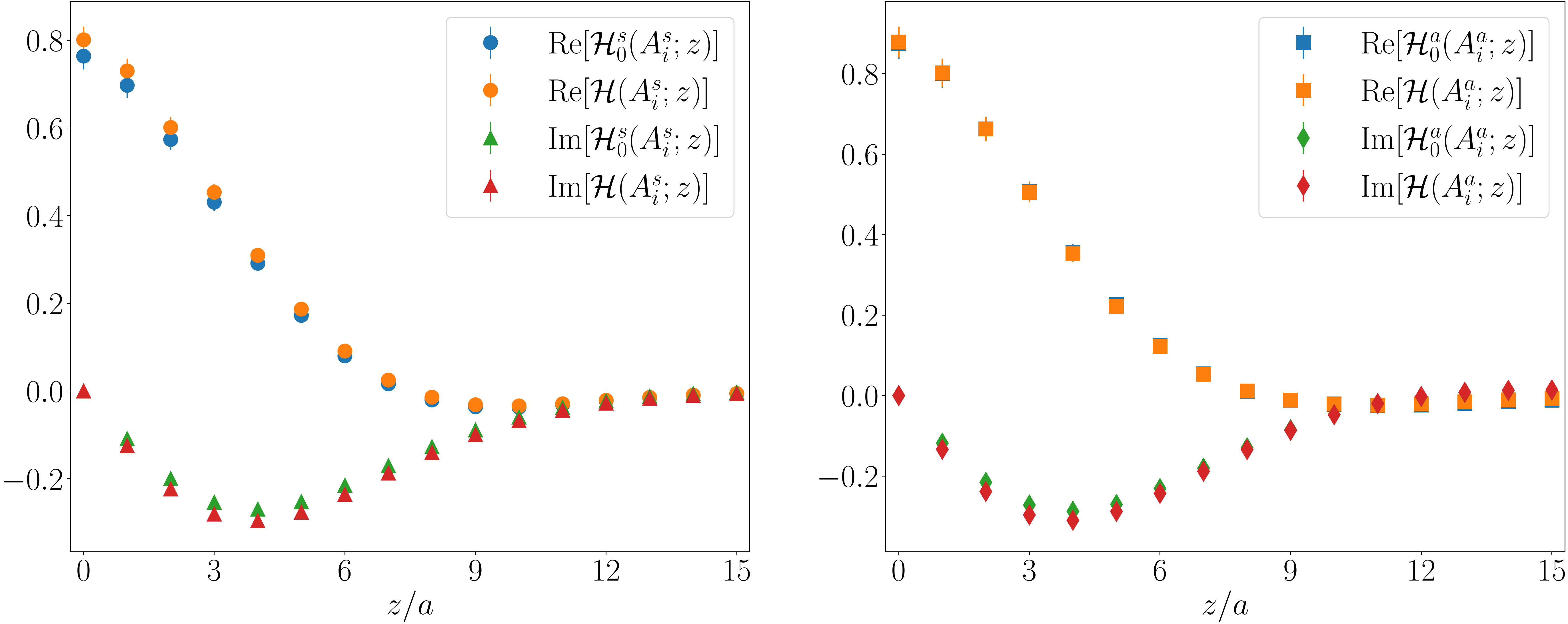}
     \vspace*{-.3cm}
    \caption{Comparison of bare ${\cal H}_0$ and ${\cal H}$ at $|P_3|=1.25$ GeV in the symmetric (left, $t=-0.69$ GeV$^2$) and asymmetric (right, $t=-0.64$ GeV$^2$) frame.}
    \label{fig:FH_a}
\end{figure}

\begin{figure}[h!]
    \centering
    \includegraphics[scale=0.25]{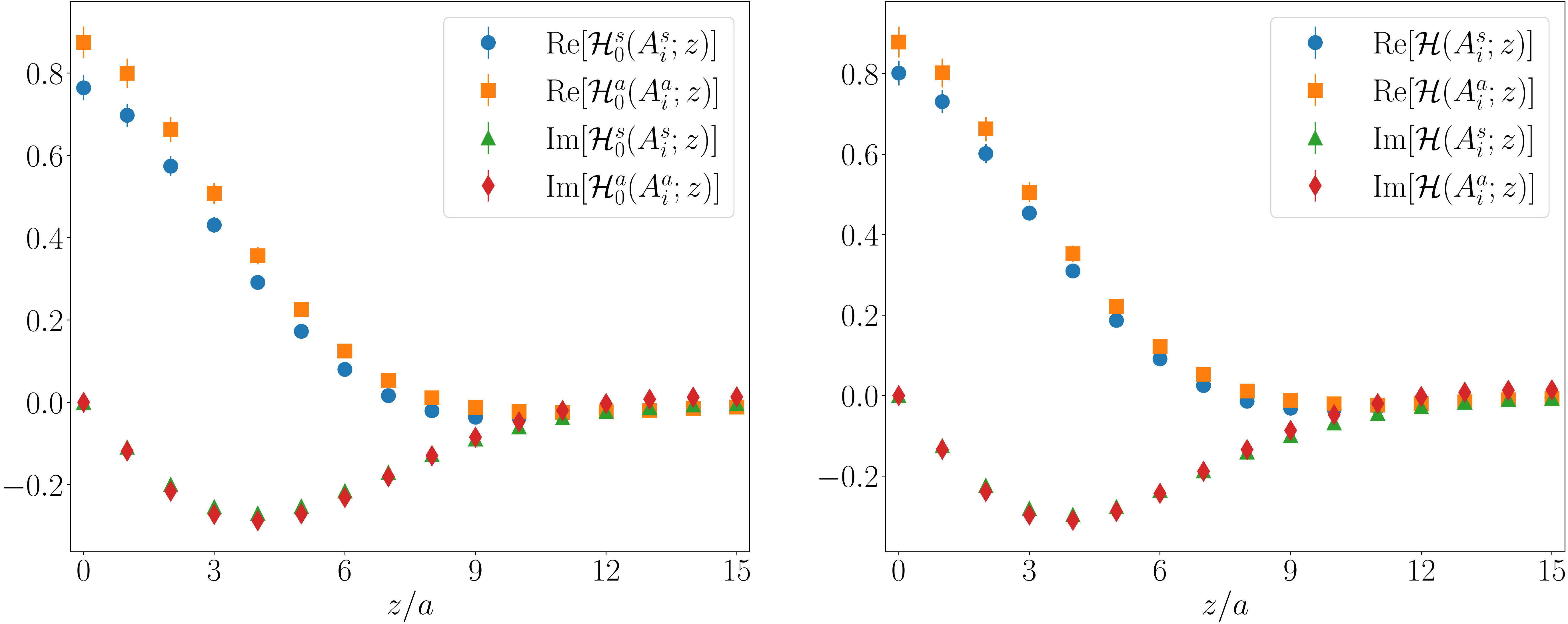}
     \vspace*{-.3cm}
    \caption{Left: Comparison of bare ${\cal H}_0$ at $|P_3|=1.25$ GeV using the symmetric ($t=-0.69$ GeV$^2$) and asymmetric ($t=-0.64$ GeV$^2$) frame matrix elements. Right: Comparison of bare ${\cal H}$ using the symmetric and asymmetric frame matrix elements.}
    \label{fig:FH_b}
\end{figure}

\begin{figure}[h!]
    \centering   
    \includegraphics[scale=0.23]{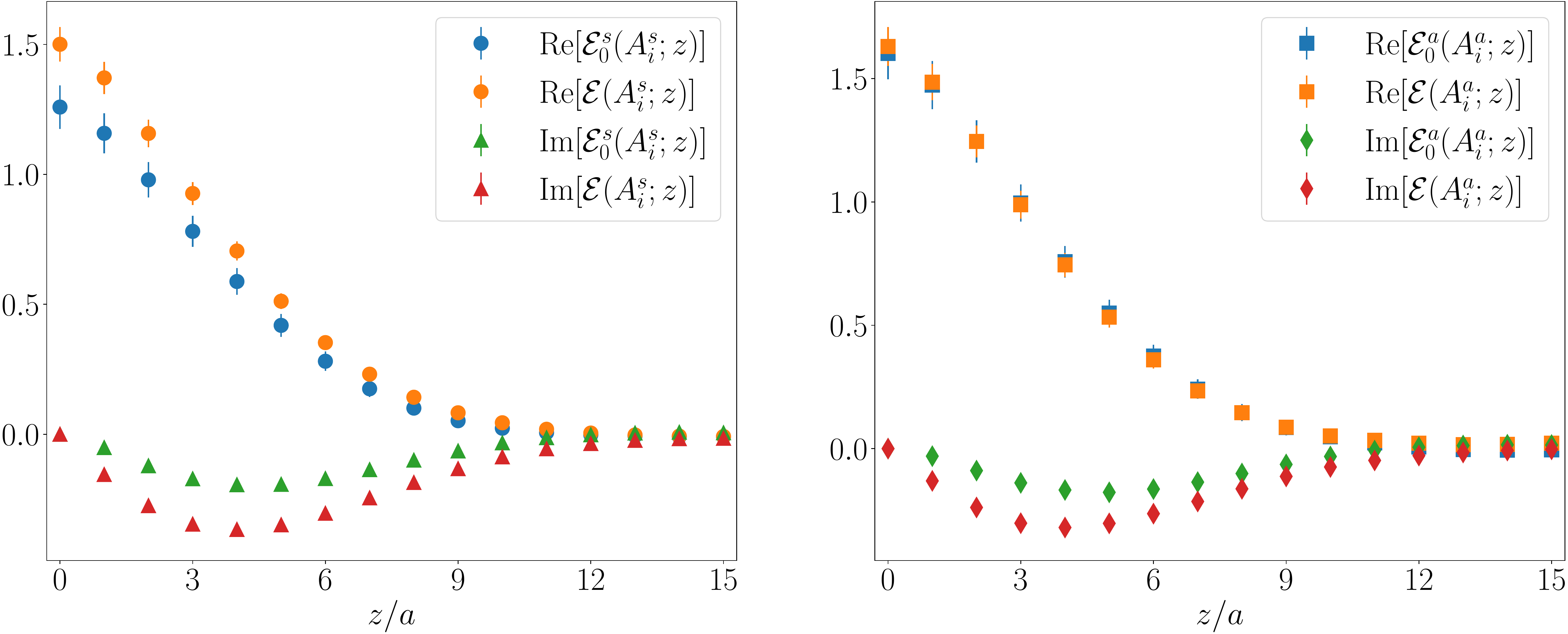}
     \vspace*{-.3cm}   
    \caption{Similar to Fig.~\ref{fig:FH_a} for ${\cal E}_0$ and ${\cal E}$ in the symmetric (left) and asymmetric (right) frame.}
    \label{fig:FE_a}
\end{figure}

\begin{figure}[h!]
    \centering   
     \includegraphics[scale=0.23]{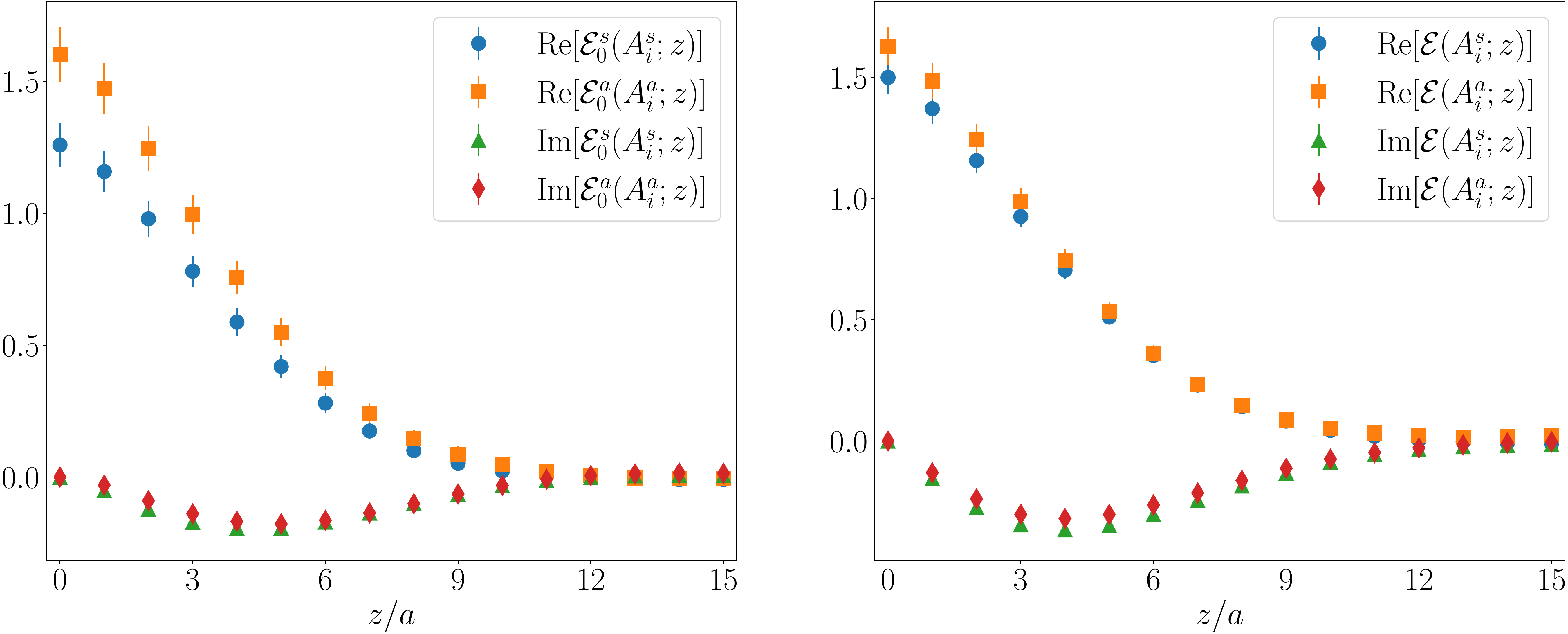}
     \vspace*{-.3cm}
    \caption{Similar to Fig.~\ref{fig:FH_b} for ${\cal E}_0$ (left) and ${\cal E}$ (right).}
    \label{fig:FE_b}
\end{figure}

In this paragraph, we use the methodology of Sec.~\ref{sec:FH_FE_setup} to calculate the GPDs based on the $\gamma_0$ operator (Lorentz non-invariant), ${\cal H}^{s/a}_0,\,{\cal E}^{s/a}_0$, as well as an alternative Lorentz-invariant operator that combines $\gamma_0,\,\gamma_1,\,\gamma_2$, defining ${\cal H},\,{\cal E}$. 
 Having the amplitudes $A_i$, one may use them for any definition of quasi-GPDs, as they contain no information on the frame and are interchangeable between the symmetric and the asymmetric frame, as long as one keeps track of the values of $P_3$ and $t$ that the quasi-GPDs correspond to.
The results for $P_3=1.25$ GeV and $t^s=-0.69$ GeV$^2$, $t^a=-0.64$ GeV$^2$ are shown in Figs.~\ref{fig:FH_a} - \ref{fig:FE_b}. 
In particular, we compare the definitions of ${\cal H}_0$ and ${\cal E}_0$, as given in Eqs.~\eqref{eq:FHs_Pi} - \eqref{eq:FEs_Pi} for the symmetric, and Eqs.~\eqref{eq:FHa_Pi} - \eqref{eq:FEa_Pi} for the asymmetric frame. 
We emphasize that defining ${\cal H}_0$ and ${\cal E}_0$ through $\gamma_0$ is frame-dependent and, thus, ${\cal H}_0^s$ and ${\cal H}_0^a$ have a different functional form; similarly for ${\cal E}_0^s$ ${\cal E}_0^a$. 
Indeed, we find numerically that the real part for both ${\cal H}_0$ and ${\cal E}_0$ is not in agreement in the two frames (see left plots of Figs.~\ref{fig:FH_b}, \ref{fig:FE_b}); agreement is found in the imaginary part.

\begin{figure}[h!]
    \centering
    \includegraphics[scale=0.23]{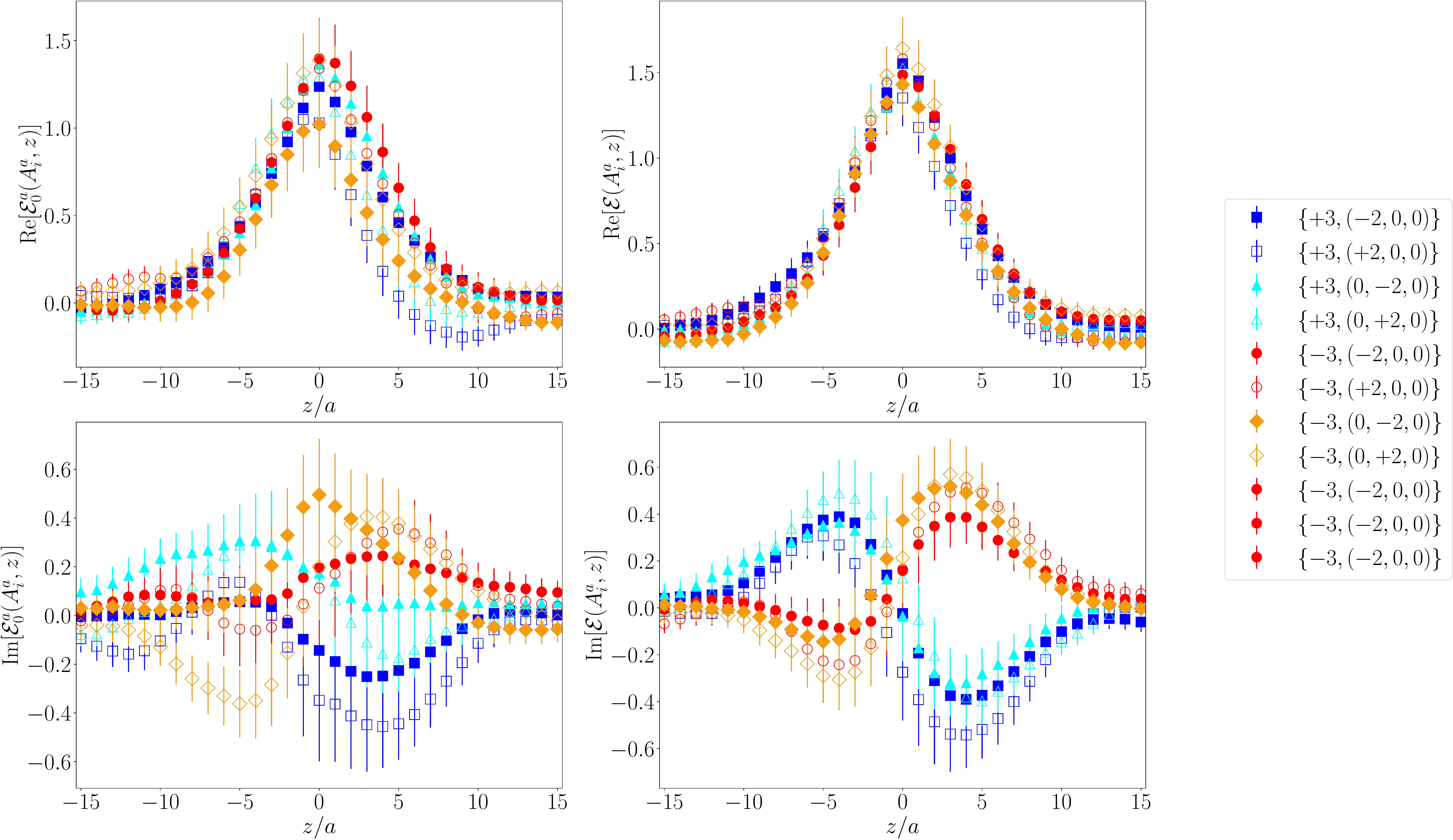}
    \caption{Bare ${\cal E}_0$ (left) and ${\cal E}$ (right) for $|P_3|=1.25$ GeV. All cases correspond to the asymmetric frame ($t=-0.64$ GeV$^2$). The top (bottom) panel corresponds to the real (imaginary) part. The notation in the legend is $\{P_3,\vec{\Delta}\}$ in units of $2\pi/L$.}
    \label{fig:E_per_dir}
\end{figure}

Another interesting comparison is for the Lorentz-invariant definitions,  ${\cal H}$ and ${\cal E}$, using matrix elements obtained in the symmetric or the asymmetric frame (see right plots of Figs.~\ref{fig:FH_b}, \ref{fig:FE_b}). 
As expected theoretically, the agreement between the two frames is significantly improved for both ${\cal H}$ and ${\cal E}$.
It is natural to also compare ${\cal H}_0$ with ${\cal H}$ and  ${\cal E}_0$ with ${\cal E}$ as extracted in each frame. 
Also in this case, an agreement is not expected at finite $P_3$, as the Lorentz-invariant and non-invariant formalism is not the same. 
However, some similarity is expected because both definitions agree at $P_3\to\infty$.
Such a comparison is shown in Figs.~\ref{fig:FH_a}, \ref{fig:FE_a}, and it is found that, for this kinematic setup, ${\cal H}_0$ is fully compatible with ${\cal H}$ in both frames; in fact, perfect agreement is found in the asymmetric case. 
An excellent agreement is found between ${\rm Re}[{\cal E}]$ and ${\rm Re}[{\cal E}_0]$ in the asymmetric frame, while in the symmetric frame there is difference. 
Finally, differences are observed between ${\rm Im}[{\cal E}]$ and ${\rm Im}[{\cal E}_0]$ for both frames. 
As previously mentioned, these quantities are not expected to be in agreement for finite momentum $P_3$. 
It is also interesting to note that the statistical errors are considerably smaller in ${\cal E}$ as compared to ${\cal E}_0$.
The origin of this behavior is illustrated in Fig.~\ref{fig:E_per_dir}, which shows the respective matrix elements separately for the eight equivalent kinematic cases.
At least for this choice of $P_3$ and $\vec{\Delta}$, the Lorentz-invariant definition improves the statistical quality of the signal, i.e.\ that these kinematic contaminations introduce additional noise to the extracted quantity. This trend holds for the symmetric frame too.
We also note that this effect does not occur, or is strongly limited, in the case of $H$ GPD.
Tracing this behavior back even further, the definition of ${\cal E}$ involves additional matrix elements that subtract the noise present in $\Pi_0(\Gamma_{1/2})$ (see Fig.~\ref{fig:Pi0G1}, particularly the imaginary part).
In turn, ${\cal H}_0$ is numerically dominated by the less noisy $\Pi_0(\Gamma_0)$ (Fig.~\ref{fig:Pi0G0}).
We remind the reader that, in general, the difference between $t^a$ and $t^s$ may be responsible for small differences between the quantities calculated in the two frames.
Also, further investigations are needed to assess the advantages and disadvantages of the various definitions for the quasi-GPDs.

\begin{figure}[h!]
    \centering
    \includegraphics[scale=0.22]{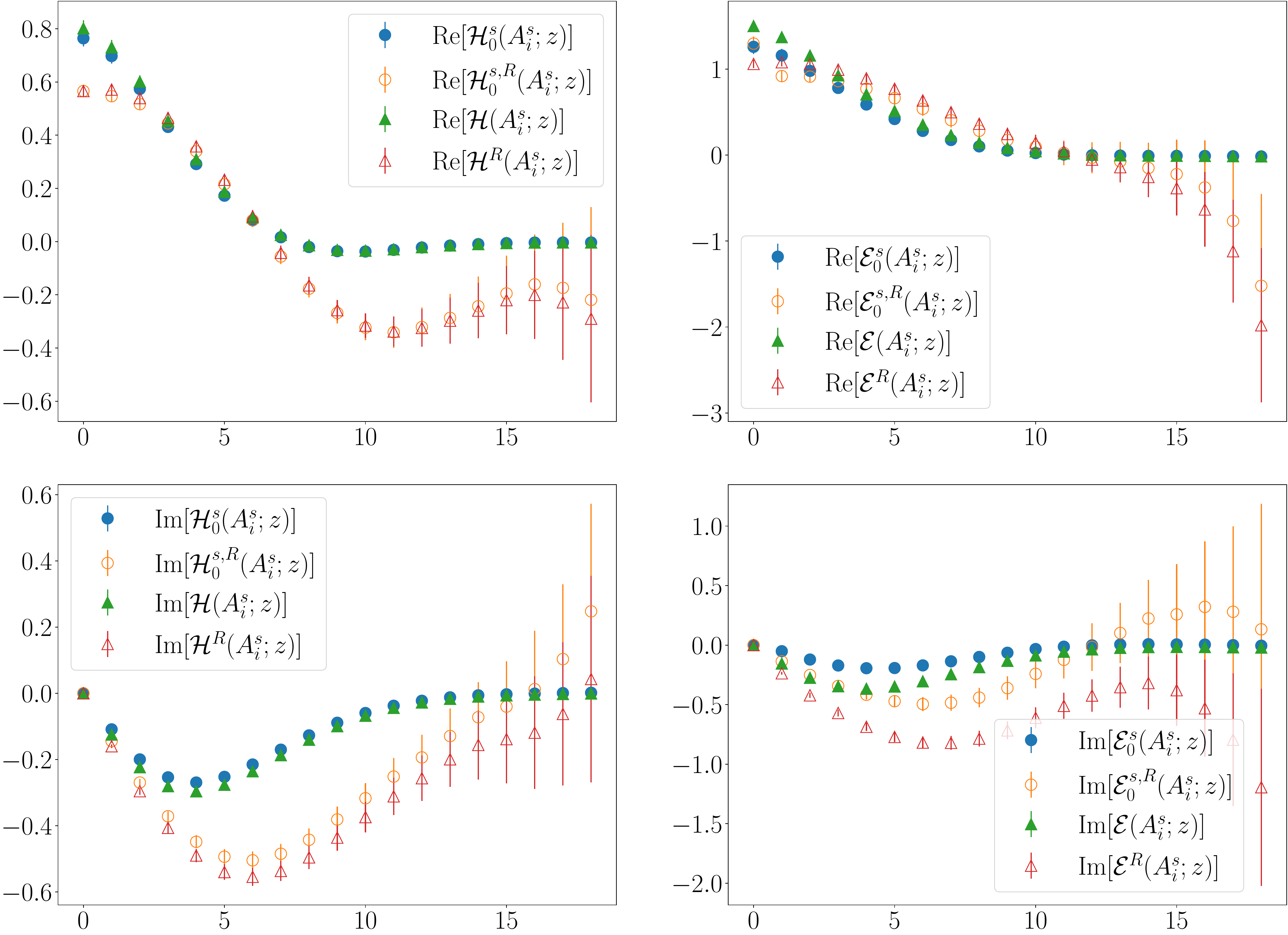}
    \vspace*{-0.3cm}
    \caption{Comparison of bare and renormalized quasi-GPDs ${\cal H}_0$, $\cal H$ (left) and ${\cal E}_0$, $\cal E$ (right) for the real (top) and the imaginary (bottom) part in coordinate space. All cases correspond to the symmetric frame ($t=-0.69$ GeV$^2$). The renormalized quantities carry a superscript $R$.}
    \label{fig:Renorm}
\end{figure}

An important component of the lattice calculation is the renormalization, which in this work is done in coordinate space using an RI prescription. 
Since this is a proof-of-concept calculation, we do not focus on the various prescriptions to improve the renormalization, such as the hybrid scheme~\cite{Ji:2020brr}, and reduction of lattice artifacts in the RI estimates~\cite{Constantinou:2022aij}, or combination of the two. 
However, we do emphasize that this is an important direction for future systematic studies of GPDs.
For compatibility with the matching formalism of Refs.~\cite{Liu:2019urm,LatticeParton:2018gjr}, we use the standard RI prescription defined on a single renormalization scale, $(a  \mu_0)^2\approx1.95$, which will also enter the matching equations. 
We find negligible dependence when varying $ \mu_0$.
As discussed in Sec.~\ref{s:renorm_match}, the appropriate renormalization for $H$ and $E$ is the one of the $\gamma_0$ operator, which is valid for both Lorentz-invariant and non-invariant quasi-GPDs. 
Details on the calculation of the renormalization functions used in this work can be found in Ref.~\cite{Alexandrou:2019lfo}. 
As an example, in Fig.~\ref{fig:Renorm} we compare the bare and renormalized values of quasi-GPDs using the symmetric frame and the two definitions, that is ${\cal H}_0^{s/a}$ and ${\cal H}$, ${\cal E}^{s/a}_0$ and ${\cal E}$. 
The plots demonstrate the challenges related to the renormalization, that is, as $z$ increases the RI prescriptions becomes less reliable. 
In practice, the value of the renormalization functions increase exponentially due to the linear divergence leading to renormalized functions that do not decay to zero. 
Such behavior can be seen in ${\cal H}^{s,R}_0$ and ${\cal H}^R$.

\subsubsection{Light-cone GPDs}

To extract the light-cone GPDs from the lattice data, one must transform the latter in momentum space to reconstruct their $x$-dependence. 
While this is necessary, the Fourier transform from a finite set of quasi-GPD matrix elements is accompanied by the so-called inverse problem\footnote{See Ref.~\cite{Karpie:2019eiq} for an extensive discussion in the context of reconstructing partonic distributions.}, which mainly affects the small-$x$ region. Nevertheless, the moderate-to-large-$x$ region is not sensitive to this inverse problem, thus allowing us to make reliable predictions.
In this work, we use the Backus-Gilbert (BG) reconstruction method~\cite{BackusGilbert}, which uses a model-independent criterion to choose the light-cone reconstructed GPDs from among the infinite set of possible solutions to the inverse problem. 
The criterion is that the variance of the solution with respect to the statistical variation of the input data should be minimal. 
We reconstruct the momentum-space distribution by applying BG separately for each value of $x$. 
It should be noted that, despite BG being model-independent and better than the naive Fourier transform, there are still limitations due the small number of lattice data sets. 
In the work presented here, we vary the number of data that enter the reconstruction, that is $z_{\rm max}=7a,\,9a,\,11a,\,13a$.

\begin{figure}[h!]
    \centering
    \includegraphics[scale=0.22]{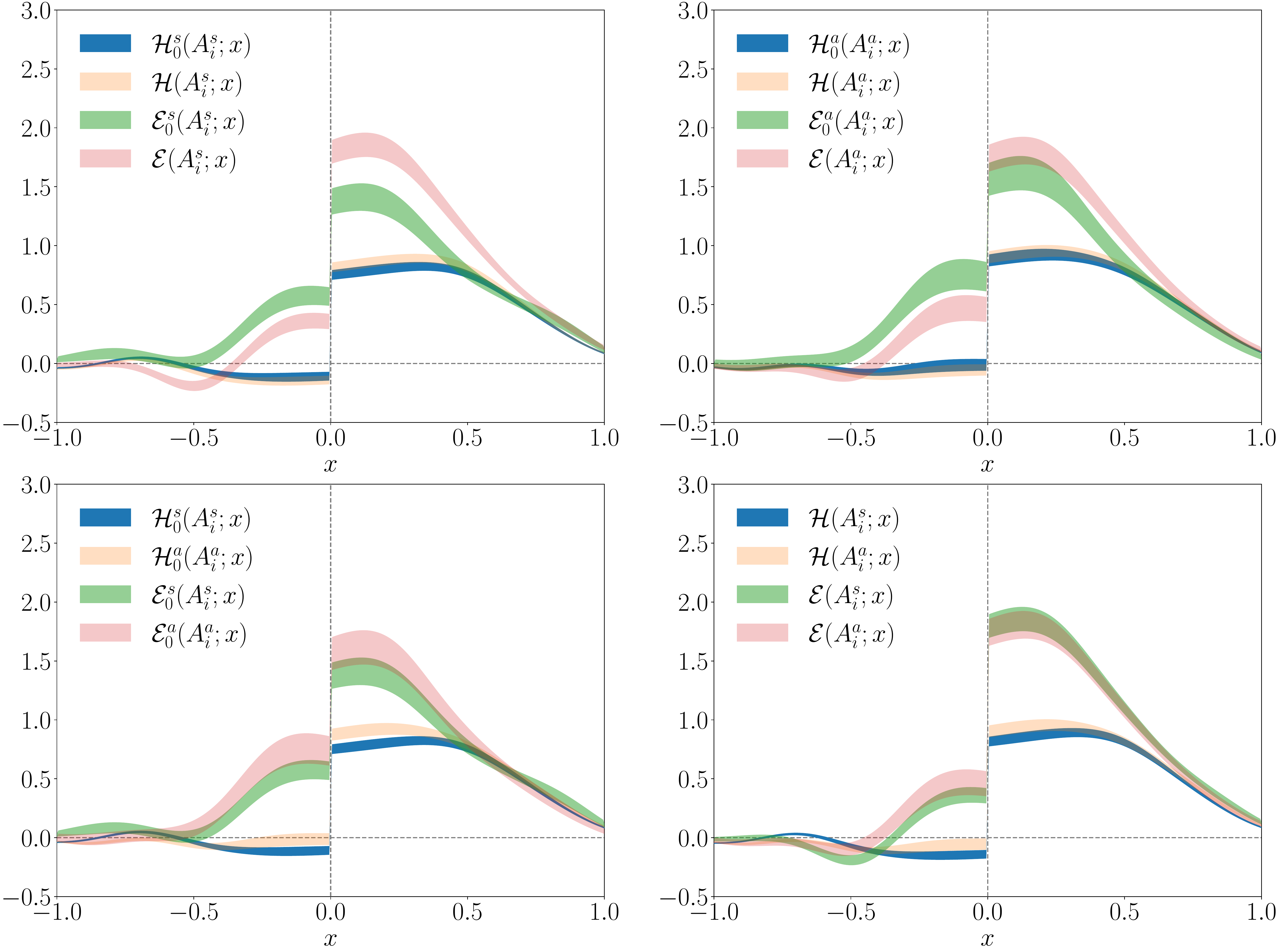}
    \vspace*{-0.3cm}
    \caption{Renormalized quasi-GPDs ${\cal H}_0$, $\cal H$ and ${\cal E}_0$, $\cal E$ in momentum space using the definitions of Eqs.~\eqref{eq:FHs} - \eqref{eq:FEs}, Eqs.~\eqref{eq:FHa} - \eqref{eq:FEa}, and Eqs.~\eqref{eq:FH_impr} - \eqref{eq:FE_impr}.}
    \label{fig:qGPDs}
\end{figure}

\begin{figure}[h!]
    \centering
    \includegraphics[scale=0.23]{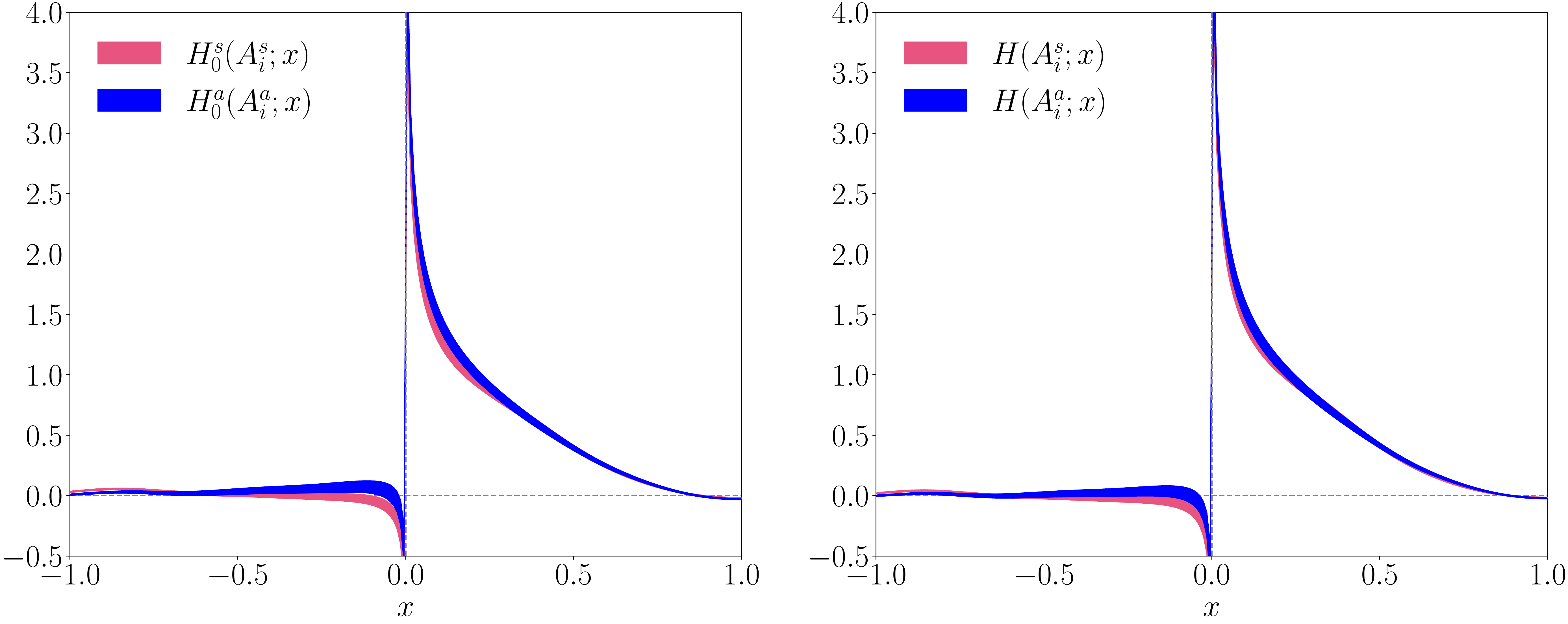}
    \vspace*{-0.3cm}
    \caption{Left: Comparison of light-cone $H_0$-GPD in the symmetric ($t=-0.69$ GeV$^2$) and asymmetric ($t=-0.64$ GeV$^2$) frame. Right: Comparison of light-cone $H$-GPD in the symmetric and asymmetric frame. Results are presented in the $\overline{\rm MS}$ scheme at 2 GeV.}
    \label{fig:Hs_Ha}
\end{figure}

\begin{figure}[h!]
    \centering
    \includegraphics[scale=0.23]{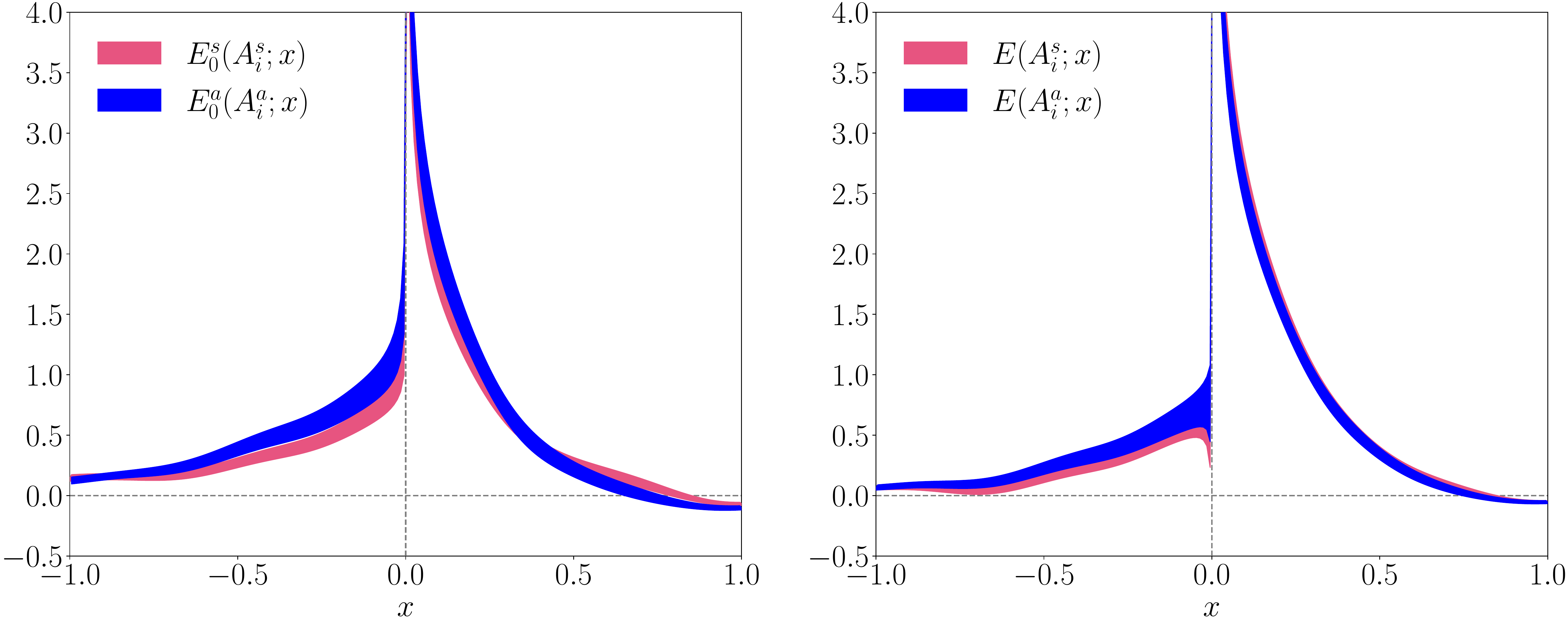}
    \vspace*{-0.3cm}
    \caption{Left: Comparison of light-cone $E_0$-GPD in the symmetric ($t=-0.69$ GeV$^2$) and asymmetric ($t=-0.64$ GeV$^2$) frame. Right: Comparison of light-cone $E$-GPD in the symmetric and asymmetric frame. Results are presented in the $\overline{\rm MS}$ scheme at 2 GeV.}
    \label{fig:Es_Ea}
\end{figure}

The $x$ dependence of quasi-GPDs for the various definitions is shown in Fig.~\ref{fig:qGPDs} using the BG reconstruction method with $z_{\rm max}=9a$. 
All definitions for the $H$ quasi-GPD are consistent, with a very mild difference between the definitions of Eq.~\eqref{eq:FHs} and Eq.~\eqref{eq:FHa} for $x<0.4$. 
Such a difference become negligible after the matching (see, e.g., Fig~\ref{fig:Hs_Ha}). 
On the contrary, the $E$ quasi GPD has a noticeable dependence on the definition. 
More precisely, the results using Eq.~\eqref{eq:FEs} and Eq.~\eqref{eq:FEa} are in agreement marginally; the agreement improves in the $x>0$ region after the matching, as seen in Fig~\ref{fig:Es_Ea}. 
Differences are also observed in Eq.~\eqref{eq:FEs} and (Eq.~\eqref{eq:FEa}) when compared to the alternative definition of Eq.~\eqref{eq:FE_impr}. 
These differences persist after the matching. 
Once again, agreement between the various definitions is not expected by construction. 
The only agreement imposed by theory is the frame independence of Eq.~\eqref{eq:FE_impr}. 
Indeed, we find that the numerical results are frame independent despite the small difference in the value of $t$ between the two frames.

The final step of the calculation is the application of the matching equations on the $x$-dependent quasi-GPDs, to connect the lattice data to the light-cone GPDs, as outlined in Sec.~\ref{s:renorm_match}. 
We use the one-loop expression of Ref.~\cite{Liu:2019urm} to relate the quasi-GPDs in the RI scheme at a scale of 1.95 GeV to the light-cone GPDs in the $\overline{\rm MS}$ scheme at 2 GeV. 
At zero skewness, the matching coefficient is exactly the same as in the quasi-PDF case~\cite{Liu:2019urm}.

\begin{figure}[h!]
    \centering
    \includegraphics[scale = 0.22]{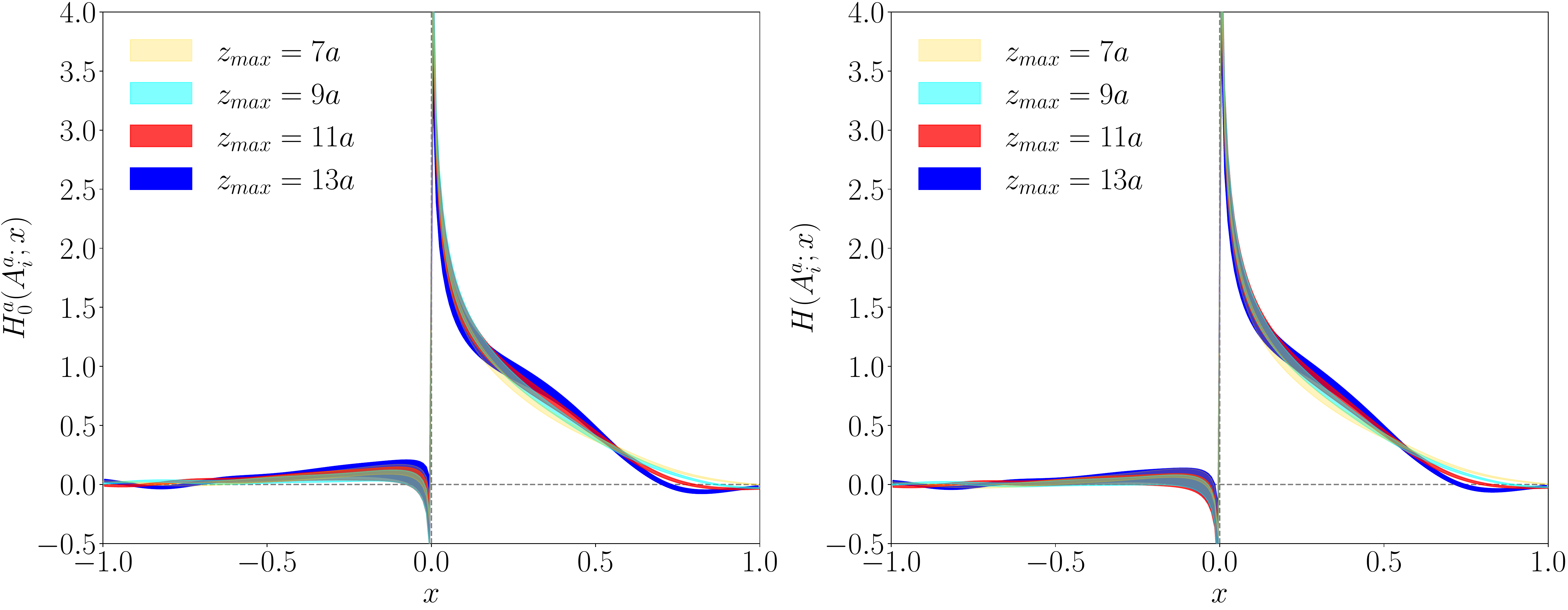}
    \vspace*{-0.3cm}
    \caption{Dependence of $H_0(A^a_i;x)$ (left) and $H(A^a_i;x)$ (right) on $z_{\rm max}$.
    The light-cone GPDs have been obtained using matrix elements in the asymmetric frame ($t=-0.64$ GeV$^2$), and are presented in the $\overline{\rm MS}$ scheme at 2 GeV.}
    \label{fig:H_zmax}
\end{figure}
\begin{figure}[h!]
    \centering
    \includegraphics[scale=0.22]{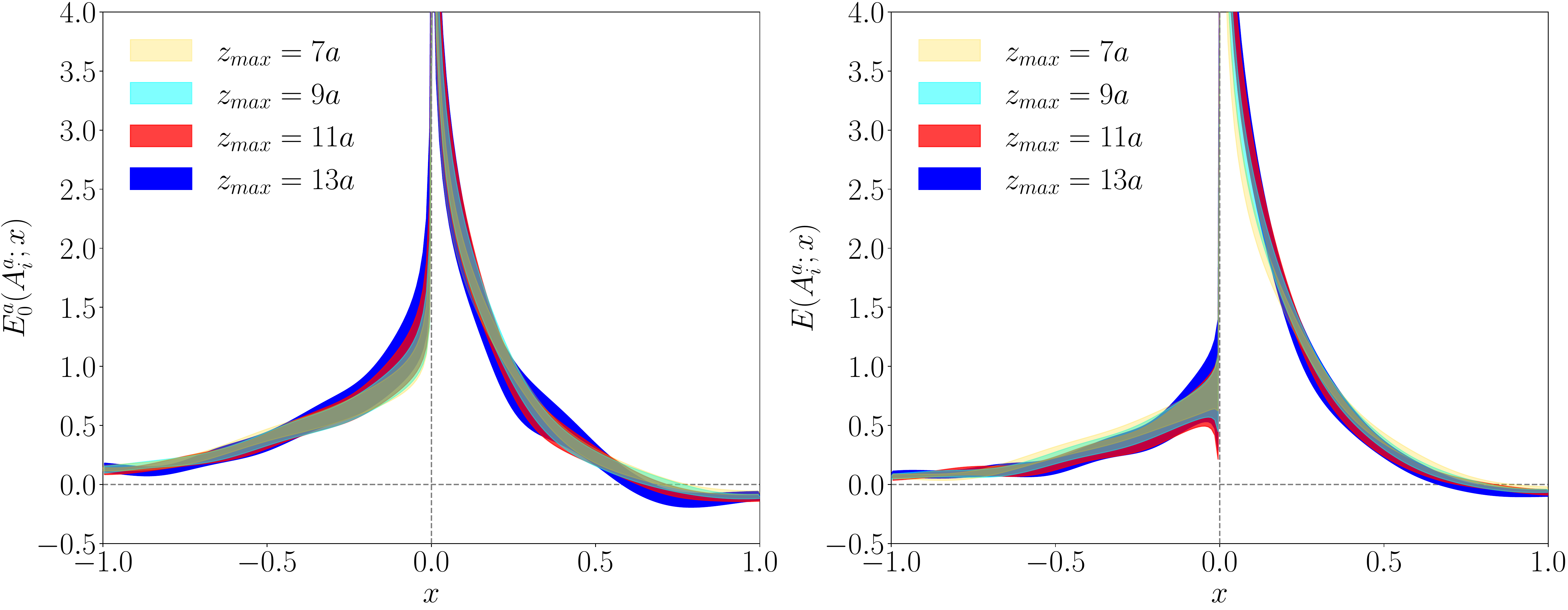}
    \vspace*{-0.3cm}
    \caption{Dependence of $E_0(A^a_i;x)$ (left) and $E(A^a_i;x)$ (right) on $z_{\rm max}$. The light-cone GPDs have been obtained using matrix elements in the asymmetric frame ($t=-0.64$ GeV$^2$), and are presented in the $\overline{\rm MS}$ scheme at 2 GeV.}
    \label{fig:E_zmax}
\end{figure}
By varying $z_{\rm max}$, we first investigate the effect of the truncation of the data set entering the reconstruction of the $x$-dependence. For simplicity, we show the effect in the light-cone GPDs.
We find very small dependence between $z_{\rm max}=7a,\,9a,\,11a,\,13a$ for all quantities calculated in both frames. 
In Fig.~\ref{fig:H_zmax}, we show the $z_{\rm max}$ dependence of $H_0^a$ and $H$ extracted from the asymmetric frame calculation. 
Similarly, in Fig.~\ref{fig:E_zmax}, we show $E_0^a$ and $E$. As can be seen, all values of $z_{\rm max}$ lead to compatible results, with the statistical uncertainties increasing with $z_{\rm max}$. 
Hence, we proceed with $z_{\rm max}=9a$ as the preferred value.

\begin{figure}[h!]
    \centering
    \includegraphics[scale=0.23]{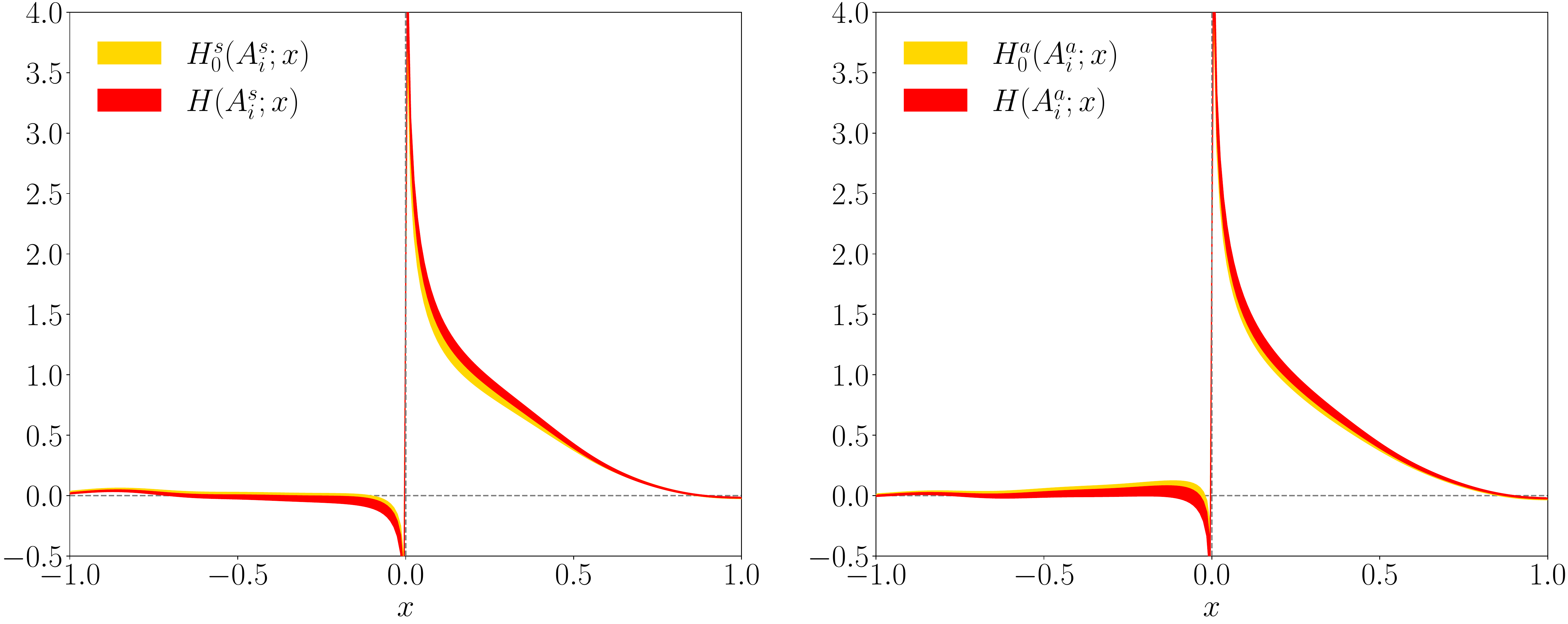}
    \vspace*{-0.3cm}
    \caption{Comparison of light-cone $H_0$ and $H$ GPDs in the symmetric (left, $t=-0.69$ GeV$^2$) and asymmetric (right, $t=-0.64$ GeV$^2$) frame. Results are presented in the $\overline{\rm MS}$ scheme at 2 GeV.}
    \label{fig:H_calH}
\end{figure}

\begin{figure}[h!]
    \centering
    \includegraphics[scale=0.23]{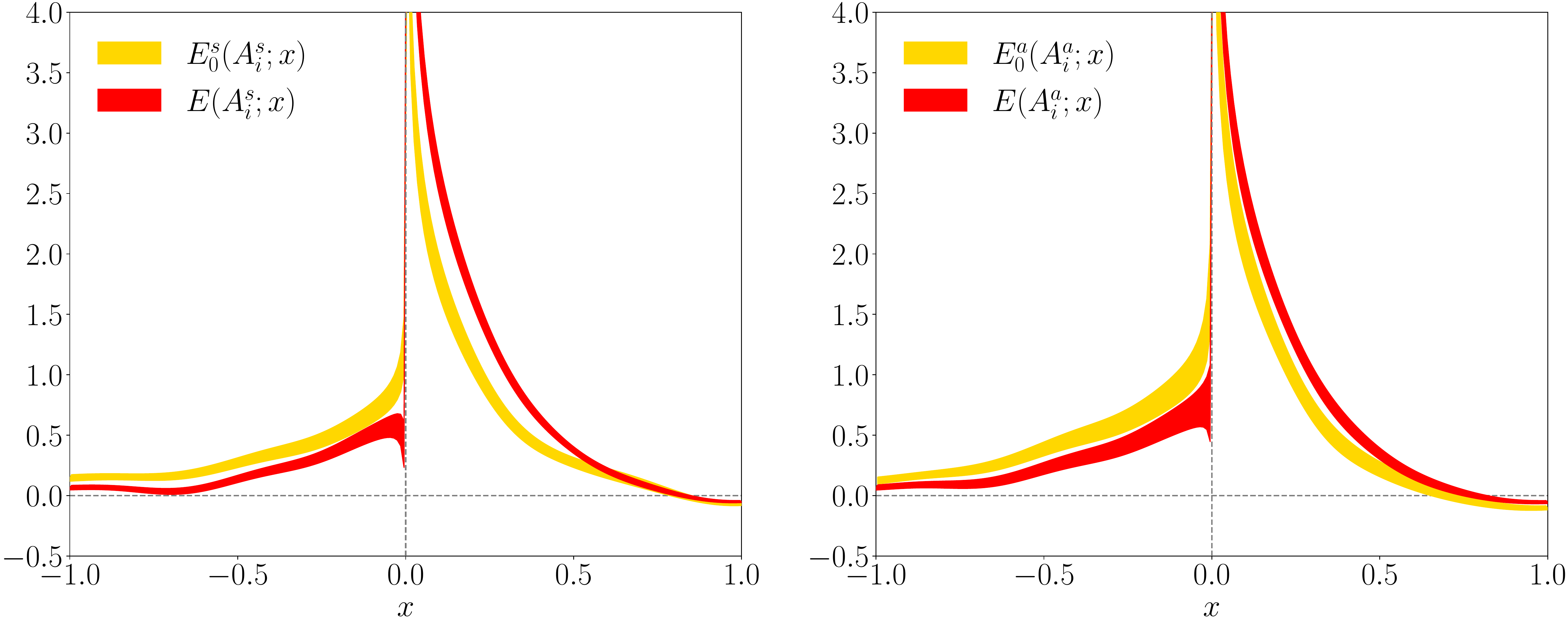}   
    \vspace*{-0.3cm}
    \caption{Comparison of light-cone $E_0$ and $E$ GPDs in the symmetric (left, $t=-0.69$ GeV$^2$) and asymmetric (right, $t=-0.64$ GeV$^2$) frame. Results are presented in the $\overline{\rm MS}$ scheme at 2 GeV.}
    \label{fig:E_calE}
\end{figure}

In Sec.~\ref{sec:res_quasi}, we compare $A_i$, ${\cal H}_0$, ${\cal E}_0$, ${\cal H}$, and ${\cal E}$, as extracted from different definitions and frames.
It is interesting to present such a comparison in the light-cone GPDs, which is summarized in Figs.~\ref{fig:H_calH} - \ref{fig:Es_Ea}. 
In particular, Fig.~\ref{fig:H_calH} demonstrates that the Lorentz invariant and non-invariant definitions for the $H$-GPD lead to the same light-cone GPDs; this holds for both the symmetric and the asymmetric frames. 
We remind the reader that the two definitions are different and such an agreement is not expected theoretically. The obtained distributions employing the $\gamma_0$ operator and symmetric frame definitions of $H_0^s$ and $E_0^s$ coincide with the results of Ref.~\cite{Alexandrou:2020zbe}.
Fig.~\ref{fig:E_calE} compares $E_0^{s/a}$ and $E$ as extracted in each frame.
Unlike the case of the $H$-GPD, here we find that the two definitions lead to GPDs that are of similar magnitude and shape, but are not in agreement for most of the $x$ region. 
Interestingly, the numerical difference between $E_0^s$ and $E$ is more prominent in the symmetric frame. 
The overall picture in comparing $H_0$ with $H$, and $E_0$ with $E$ is similar to the one for quasi-GPD matrix elements in coordinate space, as presented in Figs.~\ref{fig:FH_a} and \ref{fig:FE_a}.
Besides comparing the results from different definitions within the same frame, it is illustrative to investigate if the two frames for a given definition show any agreement. 
An agreement between different frames is expected theoretically only for the Lorentz-invariant definitions, $H$ and $E$. 
Indeed, Figs.~\ref{fig:Hs_Ha} - \ref{fig:Es_Ea} confirm perfect agreement between $H$ in the two frames; the same holds for $E$. Furthermore, such an agreement is also observed in the Lorentz non-invariant definitions $H$ and $E$.
The latter is in contrast to Figs.~\ref{fig:FH_b} and \ref{fig:FE_b}, where we observe ${\rm Re}[{\cal H}_0^s]\neq{\rm Re}[{\cal H}_0^a]$ and ${\rm Re}[{\cal E}_0^s]\neq{\rm Re}[{\cal E}_0^a]$ (all bare) at small and moderate $z$.
The agreement at the level of matched GPDs can be attributed to the enhanced errors after renormalization and non-trivial correlations between matrix elements at different $z$ entering the $x$-dependence reconstruction.
Nevertheless, the distributions employing the Lorentz-invariant definitions evince better agreement of their central values between the two frames.


\section{Summary and future prospects}
\label{sec:summary}
Lattice QCD calculations of $x$-dependent GPDs have so far been defined in the symmetric kinematic frame, which is, however, computationally very expensive to implement.
The main complication is that each value of the momentum transfer can only be accessed one at a time, as it appears in both the initial and final states. Furthermore, the symmetric frame requires two separate inversions for two separate momentum smearing at the source and sink. 
Hence, the current status of GPD calculations is still at the exploratory stage, with a very limited number of values of the momentum transfer, and, consequently, skewness.
In this work, we propose a way to resolve this issue via a new parametrization of off-forward matrix elements relevant to GPDs in terms of Lorentz-invariant amplitudes.
Specifically, the frame dependence of the matrix elements is absorbed in the kinematic factors of the parametrization, leaving the amplitudes frame independent.
Here, we present a lattice QCD calculation of off-forward matrix elements of the non-local vector operators coupled to momentum-boosted proton states. 
We observe numerically the frame independence of the amplitudes $A_i$, by comparing their estimates as extracted from the symmetric and the asymmetric frame chosen above.
Overall, we find very good agreement between the two frames.

A novel aspect of this work is that the applicability of the new parametrization in any frame has major implications in the reduction of the computational cost.
Take for instance the fixed-sink sequential inversion approach, and the asymmetric setup used in this work where the momentum transfer is assigned to the initial state, that is $\vec{p}_f=\vec{P}$,  $\vec{p}_i=\vec{P}-\vec{\Delta}$. 
The computational advantages are two-fold: (a) one can quadruple the number of measurements by adding all permutations of $\vec{\Delta}$ contributing to the same $t$; (b) several vectors $\vec{\Delta}$ may be obtained for a given $\vec{p}_f$ with an overhead of only the contraction cost. 
The asymmetric frame needs only one inversion corresponding to a single momentum smearing at the source/sink. So there is a factor of two gain in inversion even for a single momentum transfer.
Note that for both cases the momentum smearing is optimized for a selected $\vec{\Delta}$. 
However, we identify a broad range of values for the momentum smearing parameter in which the signal improvement is close to optimal.

The Lorentz-invariant amplitudes can be related to the quasi-GPDs of $H$ and $E$ in coordinate space.
The latter are not uniquely defined and here we focus on three options:
(a) definition in the symmetric frame via the $\gamma_0$ operator (${\cal H}_0^s,\,{\cal E}_0^s$);
(b) definition in the asymmetric frame via the $\gamma_0$ operator (${\cal H}_0^a,\,{\cal E}_0^a$);
(c) novel Lorentz-invariant definition ($\cal H,\,E$).
We emphasize once again that the three definitions are not equivalent; they differ by power corrections.
The first definition is of particular interest, as it has been used in previous lattice QCD calculations in the symmetric frame. 
It is still possible to extract the quasi-GPDs in the symmetric frame in a computationally less-costly way by using the Lorentz-invariant amplitudes $A_i$ obtained from the asymmetric frame. 
In such a case, the quasi-GPDs are defined at the value of $t$ corresponding to the asymmetric kinematic setup. 
The kinematic coefficients of Eqs.~\eqref{eq:FHs} - \eqref{eq:FEs} are obtained via a Lorentz transformation for a transverse boost. 
Another novel aspect in this work is the Lorentz-invariant definition of quasi-GPDs presented in Eqs.~\eqref{eq:FH_impr} - \eqref{eq:FE_impr}. 
Such a definition should be in agreement between the two frames.
We explored this direction in our lattice calculation and our findings confirm such frame independence (see, e.g., right panels of Fig.~\ref{fig:FH_b} and Fig.~\ref{fig:FE_b}).

The proposed parametrization and the introduction of the Lorentz-invariant amplitudes in not limited to the quantities presented in this work.
It is a powerful theoretical tool and has a broad range of interesting applicability that extends beyond leading twist. We believe that it has the potential to shape future calculations of GPDs from lattice QCD with a computational cost that is within reach.


\begin{acknowledgements}
This material is based upon work supported by the U.S. Department of Energy, Office of Science, Office of Nuclear Physics through Contract No.~DE-SC0012704, No.~DE-AC02-06CH11357 and within the framework of Scientific Discovery through Advance Computing (SciDAC) award Fundamental Nuclear Physics at the Exascale and Beyond (S.~B. and S.~M.).
K.~C.\ is supported by the National Science Centre (Poland) grants SONATA BIS no.\ 2016/22/E/ST2/00013 and OPUS no.\ 2021/43/B/ST2/00497. M.~C., J. D. and A.~S. acknowledge financial support by the U.S. Department of Energy, Office of Nuclear Physics, Early Career Award under Grant No.\ DE-SC0020405.
 J. D. also received support by the U.S. Department of Energy, Office of Science, Office of Nuclear Physics, within the framework of the TMD Topical Collaboration. 
The work of A.~M. has been supported by the National Science Foundation under grant number PHY-2110472, and also by the U.S. Department of Energy, Office of Science, Office of Nuclear Physics, within the framework of the TMD Topical Collaboration. 
F.~S.\ was funded by by the NSFC and the Deutsche Forschungsgemeinschaft (DFG, German Research Foundation) through the funds provided to the Sino-German Collaborative Research Center TRR110 “Symmetries and the Emergence of Structure in QCD” (NSFC Grant No. 12070131001, DFG Project-ID 196253076 - TRR 110). 
 YZ was partially supported by an LDRD initiative at Argonne National Laboratory under Project~No.~2020-0020.
Computations for this work were carried out in part on facilities of the USQCD Collaboration, which are funded by the Office of Science of the U.S. Department of Energy. 
This research used resources of the Oak Ridge Leadership Computing Facility, which is a
DOE Office of Science User Facility supported under Contract DE-AC05-00OR22725.
This research was supported in part by PLGrid Infrastructure (Prometheus supercomputer at AGH Cyfronet in Cracow).
Computations were also partially performed at the Poznan Supercomputing and Networking Center (Eagle supercomputer), the Interdisciplinary Centre for Mathematical and Computational Modelling of the Warsaw University (Okeanos supercomputer), and at the Academic Computer Centre in Gda\'nsk (Tryton supercomputer). The gauge configurations have been generated by the Extended Twisted Mass Collaboration on the KNL (A2) Partition of Marconi at CINECA, through the Prace project Pra13\_3304 ``SIMPHYS".
Inversions were performed using the DD-$\alpha$AMG solver~\cite{Frommer:2013fsa} with twisted mass support~\cite{Alexandrou:2016izb}. 
\end{acknowledgements}

\appendix

\section{Derivation of the Lorentz-invariant parameterization in terms of the $A_i$ amplitudes}
\label{s:ITDs_derivation}
In this Appendix, we outline the steps used to obtain the Lorentz-invariant amplitudes $A_i$'s that parametrize the position-space matrix element in the vector case. The starting point for a complete parametrization of the vector operator involves considering all possible Dirac bilinears consistent with the Parity constraint (see also next section where we discuss the implications of the Parity constraint), 
\begin{align}
F^{\mu} (z, P, \Delta) &= \bar{u}(p_f,\lambda') \bigg [ \dfrac{P^{\mu}}{m} A_1 + m z^\mu A_2 + \dfrac{\Delta^{\mu}}{m} A_3  + \dfrac{i\sigma^{\mu P}}{m} A_4 + m i\sigma^{\mu z} A_5 + \dfrac{i\sigma^{\mu \Delta}}{m} A_6 \nonumber \\
& \hspace{1.5cm} + \dfrac{P^{\mu} i\sigma^{z \Delta}}{m} A_7 + m z^{\mu} i\sigma^{z \Delta} A_8 + \dfrac{\Delta^{\mu} i\sigma^{z \Delta}}{m} A_9 + \dfrac{P^{\mu} i\sigma^{z P}}{m} A_{10} + m z^{\mu} i\sigma^{z P} A_{11} + \dfrac{\Delta^{\mu} i\sigma^{z P}}{m} A_{12} \nonumber \\
& \hspace{1.5cm} + \dfrac{P^{\mu} i\sigma^{P \Delta}}{m^3} A_{13} + \dfrac{z^{\mu} i\sigma^{P \Delta}}{m} A_{14} + \dfrac{\Delta^{\mu} i\sigma^{P \Delta}}{m^3} A_{15} \bigg ] u (p_i,\lambda) \, ,
\label{e:GTMD}
\end{align}
where, $A_{i} \equiv A_{i} (z \cdot P, z \cdot \Delta, t = \Delta^2, z^2)$. However, a further reduction in the number of structures is possible as shown in the following:
\begin{enumerate}
\item Using the Gordon Identity,
\begin{align}
    0 & = \bar{u} (p_f, \lambda') \bigg [ \dfrac{\Delta^{\mu}}{2m} + \dfrac{i \sigma^{\mu P}}{m}\bigg ] u (p_i, \lambda) \label{e:GI_GTMD}\\[0.2cm]
    \therefore i \sigma^{\mu P} & \rightarrow \Delta^\mu \, .
\end{align}
Hence, one can drop the term $ \propto A_4$.

\item After a multiplication by $\Delta_\mu$ in Eq.~(\ref{e:GI_GTMD}), we find,
\begin{align}
\dfrac{i \sigma^{P \Delta}}{2m} & \rightarrow \dfrac{\Delta^2}{2m} \nonumber \\[0.2cm]
 \therefore \dfrac{P^\mu i \sigma^{P \Delta}}{2m} & \rightarrow \dfrac{P^\mu\Delta^2}{2m} \, .
\end{align}
Hence, $A_{13} \rightarrow A_1$. Similarly,
\begin{align}
\therefore \dfrac{z^\mu i \sigma^{P \Delta}}{2m} & \rightarrow \dfrac{z^\mu\Delta^2}{2m} \qquad A_{14} \rightarrow A_2 \, , \\[0.2cm]
\therefore \dfrac{\Delta^\mu i \sigma^{P \Delta}}{2m} & \rightarrow \dfrac{\Delta^\mu\Delta^2}{2m} \qquad A_{15} \rightarrow A_3  \, .
\end{align}

\item After a multiplication by $z_\mu$ in Eq.~(\ref{e:GI_GTMD}), we find,
\begin{align}
\dfrac{i \sigma^{z P}}{2m} & \rightarrow - \dfrac{z \cdot \Delta}{2m} \nonumber \\[0.2cm]
 \therefore \dfrac{ P^\mu i \sigma^{z P}}{2m} & \rightarrow - \dfrac{ P^\mu (z \cdot \Delta)}{2m}  \, .
\end{align}
Hence, $A_{10} \rightarrow A_1$. Similarly,
\begin{align}
 \therefore \dfrac{ z^\mu i \sigma^{z P}}{2m} & \rightarrow - \dfrac{ z^\mu (z \cdot \Delta)}{2m}  \qquad A_{11} \rightarrow A_2 \, , \\[0.2cm]
 \therefore \dfrac{ \Delta^\mu i \sigma^{z P}}{2m} & \rightarrow - \dfrac{ \Delta^\mu (z \cdot \Delta)}{2m}  \qquad A_{12} \rightarrow A_3 \, .
\end{align}
\end{enumerate}
Therefore, in the end one is only left with 8 independent structures. 

\section{Symmetries of the amplitudes $A_i$ and checking the local case}
\label{sec:AppB}
\textbf{\textit{Symmetry of the $A_i$'s under Hermiticity}:}
The Hermitian conjugate of the correlator is, 
\begin{align}
(F^{\mu})^\dagger & = \bigg [ \langle p_f, \lambda' | \bar{q} (-\tfrac{z}{2}) \gamma^\mu q (\tfrac{z}{2}) | p_i, \lambda \rangle \bigg ]^{\dagger} = \bigg [ \langle p_f, \lambda' | q^{\dagger} (-\tfrac{z}{2}) \gamma^0 \gamma^\mu q (\tfrac{z}{2}) | p_i, \lambda \rangle \bigg ]^{\dagger} \nonumber \\[0.2cm]
& = \langle p_i, \lambda | q^{\dagger} (\tfrac{z}{2}) (\gamma^\mu)^\dagger \gamma^0 q (-\tfrac{z}{2}) | p_f, \lambda' \rangle = \langle p_i, \lambda | \bar{q} (\tfrac{z}{2}) \gamma^\mu q (\tfrac{z}{2}) | p_f, \lambda' \rangle \, .
\end{align}
For the Dirac matrices and the amplitudes $A_i$'s, this means,
\begin{align}
\bigg [ \bar{u}(p_f) P^\mu A_1 \, u(p_i)\bigg ]^\dagger & = \bar{u}(p_i) \bigg [ P^\mu A_1^{*} \bigg ] u (p_f) \, ,
\end{align}
and similarly for the structures associated with $A_{2/3}$, and
\begin{align}
\bigg [ \bar{u}(p_f) i\sigma^{\mu z} A_4 \, u(p_i)\bigg ]^\dagger & = - \bar{u}(p_i) \bigg [ i\sigma^{\mu z} A^*_4\bigg ] u (p_f) \, ,
\end{align}
and similarly for the structures associated with $A_{5/6/7/8}$. So, we get,
\begin{align}
(F^{\mu})^\dagger & =\bar{u}(p_i,\lambda) \bigg [ \dfrac{P^{\mu}}{m} A^*_1 + m z^{\mu} A^*_2 + \dfrac{\Delta^{\mu}}{m} A^*_3 - m i\sigma^{\mu z} A^*_4 - \dfrac{i\sigma^{\mu \Delta}}{m} A^*_5 \nonumber \\
& \hspace{5cm} - \dfrac{P^{\mu} i\sigma^{z \Delta}}{m} A^*_6 - m z^{\mu} i\sigma^{z \Delta} A^*_7  - \dfrac{\Delta^{\mu} i\sigma^{z \Delta}}{m} A^*_8 \bigg ] u(p_f, \lambda') \, .
\end{align}
This can be compared to $F^\mu$ after performing the transformation $p_i \leftrightarrow p_f$, which means $\Delta \rightarrow - \Delta$ and $z \rightarrow -z$. 
Then, the above equation is,
\begin{align}
(F^{\mu})^\dagger & = \bar{u}(p_f,\lambda') \bigg [ \dfrac{P^{\mu}}{m} A^*_1 - m z^{\mu} A^*_2 - \dfrac{\Delta^{\mu}}{m} A^*_3 + m i\sigma^{\mu z} A^*_4 + \dfrac{i\sigma^{\mu \Delta}}{m} A^*_5 \nonumber \\
& \hspace{5cm} - \dfrac{P^{\mu} i\sigma^{z \Delta}}{m} A^*_6 + m z^{\mu} i\sigma^{z \Delta} A^*_7  + \dfrac{\Delta^{\mu} i\sigma^{z \Delta}}{m} A^*_8 \bigg ] u(p_i, \lambda) \, .
\end{align}
The implies the following constraints on the $A_i$'s (restoring now the arguments of the $A_i$'s),
\begin{align}
A^*_1 (- z \cdot P, z \cdot \Delta, \Delta^2, z^2) & = A_1 (z \cdot P, z \cdot \Delta, \Delta^2, z^2) \, , \nonumber \\[0.2cm]
- A^*_2 (- z \cdot P, z \cdot \Delta, \Delta^2, z^2) & = A_2 (z \cdot P, z \cdot \Delta, \Delta^2, z^2) \, , \nonumber \\[0.2cm]
- A^*_3 (- z \cdot P, z \cdot \Delta, \Delta^2, z^2) & = A_3 (z \cdot P, z \cdot \Delta, \Delta^2, z^2) \, , \nonumber \\[0.2cm]
A^*_4 (- z \cdot P, z \cdot \Delta, \Delta^2, z^2) & = A_4 (z \cdot P, z \cdot \Delta, \Delta^2, z^2) \, , \nonumber \\[0.2cm]
A^*_5 (- z \cdot P, z \cdot \Delta, \Delta^2, z^2) & = A_5 (z \cdot P, z \cdot \Delta, \Delta^2, z^2) \, , \nonumber \\[0.2cm]
- A^*_6 (- z \cdot P, z \cdot \Delta, \Delta^2, z^2) & = A_6 (z \cdot P, z \cdot \Delta, \Delta^2, z^2) \, , \nonumber \\[0.2cm]
A^*_7 (- z \cdot P, z \cdot \Delta, \Delta^2, z^2) & = A_7 (z \cdot P, z \cdot \Delta, \Delta^2, z^2) \, , \nonumber \\[0.2cm]
A^*_8 (- z \cdot P, z \cdot \Delta, \Delta^2, z^2) & = A_8 (z \cdot P, z \cdot \Delta, \Delta^2, z^2) \, .
\label{e:hermicity_trans}
\end{align}

\textbf{\textit{Symmetry of the $A_i$'s under Parity}:}
We begin with,
\begin{align}
F_{ij} (p_i,p_f,z) & = \langle p_f | \bar{q}_j (-\tfrac{z}{2}) q_i (\tfrac{z}{2}) | p_i \rangle \, ,
\end{align}
where we have suppressed the helicity indices. Then, by using $U^{-1}_P U_P = 1$ we get,
\begin{align}
F_{ij} (p_i,p_f,z) & = \langle p_f| \bar{q}_j (-\tfrac{z}{2}) q_i (\tfrac{z}{2}) | p_i \rangle \nonumber \\[0.2cm]
& = \langle p_f| U^{-1}_P U_P \bar{q}_j (-\tfrac{z}{2}) U^{-1}_P U_P q_i (\tfrac{z}{2}) U^{-1}_P U_P| p_i \rangle \nonumber \\[0.2cm]
& = \langle \bar{p_f}| U_P \bar{q}_j (-\tfrac{z}{2}) U^{-1}_P U_P q_i (\tfrac{z}{2}) U^{-1}_P | \bar{p}_i \rangle \, ,
\end{align}
where we have made use of $U_P | {p}_i \rangle = | \bar{p}_i \rangle$ and $\langle p_f | U^{-1}_P = \langle \bar{p_f} |$, where $\bar{p} = (p^0, -\vec{p})$. Next, we use the parity transformation of the spinor $U_P q (z) U^{-1}_P = \gamma^0 q(\bar{z})$ or $U_P \bar{q} (z) U^{-1}_P = \bar{q}(\bar{z}) \gamma^0$ to arrive at,
\begin{align}
F_{ij} (p_i,p_f,z) & = \langle \bar{p_f}| U_P \bar{q}_j (-\tfrac{z}{2}) U^{-1}_P U_P q_i (\tfrac{z}{2}) U^{-1}_P | \bar{p}_i \rangle \nonumber \\[0.2cm]
& = \langle \bar{p_f}| \bar{q}_l (-\tfrac{z}{2}) \gamma^0_{lj} \gamma^{0}_{i i'} q_{i'} (\tfrac{z}{2}) | \bar{p}_i \rangle \nonumber \\[0.2cm]
& = \gamma^{0}_{i i'} \big ( \langle \bar{p_f}| \bar{q}_l (-\tfrac{z}{2})  q_{i'} (\tfrac{z}{2}) | \bar{p}_i \rangle \big ) \gamma^0_{lj} \, .
\end{align}
We therefore infer,
\begin{align}
F_{ij} (p_i,p_f,z) & = \gamma^{0}_{i i'} F_{i'l} (\bar{p}_i, \bar{p_f}, \bar{z}) \gamma^0_{lj} \, .
\label{e:Pcons}
\end{align}
Parity constraint also implies (see Ref.~\cite{Meissner:2009ww}),
\begin{align}
 \Gamma^\mu_V (P, \Delta, ...) & = \gamma^0 \Gamma^{\bar{\mu}}_V (\bar{P}, \bar{\Delta}, ...) \gamma^0 \, , \nonumber \\[0.2cm]
 \Gamma^{\mu \nu}_T (P, \Delta, ...) & = \gamma^0 \Gamma^{\bar{\mu} \bar{\nu}}_T (\bar{P}, \bar{\Delta}, ...) \gamma^0 \, ,
\end{align}
where, $P^{\bar{\mu}} = P_\mu$, etc. We can check the above for the structures appearing in our decomposition. For example,
\begin{align}
\gamma^ 0 \bar{P}^{\bar{\mu}} \gamma^0 A_1 (\bar{z} \cdot \bar{P}, \bar{z} \cdot \bar{\Delta}, \bar{\Delta}^2, \bar{z}^2) & = P^\mu A_1 (\bar{z} \cdot \bar{P}, \bar{z} \cdot \bar{\Delta}, \bar{\Delta}^2, \bar{z}^2) \, ,
\end{align}
and similarly for the structures associated with $A_{2/3}$, and
\begin{align}
\gamma^0 i \sigma^{\bar{\mu} \bar{z}} \gamma^0 A_4 (\bar{z} \cdot \bar{P}, \bar{z} \cdot \bar{\Delta}, \bar{\Delta}^2, \bar{z}^2) & = \gamma^0 i \sigma^{\bar{\mu} \bar{\nu}} \bar{z}_{\bar{\nu}} \gamma^0 A_4 (\bar{z} \cdot \bar{P}, \bar{z} \cdot \bar{\Delta}, \bar{\Delta}^2, \bar{z}^2) = i \sigma^{\mu z} A_4 (\bar{z} \cdot \bar{P}, \bar{z} \cdot \bar{\Delta}, \bar{\Delta}^2, \bar{z}^2)  \, ,
\end{align}
and similarly for the structures associated with $A_{5/6/7/8}$. (Note we used $\gamma^0 \sigma_{\mu \nu} \gamma^0 = \sigma^{\mu \nu}$.) So all the structures are consistent with parity.  Hence, Eq.~(\ref{e:Pcons}) implies,
\begin{align}
A_{i} (z \cdot P, z \cdot \Delta, \Delta^2, z^2) \xrightarrow{\text{P}} A_{i} (\bar{z} \cdot \bar{P}, \bar{z} \cdot \bar{\Delta}, \bar{\Delta}^2, \bar{z}^2) \, .
\label{e:parity_trans}
\end{align}

\textbf{\textit{Symmetry of the $A_i$'s under Time-reversal}:}
Time-reversal operator is anti-unitary meaning, 
\begin{align}
\langle x | U^{\dagger}_T U_T | y \rangle = \langle x | y \rangle ^* \, .
\label{e:time_reversal}
\end{align}
We begin with,
\begin{align}
F_{ij} (p_i,p_f,z) & = \langle p_f | \bar{q}_j (-\tfrac{z}{2}) q_i (\tfrac{z}{2}) | p_i \rangle \, ,
\end{align}
where we have suppressed the helicity indices. Then, by using Eq.~(\ref{e:time_reversal}) we get,
\begin{align}
F^*_{ij} (p_i,p_f,z) & = \langle p_f| \bar{q}_j (-\tfrac{z}{2}) q_i (\tfrac{z}{2}) | p_i \rangle ^* \nonumber \\[0.2cm]
& = \langle p_f| \bar{q}_j (-\tfrac{z}{2}) U^{\dagger}_T U_T q_i (\tfrac{z}{2}) | p_i \rangle \nonumber \\[0.2cm]
& = \langle {p_f}| U^{\dagger}_T U_T \bar{q}_j (-\tfrac{z}{2}) U^{\dagger}_T U_T q_i (\tfrac{z}{2}) U^{\dagger}_T U_T | {p}_i \rangle \, ,
\end{align}
where, in going from the second to the third line we have made use of the anti-unitary nature of Time-reversal operator twice (Eq.~(\ref{e:time_reversal})).
Next, we use $U_T | {p}_i \rangle = | \bar{p}_i \rangle$ and $\langle p_f | U^{\dagger}_T = \langle \bar{p_f} |$, where $\bar{p} = (p^0, -\vec{p})$,
\begin{align}
F^*_{ij} (p_i,p_f,z) & = \langle {p_f}| U^{\dagger}_T U_T \bar{q}_j (-\tfrac{z}{2}) U^{\dagger}_T U_T q_i (\tfrac{z}{2}) U^{\dagger}_T U_T | {p}_i \rangle \nonumber \\[0.2cm]
& = \langle \bar{p_f}| U_T \bar{q}_j (-\tfrac{z}{2}) U^{\dagger}_T U_T q_i (\tfrac{z}{2}) U^{\dagger}_T | \bar{p}_i \rangle
\end{align}
Next, we use the Time-reversal transformation of the spinor $U_T q (z) U^{\dagger}_T = (i \gamma^1 \gamma^3) q(-\bar{z})$ or $U_T \bar{q} (z) U^{\dagger}_T = (i \gamma^1 \gamma^3)\bar{q}(-\bar{z})$ to arrive at,
\begin{align}
F^*_{ij} (p_i,p_f,z) & = \langle \bar{p_f}| U_T \bar{q}_j (-\tfrac{z}{2}) U^{\dagger}_T U_T q_i (\tfrac{z}{2}) U^{\dagger}_T | \bar{p}_i \rangle \nonumber \\[0.2cm]
& = \langle \bar{p_f}| \bar{q}_l (\tfrac{\bar{z}}{2}) (i \gamma^1 \gamma^3)_{lj} (i \gamma^1 \gamma^3)_{i i'} q_{i'} (-\tfrac{\bar{z}}{2}) | \bar{p}_i \rangle \nonumber \\[0.2cm]
& = (i \gamma^1 \gamma^3)_{i i'} \big ( \langle \bar{p_f}| \bar{q}_l (\tfrac{\bar{z}}{2})  q_{i'} (-\tfrac{\bar{z}}{2}) | \bar{p}_i \rangle \big ) (i \gamma^1 \gamma^3)_{lj} \nonumber \\[0.2cm]
\therefore F^*_{ij} (p_i,p_f,z) & = (i \gamma^1 \gamma^3)_{i i'} F_{i'l} (\bar{p}_i, \bar{p_f}, -\bar{z}) (i \gamma^1 \gamma^3)_{lj} \, .
\end{align}
We therefore infer,
\begin{align}
\therefore F^*_{ij} (p_i,p_f,z) & = (i \gamma^1 \gamma^3)_{i i'} F_{i'l} (\bar{p}_i, \bar{p_f}, -\bar{z}) (i \gamma^1 \gamma^3)_{lj} \, .
\label{e:Tcons}
\end{align}
It is straightforward to check that all the Dirac structures appearing in our decomposition are consistent with the Time-reversal constraint. For example,
\begin{align}
(i \gamma^1 \gamma^3) 1 (i \gamma^1 \gamma^3) & \rightarrow 1 \, , \nonumber \\[0.2cm]
(i \gamma^1 \gamma^3) (i \sigma^{\bar{\mu} \bar{\nu}})^* (i \gamma^1 \gamma^3) &\rightarrow i \sigma^{\mu \nu} \, .
\end{align}
Now, keeping in mind that $z \rightarrow - \bar{z}$ under Time-reversal, we can quickly infer from Eq.~(\ref{e:Tcons}),
\begin{align}
A^*_1 (- \bar{z} \cdot \bar{P}, -\bar{z} \cdot \bar{\Delta}, \bar{\Delta}^2, \bar{z}^2) & = A_1 (z \cdot P, z \cdot \Delta, \Delta^2, z^2) \, , \nonumber \\[0.2cm]
- A^*_2 (- \bar{z} \cdot \bar{P}, -\bar{z} \cdot \bar{\Delta}, \bar{\Delta}^2, \bar{z}^2) & = A_2 (z \cdot P, z \cdot \Delta, \Delta^2, z^2) \, , \nonumber \\[0.2cm]
A^*_3 (- \bar{z} \cdot \bar{P}, -\bar{z} \cdot \bar{\Delta}, \bar{\Delta}^2, \bar{z}^2) & = A_3 (z \cdot P, z \cdot \Delta, \Delta^2, z^2) \, , \nonumber \\[0.2cm]
- A^*_4 (- \bar{z} \cdot \bar{P}, -\bar{z} \cdot \bar{\Delta}, \bar{\Delta}^2, \bar{z}^2) & = A_4 (z \cdot P, z \cdot \Delta, \Delta^2, z^2) \, , \nonumber \\[0.2cm]
A^*_5 (- \bar{z} \cdot \bar{P}, -\bar{z} \cdot \bar{\Delta}, \bar{\Delta}^2, \bar{z}^2) & = A_5 (z \cdot P, z \cdot \Delta, \Delta^2, z^2) \, , \nonumber \\[0.2cm]
- A^*_6 (- \bar{z} \cdot \bar{P}, -\bar{z} \cdot \bar{\Delta}, \bar{\Delta}^2, \bar{z}^2) & = A_6 (z \cdot P, z \cdot \Delta, \Delta^2, z^2) \, , \nonumber \\[0.2cm]
A^*_7 (- \bar{z} \cdot \bar{P}, -\bar{z} \cdot \bar{\Delta}, \bar{\Delta}^2, \bar{z}^2) & = A_7 (z \cdot P, z \cdot \Delta, \Delta^2, z^2) \, , \nonumber \\[0.2cm]
- A^*_8 (- \bar{z} \cdot \bar{P}, -\bar{z} \cdot \bar{\Delta}, \bar{\Delta}^2, \bar{z}^2) & = A_8 (z \cdot P, z \cdot \Delta, \Delta^2, z^2) \, .
\label{e:time_trans}
\end{align}
The above expressions remain valid if Parity is applied along with Time-reversal.

{\textbf{\textit{Symmetry of the $A_i$'s under Hermiticity and Time-reversal}}}: The symmetry property of the $A_i$'s under $\Delta \rightarrow - \Delta$ can be understood through their (simultaneous) behavior under Hermiticity and time-reversal transformations,
\begin{align}
A_1 (\bar{z} \cdot \bar{P}, -\bar{z} \cdot \bar{\Delta}, \bar{\Delta}^2, \bar{z}^2) & = A_1 (z \cdot P, z \cdot \Delta, \Delta^2, z^2) \, , \nonumber \\[0.2cm]
A_2 (\bar{z} \cdot \bar{P}, -\bar{z} \cdot \bar{\Delta}, \bar{\Delta}^2, \bar{z}^2) & = A_2 (z \cdot P, z \cdot \Delta, \Delta^2, z^2) \, , \nonumber \\[0.2cm]
- A_3 (\bar{z} \cdot \bar{P}, -\bar{z} \cdot \bar{\Delta}, \bar{\Delta}^2, \bar{z}^2) & = A_3 (z \cdot P, z \cdot \Delta, \Delta^2, z^2) \, , \nonumber \\[0.2cm]
- A_4 (\bar{z} \cdot \bar{P}, -\bar{z} \cdot \bar{\Delta}, \bar{\Delta}^2, \bar{z}^2) & = A_4 (z \cdot P, z \cdot \Delta, \Delta^2, z^2) \, , \nonumber \\[0.2cm]
A_5 (\bar{z} \cdot \bar{P}, -\bar{z} \cdot \bar{\Delta}, \bar{\Delta}^2, \bar{z}^2) & = A_5 (z \cdot P, z \cdot \Delta, \Delta^2, z^2) \, , \nonumber \\[0.2cm]
A_6 (\bar{z} \cdot \bar{P}, -\bar{z} \cdot \bar{\Delta}, \bar{\Delta}^2, \bar{z}^2) & = A_6 (z \cdot P, z \cdot \Delta, \Delta^2, z^2) \, , \nonumber \\[0.2cm]
A_7 (\bar{z} \cdot \bar{P}, -\bar{z} \cdot \bar{\Delta}, \bar{\Delta}^2, \bar{z}^2) & = A_7 (z \cdot P, z \cdot \Delta, \Delta^2, z^2) \, , \nonumber \\[0.2cm]
- A_8 (\bar{z} \cdot \bar{P}, -\bar{z} \cdot \bar{\Delta}, \bar{\Delta}^2, \bar{z}^2) & = A_8 (z \cdot P, z \cdot \Delta, \Delta^2, z^2) \, .
\label{e:HTR_cons}
\end{align}

{\textbf{\textit{Consistency in the local case $\boldsymbol{z = 0}$:}}} It is interesting to check if our decomposition is consistent with the local vector current. We recall that the local vector operator that defines the Dirac ($F_{1}$) and the Pauli ($F_{2}$) form factors are,
\begin{align}
\langle p_f,\lambda '| \bar{q}(0)\gamma^{\mu} q(0) |p_i,\lambda \rangle &= \bar{u} (p_f,\lambda ') \bigg [ \gamma^{\mu} F^{q}_{1} (t)  + \dfrac{i \sigma^{\mu \alpha}\Delta_{\alpha}}{2m} F^{q}_{2}(t) \bigg ] u (p_i,\lambda ) \, .
\end{align}
So there are two independent Form Factors. The Form Factors are real functions. On the other hand, our decomposition in the local limit reduces to
\begin{align}
F^{\mu} \big |_{z=0} & = \bar{u}(p_f,\lambda') \bigg [ \dfrac{P^{\mu}}{m} A_1 (\Delta^2)  +  \dfrac{\Delta^{\mu}}{m} A_3 (\Delta^2) + \dfrac{i\sigma^{\mu \Delta}}{m} A_5 (\Delta^2) \bigg ] u(p_i, \lambda) \,.
\end{align}
Now, note that for us the $A_i$'s are ``complex amplitudes", but for consistency with the local vector operator we must be able to show that the $A_i$'s surviving are real (that is, either the real or its imaginary part survives). Recall that Hermiticity leads to,
\begin{align}
A^*_1 & = A_1 \, , \qquad
- A^*_3  = A_3 \, , \qquad
A^*_5  = A_5 \, .
\end{align}
Then these constraints on these $A_i$'s at $z=0$ leads to,
\begin{align}
\textrm{Im.} (A_1)  = 0 \, , \qquad \textrm{Re.} (A_3)  = 0 \, , \qquad \textrm{Im.} (A_5)  = 0 \, .
\label{e:c1}
\end{align}
This doesn't help fully because we are still left with three $A_i$'s. We must be able to show that we are left with only two $A_i$'s. For this, we turn to Time-reversal transformation to check if it poses any additional constraint on the $A_i$'s. Recall that Time-reversal leads to,
\begin{align}
A^*_1  & = A_1  \, , \qquad
A^*_3   = A_3 \, , \qquad
A^*_5   = A_5  \, .
\end{align}
Then these constraints on these $A_i$'s at $z=0$ leads to,
\begin{align}
\textrm{Im.} (A_1)  = 0 \, , \qquad \textrm{Im.} (A_3)  = 0 \, , \qquad \textrm{Im.} (A_5) = 0 \, .
\label{e:c2}
\end{align}
Combining Eqs.~(\ref{e:c1}) and~(\ref{e:c2}) we conclude, 
\begin{align}
\textrm{Re.} (A_3) & = 0 \, , \qquad \textrm{Im.} (A_3) = 0 \nonumber \\[0.2cm]
\therefore A_3 & = 0 \, .
\end{align}
Therefore Eqs.~(\ref{e:c1}) and~(\ref{e:c2}) tell us that the only contribution at $z=0$ comes from,
\begin{align}
\textrm{Re.} (A_1) \neq  0 \, , \qquad \textrm{Re.} (A_5)\neq 0 \, .
\end{align}
Hence our decomposition is consistent with the local vector current.

\section{Euclidean-space expressions for the traces for any skewness}
\label{app:C}
Here we provide the general expressions for the traces for any frame and for any skewness.
\begin{itemize}
\item {\textbf{\textit{$F_{0}$ with unpolarized projector}}}:
\begin{align}
\Pi_{0} (\Gamma_0) & = \dfrac{-iK}{4m^3} P_0 \bigg ( - p_i \cdot p_f + m^2 -i m (E_i + E_f) \bigg ) A_1 - \dfrac{iK}{4m^3} \Delta_0 \bigg ( - p_i \cdot p_f + m^2 -i m (E_i + E_f) \bigg ) A_3 \nonumber \\
& - \dfrac{iK}{4m^3} \bigg ( m^2 (m - i E_f) p_i \cdot z - m^2 (m-i E_i) p_f \cdot z \bigg ) A_4 \nonumber \\
& - \dfrac{i K}{4m^3} \bigg ( (m - i E_f) p_i \cdot \Delta - (m-i E_i) p_f \cdot \Delta + \Delta_0 m (E_f - E_i) \bigg ) A_5 \nonumber \\
& - \dfrac{K}{4m^3} P_0 \bigg ( i (p_f \cdot \Delta) (p_i \cdot z) - i (p_i \cdot \Delta) (p_f \cdot z)  + \Delta_0 m \big ( (p_f \cdot z) - p_i \cdot z\big ) \bigg ) A_6 \nonumber \\
& - \dfrac{K}{4m^3} \Delta_0 \bigg ( i (p_f \cdot \Delta) (p_i \cdot z) - i (p_i \cdot \Delta) (p_f \cdot z)  + \Delta_0 m \big ( (p_f \cdot z) - p_i \cdot z\big ) \bigg ) A_8 \, ,
\end{align}
where the kinematic factor $K$ is defined as,
\begin{align}
K & = \dfrac{2 M^2}{\sqrt{E_f E_i (E_f + M) (E_i + M)}} \, .
\end{align}

\item {\textbf{\textit{$F_{0}$ with polarized projector}}}:
\begin{align}
\Pi_{0} (\Gamma_j) & = \dfrac{i K}{4m^3} P_{0} \epsilon_{j p_f p_i 0} A_1 + \dfrac{i K}{4m^3} \Delta_{0} \epsilon_{j p_f p_i 0} A_3 \nonumber \\
& + \dfrac{i K}{4m^3} \bigg (m^2(m-2iE_f) \epsilon_{j p_i z 0} + m^3 \epsilon_{j p_f z 0} + m^2i\epsilon_{j p_f p_i z} \bigg ) A_4 \nonumber \\
& + \dfrac{i K}{4m^3} \bigg ( (m-2i E_f) \epsilon_{j p_i \Delta 0} + m \epsilon_{j p_f \Delta 0} + i \epsilon_{j p_f p_i \Delta} -i \Delta_0 \epsilon_{j p_f p_i 0}\bigg ) A_5 \nonumber \\
& - \dfrac{K}{4m^3} i P_0 \bigg ( m^2 \epsilon_{j z \Delta 0} + (-E_f-i m) \epsilon_{j p_i z \Delta} - i m \epsilon_{j p_f z \Delta} - p_{j,f} \boldsymbol{\epsilon_{p_i z \Delta}}  + (p_i \cdot z) \epsilon_{j p_f \Delta 0} - (p_i \cdot \Delta) \epsilon_{j p_f z 0}\bigg ) A_6 \nonumber \\
& - \dfrac{K}{4m^3}  i \Delta_0 \bigg ( m^2 \epsilon_{j z \Delta 0} + (-E_f-i m) \epsilon_{j p_i z \Delta} - i m \epsilon_{j p_f z \Delta} - p_{j,f} \boldsymbol{\epsilon_{p_i z \Delta}} + (p_i \cdot z) \epsilon_{j p_f \Delta 0} - (p_i \cdot \Delta) \epsilon_{j p_f z 0}\bigg ) A_8  \, ,
\end{align}
where $\epsilon_{a b c d}$ is a 4-dimensional Levi-Civita tensor.

\item \textbf{\textit{$F_{i}$ with unpolarized projector}}:
\begin{align}
\Pi_{i} (\Gamma_0) & = \dfrac{K}{4m^3} P_{i}  \bigg ( -p_i \cdot p_f + m^2 - i m (E_i + E_f) \bigg ) A_1 + \dfrac{K}{4m^3} \Delta_{i} \bigg ( - p_i \cdot p_f + m^2 - i m (E_i + E_f) \bigg ) A_3 \nonumber \\
& + \dfrac{K}{4m^3} \bigg ( m^2i p_{i,i} (p_f \cdot z) - m^2i p_{i,f} (p_i \cdot z) \bigg ) A_4 + \dfrac{K}{4m^3} \bigg ( i p_{i,i} (p_f \cdot \Delta + i \Delta_0 m) + p_{i,f} ( - ip_i \cdot \Delta +  \Delta_0 m) \bigg ) A_5 \nonumber \\
& - \dfrac{i K}{4m^3} P_{i} \bigg ( i (p_f \cdot \Delta) (p_i \cdot z) - i (p_i \cdot \Delta) (p_f \cdot z) +  \Delta_0 m \big ( (p_i \cdot z) - (p_f \cdot z) \big ) \bigg ) A_6 \nonumber \\
& - \dfrac{K}{4m^3} \Delta_{i} \bigg ( i (p_f \cdot \Delta) (p_i \cdot z) - i (p_i \cdot \Delta) (p_f \cdot z) +  \Delta_0 m \big ( (p_i \cdot z) - (p_f \cdot z) \big ) \bigg ) A_8 \, .
\end{align}

\item \textbf{\textit{$F_{i}$ with polarized projector}}:
\begin{align}
\Pi_{i} (\Gamma_j) & = - \dfrac{K}{4m^3} P_{i} \epsilon_{j p_f p_i 0} A_1 - \dfrac{K}{4m^3} \Delta_{i} \epsilon_{j p_f p_i 0} A_3  \nonumber \\
& - \dfrac{i K}{4m^3} \bigg ( m^4 \epsilon_{ijz0} + m^2(E_f + i m) \epsilon_{ijp_i z} + i m^3 \epsilon_{ij p_f z} - m^2 p_{i,i} \epsilon_{j p_f z 0} - m^2 p_{j,f} \epsilon_{i p_i z 0} + m^2 (p_i \cdot z) \epsilon_{ij p_f 0} \bigg ) A_4 \nonumber \\
& - \dfrac{i K}{4m^3} \bigg ( m^2 \epsilon_{ij\Delta 0} + ( E_f + i m) \epsilon_{ijp_i \Delta} + i m \epsilon_{ij p_f \Delta} - p_{i,i} \epsilon_{j p_f \Delta 0} - p_{j,f} \epsilon_{i p_i \Delta 0} +(p_i \cdot \Delta) \epsilon_{ij p_f 0} \bigg ) A_5 \nonumber \\
& + \dfrac{K}{4m^3}  P_{i} \bigg ( m^2 \epsilon_{j z \Delta 0} + (-E_f - i m) \epsilon_{j p_i z \Delta} - i m \epsilon_{j p_f z \Delta} - p_{j,f} \boldsymbol{\epsilon_{p_i z \Delta}}  + (p_i \cdot z) \epsilon_{j p_f \Delta 0} -(p_i \cdot \Delta) \epsilon_{j p_f z 0}\bigg ) A_6 \nonumber \\
& + \dfrac{K}{4m^3}  i \Delta_i \bigg ( m^2 \epsilon_{j z \Delta 0} + (-E_f - i m) \epsilon_{j p_i z \Delta} - i m \epsilon_{j p_f z \Delta} - p_{j,f} \boldsymbol{\epsilon_{p_i z \Delta}}  + (p_i \cdot z) \epsilon_{j p_f \Delta 0} -(p_i \cdot \Delta) \epsilon_{j p_f z 0}\bigg ) A_8 \, .
\end{align}

\item {\textbf{\textit{$F_{3}$ with unpolarized projector}}}:
\begin{align}
\Pi_{3} (\Gamma_0) & = \dfrac{K}{4m^3} P_3 \bigg ( - p_i \cdot p_f + m^2 - i m (E_i + E_f) \bigg ) A_1 + \dfrac{K}{4m^3} z_3 \bigg ( - m^2 p_i \cdot p_f + m^4 - i m^3 (E_i + E_f) \bigg ) A_2 \nonumber \\
& + \dfrac{K}{4m^3} \Delta_3 \bigg ( - p_i \cdot p_f + m^2 - i m (E_i + E_f) \bigg ) A_3 + \dfrac{K}{4m^3} \bigg ( i p_{3,i} (p_f \cdot \Delta + i \Delta_0 m) + p_{3,f} (- i p_i \cdot \Delta + \Delta_0 m) \bigg ) A_5 \nonumber \\
& - \dfrac{iK}{4m^3} P_3 \bigg ( i (p_f \cdot \Delta) (p_i \cdot z) - i (p_i \cdot \Delta) (p_f \cdot z) +  \Delta_0 m\big ( (p_i \cdot z) - (p_f \cdot z) \big ) \bigg ) A_6 \nonumber \\
& - \dfrac{iK}{4m^3} z_3 \bigg ( m^2 i (p_f \cdot \Delta) (p_i \cdot z) - m^2 i (p_i \cdot \Delta) (p_f \cdot z) +  \Delta_0 m^3 \big ( (p_i \cdot z) - (p_f \cdot z) \big ) \bigg ) A_7 \nonumber \\
& - \dfrac{iK}{4m^3} \Delta_3 \bigg ( i (p_f \cdot \Delta) (p_i \cdot z) - i (p_i \cdot \Delta) (p_f \cdot z) +  \Delta_0 m\big ( (p_i \cdot z) - (p_f \cdot z) \big ) \bigg ) A_8 \, .
\end{align}

\item \textbf{\textit{$F_{3}$ with polarized projector}}:
\begin{align}
\Pi_{3} (\Gamma_j) & = - \dfrac{K}{4m^3} P_{3} \epsilon_{j p_f p_i 0} A_1  - \dfrac{K}{4m^3} m^2 z_3 \epsilon_{j p_f p_i 0} A_2 - \dfrac{ K}{4m^3} \Delta_{3} \epsilon_{j p_f p_i 0}  A_3 \nonumber \\
& - \dfrac{i K}{4m^3} \bigg ( m^2 \epsilon_{3j\Delta 0} + ( E_f + i m) \epsilon_{3jp_i \Delta} + m \epsilon_{3j p_f \Delta} - p_{3,i} \epsilon_{j p_f \Delta 0} - p_{j,f} \epsilon_{3 p_i \Delta 0} + (p_i \cdot \Delta) \epsilon_{3j p_f 0} \bigg ) A_5 \nonumber \\
& - \dfrac{K}{4m^3}  i P_3 \bigg ( m^2 \epsilon_{j z \Delta 0} + (-E_f - i m) \epsilon_{j p_i z \Delta} - i m \epsilon_{j p_f z \Delta} - p_{j,f} \boldsymbol{\epsilon_{p_i z \Delta}}  + (p_i \cdot z) \epsilon_{j p_f \Delta 0} -(p_i \cdot \Delta) \epsilon_{j p_f z 0}\bigg ) A_6 \nonumber \\
& - \dfrac{K}{4m^3} i z_3 \bigg ( m^4 \epsilon_{j z \Delta 0} + m^2 (-E_f - i m) \epsilon_{j p_i z \Delta} - i m^3 \epsilon_{j p_f z \Delta} - m^2 p_{j,f} \boldsymbol{\epsilon_{p_i z \Delta}}  \nonumber \\
& \hspace{8cm} + m^2 (p_i \cdot z) \epsilon_{j p_f \Delta 0} - m^2(p_i \cdot \Delta) \epsilon_{j p_f z 0}\bigg )  A_7 \nonumber \\
& - \dfrac{K}{4m^3}  i \Delta_3 \bigg ( m^2 \epsilon_{j z \Delta 0} + (-E_f - i m) \epsilon_{j p_i z \Delta} - i m \epsilon_{j p_f z \Delta} - p_{j,f} \boldsymbol{\epsilon_{p_i z \Delta}}  + (p_i \cdot z) \epsilon_{j p_f \Delta 0} -(p_i \cdot \Delta) \epsilon_{j p_f z 0}\bigg ) A_8 \, .
\end{align}
\end{itemize}

\bibliography{references.bib}

\end{document}